\documentclass[12pt]{article}
\usepackage{graphicx}
\usepackage{hep}
\usepackage{amsmath}
\usepackage{amssymb}
\usepackage{cite}
\usepackage{a4wide}

\newcommand{\sfL}{\ensuremath{\tilde{f}_\textrm{L}}}
\newcommand{\sfR}{\ensuremath{\tilde{f}_\textrm{R}}}

\newcommand{\msq}{\ensuremath{m_{\squark}}}
\newcommand{\MsfL}{\ensuremath{M_{\sfL}}}
\newcommand{\MsfR}{\ensuremath{M_{\sfR}}}

\newcommand{\msusy}{\ensuremath{M_{\mathrm{SUSY}}}}

\newcommand{\GeV}{\mbox{\,GeV}}
\newcommand{\TeV}{\mbox{\,TeV}}

\newcommand{\sbottom}{\ensuremath{\tilde{b}}}

\newcommand{\squark}{\ensuremath{\tilde{q}}}
\newcommand{\stopp}{\ensuremath{\tilde{t}}}

\newcommand{\al}{\ensuremath{\alpha_{s}}}

\newcommand{\hp}{\ensuremath{H_+}}
\newcommand{\hm}{\ensuremath{H_-}}

\def\Title#1{\begin{center} {\Large\bf #1 } \end{center}}
\def\Author#1{\begin{center}{ \sc #1} \end{center}}
\def\Address#1{\begin{center}{ \it #1} \end{center}}

\newcommand\pubblock{\rightline{\begin{tabular}{l} \pubnumber\\
        \pubdate\\ \hepnumber \end{tabular}}}
\newcommand\pubnumber{ECM-UB-PF 08/22}
\newcommand\pubdate{December 2008}
\newcommand\hepnumber{arXiv:0812.1114 [hep-ph]}

\newcommand{\SUSY}{Nilles:1984ex,Haber:1985rc,Lahanas:1987uc,Ferrara87}
\newcommand{\TESLA}{Aguilar-Saavedra:2001rg}
\newcommand{\LHC}{ATLAS,Ball:2007zza}
\newcommand{\FeynArts}{Kublbeck:1990xc,vanOldenborgh:1989wn,Hahn:1998yk,Hahn:2000kx,Hahn:2001rv,FAFCuser}

\newcommand{\previous}{Kraml:1996kz,Djouadi:1996wt,Guasch:2002ez,Guasch:2002qa}

\newcommand{\vacuum}{Frere:1983ag,Claudson:1983et,Kounnas:1983td,Gunion:1987qv}

\begin{document}

\pubblock

\vfill
\def\thefootnote{\fnsymbol{footnote}}
\Title{Effective description of squark interactions}
\Author{Jaume Guasch$^{a,b}$, Siannah Pe{\~n}aranda$^c$, Ra{\"u}l S{\'a}nchez-Florit$^{d,b}$}
\Address{\textsl{$^a$ Departament de F{\'\i}sica Fonamental,\\ Universitat de Barcelona,
    Diagonal 647, E-08028 Barcelona, Catalonia, Spain\\
$^b$ Institut de Ci{\`e}ncies del Cosmos de la Universitat de Barcelona,\\ 
    Diagonal 647, E-08028 Barcelona, Catalonia, Spain\\
$^c$        Departamento de F{\'\i}sica Te{\'o}rica,
Facultad de Ciencias,\\ Universidad de Zaragoza,
            E-50009 Zaragoza, Spain\\
$^d$ Departament d'Estructura i Constituents de la Mat{\`e}ria,\\ Universitat de Barcelona,
    Diagonal 647, E-08028 Barcelona, Catalonia, Spain\\
E-mails: jaume.guasch@ub.edu, siannah@unizar.es, florit@ffn.ub.es}
} \vspace{1cm}

\begin{abstract}
We propose an effective description of squark interactions with
charginos/neutra\-linos. We recompute the strong corrections to 
squark partial decay widths, 
and compare the full one-loop computation with the
effective description. The effective description includes the effective
Yukawa couplings, and another logarithmic term which encodes the
supersymmetry-breaking. The proposed effective couplings reproduce
correctly the radiative-corrected partial decay widths of the squark decays
into charginos and neutralinos in all relevant regions of the parameter
space.
\end{abstract}
\vfill

\def\thefootnote{\arabic{footnote}}
\newpage

\section{Introduction}
The Standard Model (SM) of the strong and electroweak interactions is the
present paradigm of particle physics. Its validity has been tested to a
level better than one per mille at particle
accelerators~\cite{Amsler:2008zz}. 
Nevertheless, there are
arguments against the SM being the fundamental model of
particle interactions~\cite{Haber:1993kz}, giving rise to the
investigation of competing alternative or extended models, which
can be tested at high-energy colliders, such as the Large Hadron Collider
(LHC)~\cite{\LHC}, or a $500-1000\GeV$, $e^+e^-$ International Linear
Collider (ILC)~\cite{\TESLA,Weiglein:2004hn}. 
One of the most promising possibilities for physics beyond
the SM is the incorporation of Supersymmetry (SUSY), which leads to
a  renormalizable field theory with precisely calculable
predictions to be tested in present and future experiments. 
The simplest supersymmetric extension of the SM is the Minimal Supersymmetric
Standard Model (MSSM)~\cite{\SUSY}. Among the most important
phenomenological consequences of SUSY models, is the prediction of new
particles. There is much excitement for the possibility of discovering
these new particles at the recently built LHC~\cite{deJong:2008uj,Ellis:2008di}, 
and their properties
will need to be precisely measured to confirm (or refute) that they belong to
a SUSY model. This last effort might be better suited for the
ILC~\cite{\TESLA,Weiglein:2004hn,Feng:1995zd,Cheng:1997vy,Freitas:2002gh}, currently being
projected. This job needs the performance of precision measurements, but
also of precision computations which are well suited for experimental
comparisons. In the present work we will focus on the properties of the
SUSY partners of the SM quarks -- the squarks. 

Once produced,  squarks will decay in a way dependent
on the
model parameters  (see e.g.~\cite{Bartl:1994bu}). If gluinos
(the fermionic SUSY partners of 
gluons) are light enough, squarks will mainly decay into gluinos
and quarks ($\squark\to q\tilde{g}$)~\cite{Beenakker:1996dw,Beenakker:1996de}, which proceeds trough a coupling
constant of strong strength. If the mass difference among different
squarks is large enough, some squarks can decay via a bosonic channel
into an electroweak gauge boson and another squark 
($\squark_a \to \squark_b'(Z,\,W^\pm)$), and if Higgs bosons are
light enough, also the scalar decay channels are available ($\squark_a
\to \squark_b'(h^0,\,H^0,\,A^0,\,H^\pm)$)~\cite{Bartl:1997pb,Bartl:1998xk,Bartl:1998xp,Bartl:1999bg}, which can be dominant for
third generation squarks due to the large Yukawa couplings. Otherwise,
the main decay channels of squarks are their partial decays into
charginos/neutra\-linos (the fermionic SUSY partners of the electroweak
gauge and Higgs bosons) and quarks ($\squark \to q'\chi$). Some of those
channels are expected to be always open, given the large mass
difference between quarks and squarks, and that the
charginos/neutra\-linos are expected to be lighter than most of squarks in
the majority of SUSY-breaking models. In the few cases in which these
channels are closed, the squarks will decay through flavour changing
neutral channels~\cite{Hikasa:1987db,Han:2003qe,delAguila:2008iz}, or
through three- or four-body  decay channels involving a non-resonant
SUSY particle~\cite{Porod:1996at,Porod:1998yp,Boehm:1999tr,Djouadi:2000aq,Das:2001kd,Djouadi:2000bx}.

Here we will concentrate on the squark decay channels involving
  charginos and neutralinos.
Their partial decay widths were computed some time ago,
including the radiative corrections due to the strong
(QCD)~\cite{Hikasa:1995bw,Kraml:1996kz,Djouadi:1996wt}, and the electroweak
(EW)\cite{Guasch:1998as,Guasch:2001kz,Guasch:2002ez,Guasch:2002qa} sectors of the
theory. These radiative corrections are large in certain regions of the
parameter space\cite{Guasch:2002ez}, and their complicated expressions
are not suitable for their introduction in the monte-carlo programs used
for experimental analyses. In this work we present
approximations for the partial decay widths of squarks into charginos and neutralinos,
 including the QCD
corrections, and compare these approximations against the fixed-order
one-loop corrected partial widths. 

In section~\ref{sec:notation} we introduce our notation and
conventions for particles and couplings, and set up the numerical
values that we will use in our analysis,
section~\ref{sec:qcd-corrections} presents the QCD one-loop computation
of the partial decay widths and shows some numerical examples,
in section~\ref{sec:RGE} we perform a renormalization group analysis of
the partial decay widths, in section~\ref{sec:numerical} we perform a
numerical comparison of the one-loop and renormalization group
computations, and finally section~\ref{sec:conclusions} shows our
conclusions.
\section{Notation, conventions and numerical setup}
\label{sec:notation}
To describe the computation of the partial decay widths, we will follow
the conventions of
Ref.~\cite{Hahn:2001rv}. Throughout this work we will use a
third-generation notation to describe quarks and squarks, but the analytic
results and conclusions are completely general, and can be used for
quarks-squarks of any generation. We will show numerical results only
for third generation quarks/squarks
(top $t$/stop $\stopp$/bottom $b$/sbottom $\sbottom$), since their decay
widths are the ones that present the most interesting properties.

We will study the partial decay widths of sfermions into fermions and
charginos/neutra\-linos,
\begin{equation}
\Gamma(\tilde{f} \to f' \chi)\,\,.
\label{eq:gammadef}
\end{equation}

We denote {the two sfermion-mass eigenvalues
by $m_{\tilde{f}_a}\,(a=1,2)$, with
$m_{\tilde{f}_1}<m_{\tilde{f}_2}$}.
The sfermion-mixing angle $\osf$
is defined by the transformation relating the weak-interaction
($\sfr^\prime_a=\sfr_L, \sfr_R$) and the mass eigenstate
 ($\sfr_a=\sfr_1, \sfr_2$) sfermion bases:
\begin{equation}
\label{eq:defsq}
  \sfr_a=R_{ab}^{(f)}\, \sfr^\prime_b\,\,; \,\,\,\,
  R^{(f)}=\left(\begin{array}{cc}
      \cos\osf&-\sin\osf \\
      \sin\osf &\cos\osf
    \end{array}\right)\,.
\end{equation}
By this  basis transformation, the sfermion mass matrix,
\begin{equation}
{\cal M}_{\tilde{f}}^2 =\left(\begin{array}{cc}
\MsfL^2+m_f^2+c_{2\beta}(T_3-Q\,s_W^2)\,M_Z^2 
 &  m_f\, M^{LR}_f\\
 m_f\, M^{LR}_f &
 \MsfR^2+m_f^2+Q\,c_{2\beta} \,s_W^2\,M_Z^2  
\end{array} \right)\,,
\label{eq:sbottommatrix}
\end{equation}
becomes diagonal: 
$R^{(f)}\,{\cal M}_{\tilde{f}}^2\,R^{(f)\dagger}=
{\rm diag}\left\{m_{\tilde{f}_1}^2,
  m_{\tilde{f}_2}^2\right\}$. $\MsfL^2$ is the
soft-SUSY-breaking mass parameter of the $SU(2)_L$
doublet\footnote{{With $M_{\tilde{t}_L}=M_{\tilde{b}_L}$ due to $SU(2)_L$ gauge invariance.}}, whereas
$\MsfR^2$ is the soft-SUSY-breaking mass parameter of the
singlet. $T_3$ and 
$Q$ are the usual third component of the isospin and the
electric 
charge respectively, $m_f$ is the corresponding fermion mass, and $s_W$
is the sinus of the weak mixing angle.\footnote{We abbreviate trigonometric functions by their initials,
 like $s_W\equiv\sin \theta_W$, $c_{2\beta}\equiv\cos (2\beta)$, $t_W\equiv s_W/c_W$, etc.} 
The mixing parameters in the non-diagonal entries read
$$
M^{LR}_b=A_b-\mu\tb\ \ \ ,\ \ \ M^{LR}_t=A_t-\mu/\tb\,\,.
$$
$A_{b,t}$ are the trilinear soft-SUSY-breaking couplings, $\mu$ is the
higgsino mass parameter, and $\tb$ is the ratio between the vacuum
expectation values of the two Higgs doublets $\tb=v_2/v_1$. The input
parameters in the sfermion sector are then:
\begin{equation}
(\MsfL,M_{\tilde{b}_R},M_{\tilde{t}_R},A_b,A_t,\mu,\tb)\ \ ,
\label{eq:inputsf}
\end{equation}
for each sfermion doublet. From them, we can derive the masses and
mixing angles:
\begin{equation}
(\msbo,\msbt,\osb)\ ,\ (\msto,\mstt,\ost)\ \ .
\label{eq:outputsf}
\end{equation}
For the trilinear couplings, we require the approximate (necessary) condition
\begin{equation}
A_q^2<3\,(m_{\tilde{t}}^2+m_{\tilde{b}}^2+M_H^2+\mu^2)\,,
\label{eq:necessary}
\end{equation}
where $m_{\tilde{q}}$ is of the order of the average squark masses
for $\tilde{q}=\tilde{t},\tilde{b}$, to avoid colour-breaking minima 
in the MSSM Higgs potential\,\cite{\vacuum}.  

Although the tree-level chargino ($\cplus$)-neutralino ($\neut$) sector is well known, we
give here a short description, in order to set our conventions. 
We start by constructing the following set of Weyl spinors:
\begin{equation}
\begin{array}{lcl}
\Gamma^+&\equiv&(-i \tilde W^+,\tilde H_2^+) \,\,,\\
\Gamma^-&\equiv&(-i \tilde W^-,\tilde H_1^-) \,\,,\\
\Gamma^0&\equiv&(-i \tilde B^0,-i \tilde W_3^0,\tilde H_1^0,\tilde H_2^0) \,\,.
\end{array}
\label{eq:inosweak}
\end{equation}
The mass Lagrangian in this basis reads
\begin{equation}
{\cal L}_M=-\frac{1}{2}\begin{pmatrix}\Gamma^+,\Gamma^-\end{pmatrix}
\begin{pmatrix}0&{\cal M}^T\\
{\cal M}&0\end{pmatrix}
\begin{pmatrix} \Gamma^+\\ \Gamma^-\end{pmatrix}
-\frac{1}{2} \begin{pmatrix}\Gamma_1,\Gamma_2,\Gamma_3,\Gamma_4\end{pmatrix}
{\cal M}^0 \begin{pmatrix}\Gamma_1\\ \Gamma_2 \\ \Gamma_3 \\ \Gamma_4\end{pmatrix}
+\mbox{ h.c.}\,\,,
\end{equation}
where we have defined
\begin{eqnarray}
{\cal M}&=&\begin{pmatrix}
M&\sqrt{2} M_W \sbtt \cr
\sqrt{2} M_W \cbtt&\mu
\end{pmatrix}\,\,,\nonumber\\
{\cal M}^0&=&
\begin{pmatrix}
M^\prime&0&M_Z\cbtt\swp&-M_Z\sbtt\swp \cr
0&M&-M_Z\cbtt\cwp&M_Z\sbtt\cwp\cr
M_Z\cbtt\swp&-M_Z\cbtt\cwp&0&-\mu \cr
-M_Z\sbtt\swp&M_Z\sbtt\cwp&-\mu&0
\end{pmatrix}\,\,\,,
\label{eq:massacplusneut}\end{eqnarray}
{with $M$ and $M'$} the $SU(2)_L$ and $U(1)_Y$
soft-SUSY-breaking gaugino 
masses. 
The four-component mass-eigenstate fields are related to the ones
in~(\ref{eq:inosweak}) by 
\begin{equation}
  \chi_i^{+}= \begin{pmatrix}
    V_{ij}\Gamma_j^{+} \cr U_{ij}^{*}\bar{\Gamma}_j^{-}
  \end{pmatrix}
  \; \;\; \;\;,\;\;\;\;\;
  \chi_i^{-}= {\cal C}\bar{\chi_i}^{+T} =\begin{pmatrix}
    U_{ij}\Gamma^{-}_j \cr V_{ij}^{*}\bar{\Gamma}_j^{+} 
  \end{pmatrix}
\ \ , \ \  \chi_{\alpha}^0= \begin{pmatrix}N_{\alpha\beta}\Gamma_{\beta}^0 \cr 
    N_{\alpha\beta}^{*}\bar{\Gamma}_{\beta}^0
  \end{pmatrix}=  
  {\cal C}\bar{\chi}_{\alpha}^{0T}\ ,
  \label{eq:ninos} 
  \nonumber
\end{equation}
where $U$, $V$ and $N$ are in general complex matrices that 
diagonalize the mass-matrices~(\ref{eq:massacplusneut}):
\begin{equation}
\begin{array}{lcccl}
U^* {\cal M} V^\dagger&=&{\cal M}_D&=&{\rm diag}\left(M_1,M_2\right)\,\,(0<M_1<M_2)\,\,,\\
 N^*{\cal M}^0 N^\dagger &=&{\cal M}^0_D&=&
{\rm diag}\left(M_1^0,M_2^0,M_3^0,M_4^0\right)\,\,(0<M_1^0<M_2^0<M_3^0<M_4^0)\,\,.
\end{array}\label{eq:defUVN}
\end{equation}

Using this notation, the tree-level
interaction Lagrangian between fermion-sfermion-(chargino or neutralino)
reads~\cite{Guasch:2002ez}
    \begin{eqnarray}
      \label{eq:Lqsqcn}
      {\cal L}_{\chi \tilde{f} f'}&=&\sum_{a=1,2}\sum_{r} {\cal L}_{\chi_r
      \tilde{f}_a f'} + \mbox{ h.c.}\,\,,\nonumber\\
     {\cal L}_{\chi_r \tilde{f}_a f'}&=& -g\,\tilde{f}_a^* \bar{\chi}_r
      \left(A_{+ar}^{(f)}\pl +  A_{-ar}^{(f)}\pr\right) f'\,\,.
    \end{eqnarray}
Here we have adopted a compact notation, where
$f'$ is either $f$ or its 
$SU(2)_L$ partner for $\chi_r$ being a neutralino or a chargino,
respectively. Roman characters 
   $a,b\ldots$ are reserved for sfermion indices and 
   $i,j,\ldots$ for chargino indices; 
   Greek indices $\alpha,\beta,\ldots$ denote neutralinos;
   Roman indices $r,s\ldots$ indicate either a chargino or a
   neutralino. For example, the top-squark interactions with charginos
   are obtained by replacing $f\to t$, $f'\to b$, $\chi_r\to \cmin_r$,
   $r=1,2$. The coupling matrices that encode the dynamics are given by
    \begin{eqnarray}
        \label{V1Apm}
        \Apit &=& \Rot\Vo^*-\lt\Rtt\Vt^*\, ,\nonumber\\
        \Amit &=& -\lb\Rot\Ut\, ,\nonumber\\
        \Apat &=&\frac{1}{\sqrt{2}} \left(
            \Rot\left(\Nt^*+\YL\tw\No^*\right)
            +\sqrt{2}\lt\Rtt\Nf^*
          \right)\, ,\nonumber\\
        \Amat &=& \frac{1}{\sqrt{2}} \left(
            \sqrt{2}\lt\Rot\Nf
            -\YRt\tw\Rtt\No
            \right)\, ,\nonumber\\
        \Apib &=& \Rob\Uo^*-\lb\Rtb\Ut^*\, ,\nonumber\\
        \Amib &=& -\lt\Rob\Vt\, ,\nonumber\\
        \Apab &=& -\frac{1}{\sqrt{2}} \left(
          \Rob\left(\Nt^*-\YL\tw\No^*\right)
          -\sqrt{2}\lb\Rtb\Nth^*
        \right)\, ,\nonumber\\
        \Amab &=& -\frac{1}{\sqrt{2}} \left(
          -\sqrt{2}\lb\Rob\Nth
          +\YRb\tw\Rtb\No
          \right) \, ,
     \end{eqnarray}
with $\YL$ and $Y_R^{t,b}$ the weak hypercharges of the left-handed
$SU(2)_L$ doublet and right-handed singlet fermion, and
$\lt=\mt/(\sqrt{2}\mw\sin\beta)$ and $\lb=\mb/(\sqrt{2}\mw\cos\beta)$
are the Yukawa couplings normalized to the $SU(2)_L$ gauge coupling
constant $g$. Note the following, each coupling is formed by two parts:
the gaugino part, formed exclusively by gauge couplings, and the
higgsino part, which contains factors of the quark masses, each of these
parts will receive different kinds of corrections (see below). 

Using these definitions, the tree-level partial decay widths read
\begin{eqnarray}
  \Gamma^{tree}_{ar}&=&\Gamma^{tree}(\sfra \to f' \chi_r)=
  \frac{g^2}{16\,\pi\,\msfa^3}\,\lambda
  (\msfas,M_r^2,m_{f'}^2)\times\,  \nonumber\\ 
  &&\times\left[
    (\msfas-M_r^2-m_{f'}^2)
    \left(|A^{(f)}_{+ar}|^2+|A^{(f)}_{-ar}|^2 \right)
    -4\,m_{f'}\,M_r {\rm Re}\left(A^{(f)}_{+ar}\,A^{(f)*}_{-ar}\right)
  \right]\,\,,
  \label{eq:treleevelgamma}
\end{eqnarray}
with $\lambda(x^2,y^2,z^2)=\sqrt{ [x^2-(y-z)^2][x^2-(y+z)^2]}$.

\subsection{Numerical setup}
\label{sec:numsetup}
For the numerical analysis and plots we will use fixed values for the
SUSY parameters, and make plots by changing one parameter at a time. For
the central values of the parameters we take:
\begin{equation}
\begin{array}{c}
\tb=5\ ,\ 
\mu=300\GeV\ ,\ 
M=200 \GeV\ ,\ 
\MsfL=800\GeV\ ,\ \mg=3000\GeV \ ,\\
\msusy\equiv \MsfR=1000\GeV\ ,\ 
A_{t}=A_{b}=2\MsfL+\mu/\tb=1660\GeV\ ,
\end{array}
\label{eq:SUSYparams}
\end{equation} 
where we have introduced a parameter $\msusy$ as a shortcut for all the
SUSY mass parameters which are not explicitly given. 
We use the GUT relation $M'=5/3 \,\tw^2 \,M$ for the bino mass parameter. 
For the SM
parameters we use $\mt=171.2\GeV$, $\mb=4.7\GeV$, $\al(\mz)=0.1172$,
 $s_{W}^2=0.221$, $\mz=91.1875\GeV$, $1/\alpha=137.035989$. 
The renormalization scale $Q$ is taken to be the physical mass of the
decaying squark.
The value of the trilinear couplings $A_{b,t}$ is given by the algebraic
expression, the given numerical value corresponds to
the default values of the other parameters, this numerical value will
change in the plots, the chosen expression allows to show plots with a
significant parameter variation avoiding colour-breaking-vacuum conditions~(\ref{eq:necessary}).
With
these input parameters, the central values for the physical SUSY particle
masses are:
\begin{eqnarray}
  M_{\cplus}&=&(170.40, 337.50)\GeV\ \ ,\nonumber\\
  M_{\neut}&=&(89.52, 172.28, 305.46, 338.58)\GeV\ \ ,\nonumber\\
  m_{\sbottom}&=&(802.05, 1000.30)\GeV \ \ , \nonumber\\
  m_{\stopp}&=&(720.55, 1084.25)\GeV \ \ .
  \label{eq:physicalmasses}  
\end{eqnarray}
It is illustrative to identify the largest EW-basis component in each
physical state. Of course, we have performed our computation using the
full numerical mixing among the EW-basis and the physical-basis
components, but this identification will help us to analyze the
numerical results. The lightest squarks ($\stopp_1,\sbottom_1$) are
predominantly left-handed, the lightest chargino and neutralinos
($\cplus_1,\neut_1,\neut_2$) are predominantly gaugino-like, whereas the
heaviest ones ($\cplus_2,\neut_3,\neut_4$) are predominantly higgsino-like.
Of course, the parameters in eq.~(\ref{eq:SUSYparams}) are just and
example for illustrative purposes, we have checked that our conclusions
hold for a wide range of the parameter space.

\section{QCD Corrections}
\label{sec:qcd-corrections}
Following this setup, we have computed the full one-loop QCD corrections
to the squark partial decay widths into charginos and
neutralinos~(\ref{eq:gammadef}). The renormalization prescriptions
follow that of Ref.\cite{Guasch:2002ez}. 
The QCD corrections include contributions
from gluon loops, gluino loops, and gluon bremsstrahlung. The full
one-loop corrections have been performed using the
FeynArts/FormCalc/LoopTools packages~\cite{\FeynArts}. We have used
dimensional reduction for the regularization of ultraviolet (UV)
divergences, and a small gluon mass to regularize the infrared (IR)
divergences. The three-body phase-space integration of the real gluon
emission is performed analytically over the full energy range, and the
dependence on the gluon mass is seen to cancel between the virtual and
the real corrections. We have found full agreement with previous
works~\cite{\previous}, and will not repeat the full lengthy formulae
here. The corrections are seen to be numerically large, specially in
certain regions of the parameter space~\cite{Guasch:2002ez}, specially
those involving processes with a bottom-squark in the initial state, and
in a regime of large $\tb$ values. 

We follow the hints from Higgs-boson
physics~\cite{Coarasa:1996qa,Carena:1999py,Guasch:2001wv,Guasch:2003cv},
and define effective Yukawa couplings which should encode the leading
part of the corrections\cite{Carena:1999py}:
\begin{eqnarray}
\lb^{eff.}\equiv \frac{\mb^{eff.}}{v_{1}}\equiv
\frac{\mb(Q)}{v_{1}(1+\Delta \mb)}\ \ , \nonumber\\
\lt^{eff.}\equiv \frac{\mt^{eff.}}{v_{2}}\equiv \frac{\mt(Q)}{v_{2}(1+\Delta \mt)}\ \ ,\label{eq:hbhbteff1}
\end{eqnarray}
where $m_{q}(Q)$ is the running quark mass and $\Delta m_{q}$ is the
finite threshold correction. The SUSY-QCD contributions to $\Delta m_{q}$ are:
\begin{eqnarray}
\Delta \mb^{SQCD}&=&\frac{2\al}{3\pi}\mg\mu \tb\, I(m_{\tilde{b_{1}}},m_{\tilde{b_{2}}}, \mg)\ \ ,\nonumber\\
\Delta \mt^{SQCD}&=&\frac{2\al}{3\pi}\mg\frac{\mu}{\tb}\, I(m_{\tilde{t_{1}}},m_{\tilde{t_{2}}}, \mg)\ \ ,
\label{eq:deltamq}
\end{eqnarray} 
where the function $I(a,b,c)$ is the scalar three-point function at zero
momentum transfer, and reads:
$$
I(a,b,c)=\frac{a^2 b^2\ln(a^2/b^2) + b^2c^2\ln(b^2/c^2) +
  c^2a^2\ln(c^2/a^2)}{(a^2-b^2) (b^2-c^2) (a^2-c^2)}\ \ .
$$
The effective description of squark decays consists in replacing the
tree-level quark masses in the couplings~(\ref{V1Apm}) by the effective
Yukawa couplings of eq.~(\ref{eq:hbhbteff1}), and use this lagrangian to
compute the partial decay width, schematically:
\begin{equation}
\Gamma^{Yuk.-eff.}=\Gamma^{tree}(m_q^{eff.})\ \ .
\label{eq:effdefin1}
\end{equation}
This expression contains the large one-loop corrections from the
finite threshold corrections~(\ref{eq:hbhbteff1}), but it also contains
higher order corrections. At this point we can make the following:
make a computation that combines the higher order effects (which ignore
the effects of external momenta) and the fixed one-loop (which ignore
the higher order effects). At the same time, this will allow us to
quantify the degree of accuracy obtained by the effective
description. We define a \textit{Yukawa-improved decay width computation}:
\begin{equation}\label{improved}
\Gamma^{Yuk.-imp.}\equiv\Gamma^{tree}(m_{q}^{eff.})+(\Gamma^{1-loop}-\Gamma^{1-loop\;Yuk.-eff.})\equiv \Gamma^{tree}(m_{q}^{eff.})(1+\delta^{Yuk.-rem.})
\end{equation} 
where
\begin{eqnarray}
\Gamma^{1-loop}&=&\Gamma^{tree}+\delta\Gamma^{1-loop}\nonumber\\
\delta^{1-loop}&=&\frac{\delta\Gamma^{1-loop}}{\Gamma^{tree}}\nonumber\\
\delta^{Yuk.-rem.}&=&\frac{\Gamma^{1-loop}-\Gamma^{1-loop\;Yuk.-eff.}}{\Gamma^{tree}(m_{q}^{eff.})}\label{eq:defremainder}
\end{eqnarray}
Here $\Gamma^{1-loop}$ is the one-loop fixed order prediction for the
partial decay width,  $\Gamma^{ 1-loop\; Yuk.-eff.}$ is the one-loop
expansion of the prediction using effective couplings, and therefore,
the \textit{remainder} contribution ($\delta^{Yuk.-rem.}$) is the part of the
one-loop contribution that can not be described by the Yukawa effective
couplings, it quantifies the approximation done by the effective
description. 

The one-loop effective prediction $\Gamma^{ 1-loop\; Yuk.-eff.}$ is computed
by taking the computation using effective couplings~(\ref{eq:effdefin1}), expanding it in series, and keeping only the
one-loop terms. Specifically:
\begin{equation}
m(Q)=m(m)\left[1-\frac{2}{\pi}\al(Q)\log(\frac{Q}{m})+\ldots\right]\
\ ,
\end{equation} 
and \begin{equation}
m^{eff.}=m(Q)(1-\Delta m)\ \ ,
\end{equation} 
therefore, the part of the one-loop effective mass is:
\begin{equation}
\delta m^{eff.}=m(m)\left[-\frac{2}{\pi}\al(Q)\log
  (\frac{Q}{m})-\Delta m\right]\ \ ,
\end{equation} 
this is the mass that will be used in the effective Yukawa couplings to
compute $\Gamma^{1-loop\; Yuk.-eff.}$, and $m(m)$ is the running quark
mass at the quark mass scale.
Finally, we define a \textit{Yukawa-improved} correction factor in the
following way:
\begin{equation}
\delta^{Yuk.-imp.}=\frac{\Gamma^{Yuk.-imp.}-\Gamma^{tree}(m_{q})}{\Gamma^{tree}(m_{q})}\ \ .
\label{eq:defimproved}
\end{equation}

\begin{figure}[tbp]
\centering
\begin{tabular}{cc}
\includegraphics*[width=6cm]{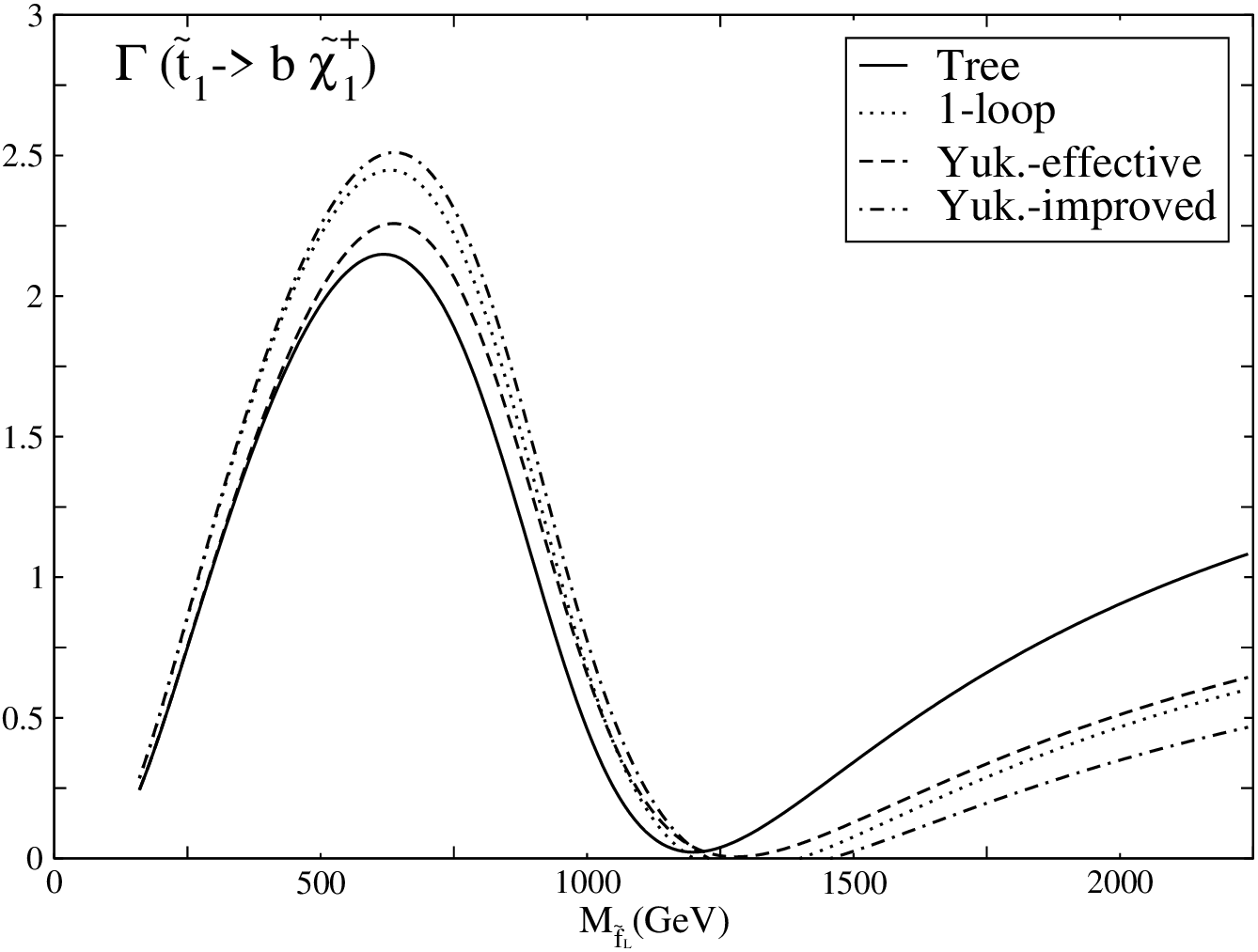}&
\includegraphics*[width=6cm]{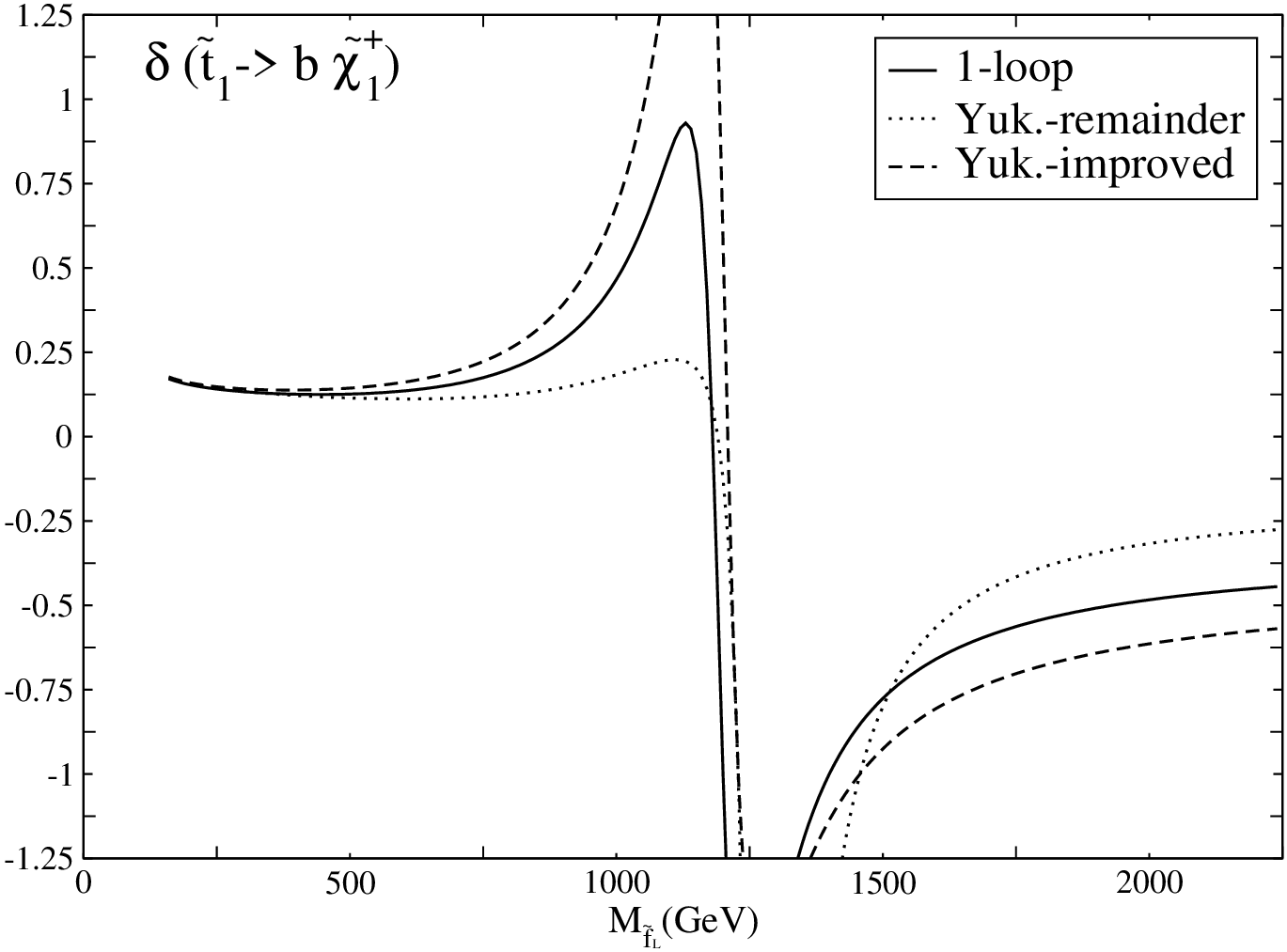}
\\
(a) & (b)
\end{tabular}
\caption{\textbf{a)} Partial decay width and \textbf{b)} relative
  corrections to the top-squark decay width into the lightest chargino
  as a function of the $SU(2)_L$ squark mass scale $\MsfL$,
  for the parameters of eq.~(\ref{eq:SUSYparams}).}
\label{fig:stopdeccormsq1}
\end{figure}

All these definitions will allow us to precisely analyze the
approximations. As an example, Fig.~\ref{fig:stopdeccormsq1} shows the
partial decay width (and the relative correction) of a
top-squark decaying into
the lightest chargino, as a function of the 
$SU(2)_L$ squark mass scale  $\MsfL$~(\ref{eq:sbottommatrix}),
the rest of the parameters are given in~(\ref{eq:SUSYparams}). 
We see a big dip in the corrections for
squark masses around $1250\GeV$, with negative corrections surpassing
$-100\%$ -- which would mean a \textit{negative} decay width, which
obviously does not make sense. What happens is that, for this very
special setup of parameters, the tree-level computation of the partial
decay width vanishes, so the one-loop contribution exceeds the
tree-level prediction. Under these circumstances one-loop perturbation theory
does not hold, and we can not claim the validity of any result obtained
by the one-loop perturbative expansion, that is: we can not give a prediction for the
decay width in those parts of the parameter space in the present approximation. 
Note, also, that the effective prediction
$\Gamma^{Yuk.-eff.}$~(\ref{eq:effdefin1}) is (by definition) a positive
quantity, therefore the effective description can not reproduce the
one-loop result at all, which means a large remainder
$\delta^{Yuk.-rem.}$~(\ref{eq:defremainder}). 
However, there are a
couple of circumstances surrounding these situations: first of all, they
appear in tiny regions of the parameter space; second, and more
important, this effect occurs precisely on decay channels that
have a negligibly small branching ratio, and therefore are
phenomenologically irrelevant. We can see that the dip in
Fig.~\ref{fig:stopdeccormsq1}b around $1250\GeV$ coincides with the
minimum of the partial decay width in Fig.~\ref{fig:stopdeccormsq1}a. For
these reasons we will not try to 
give a reasonable prediction for these decay widths in those corners of
the parameter space. From now on we will limit ourselves to
point out where
they appear, so that the reader is warned that we can not trust the
results in those cases. Outside of this dip, there are two different
regions. For squark masses larger than $1250\GeV$ the one-loop correction
is around 
$-45\%$ 
whereas the \textit{remainder} correction~(\ref{eq:defremainder}) is around
$-28\%$
that means that, roughly, one third of the one-loop corrections can
be described as coming from the effective
couplings~(\ref{eq:hbhbteff1}). Since the corrections (from both:
one-loop and effective couplings) are quite large, one can provide the
\textit{improved}~(\ref{eq:defimproved}) description. In the present
situation $\delta^{Yuk.-imp.}$ is larger than the corrections from the
effective couplings and the fixed-order one-loop corrections, 
$\delta^{Yuk.-imp.}\sim -57\%$
but it
accounts for two kind of effects: the contribution of higher order
terms, and the dependence of the radiative corrections on the external
momenta. On the other side of the plot, for squark masses below
$1250\GeV$, the situation is quite different. The one-loop corrections
are relatively small
($15\%$),
whereas the effective description gives a slightly smaller result,
this region has light particles running in the loops, and the one-loop
functions are expected to depend much on the external momenta. In this
situation the effective description can not describe properly
  the radiative corrections.  The improved
  description~(\ref{eq:defimproved}), on the other hand, 
basically coincides with the fixed order one-loop result. In summary:
our improved description includes the higher order terms of the
Yukawa-effective couplings, and the external momenta dependence of the
one-loop corrections. It is able to describe both situations: when the
one-loop computation gives a sufficient approximation, and when 
higher order corrections are important and should be taken into account.

\begin{figure}[tbp]
\centering
\begin{tabular}{cc}
\includegraphics*[width=6cm]{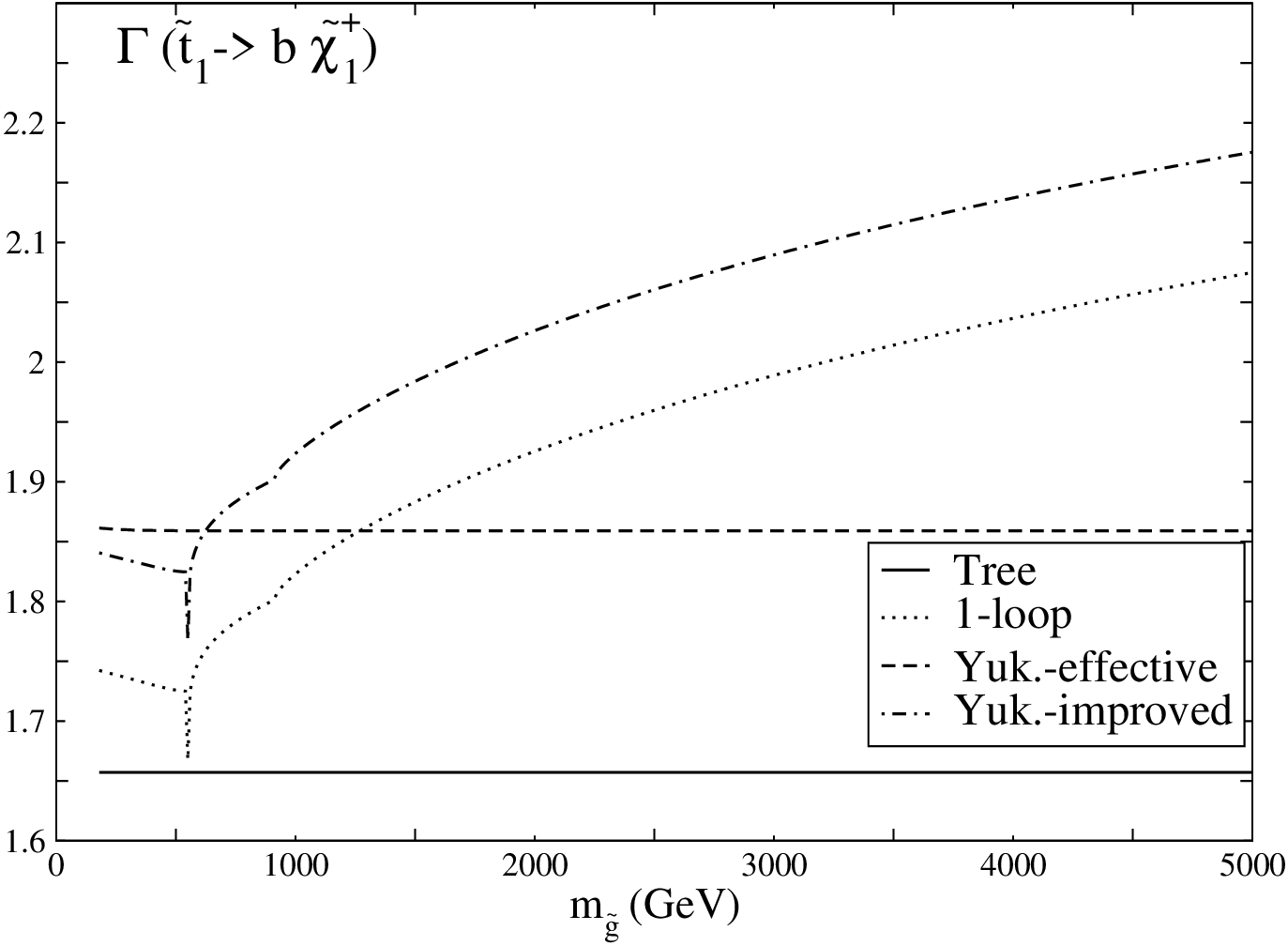}&
\includegraphics*[width=6cm]{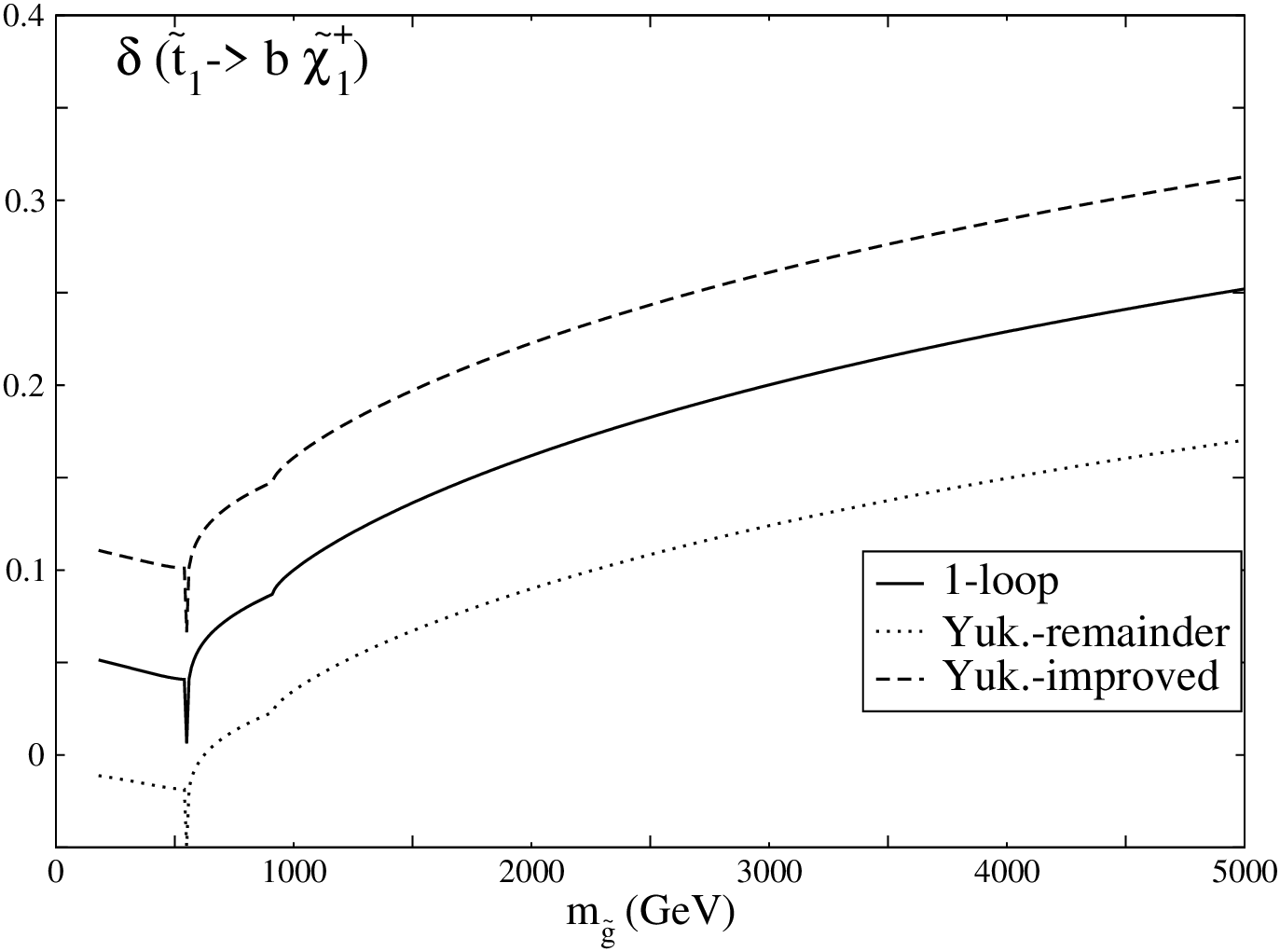}
\\
(a) & (b) 
\end{tabular}
\caption{\textbf{a)} Top-squark partial decay width into the lightest
  chargino and \textbf{b)} relative corrections, as a
  function of the gluino mass. Shown are the different approximations:
  tree-level; one-loop; Yukawa-effective coupling; Yukawa-improved.
  The input parameters are given in eq.~(\ref{eq:SUSYparams}).}
\label{fig:stopdeccormgl1}
\end{figure}

Be as it may, it turns out that the effective description using just
the Yukawa threshold corrections~(\ref{eq:hbhbteff1}) is not enough for the squark decay
widths description. The one-loop corrections develop a term which grows
as the gluino mass $\mg$~\cite{Hikasa:1995bw}, which is absent in the effective Yukawa
couplings~(\ref{eq:hbhbteff1}). Fig.~\ref{fig:stopdeccormgl1} shows the comparison of
the tree-level, one-loop, and Yukawa-effective computations of the top-squark
decay width into charginos, as a function of the gluino mass, for the
input parameters of eq.~(\ref{eq:SUSYparams}). One can
clearly see the log-like behaviour of the one-loop corrections, which
can not be reproduced using the effective 
coupling\footnote{The variation of the Yukawa-effective
    corrections on the gluino mass~(\ref{eq:deltamq}) in
    Fig.~\ref{fig:stopdeccormgl1}a is of a $0.13\%$, which can not be
    appreciated in the plot.}.
The logarithmic terms in the gluino mass are a clear
example of the non-decoupling effects. To understand their origin, one
can think on the following: in a process where all external (initial and
final state) particles belong to the SM sector, one can separate the
loop contributions in two kinds: SM-like (with only SM and Higgs boson
particles running inside the loops), and non-SM like (with only SUSY
particles running inside the loops). Each kind is UV
(and IR) finite by itself, so one can remove the non-SM-like part from
the computation -- or take all non-SM-like particle masses very large --
without any ill effect. In the present case, however, there are SM and
SUSY particles as external states, the Feynman diagrams can no longer be
divided into SM-like and non-SM like. The UV divergences from the SM
sector are cancelled against UV-divergences of the SUSY-sector, and as a
consequence if we remove a SUSY particle from the computation, the
computation is UV-divergent -- and therefore meaningless. If we try to
\textit{remove} a particle by setting its mass to very large values,
this divergence appears as a logarithm of the corresponding particle
mass, and we obtain the aforementioned non-decoupling effects. 
In summary: the QCD corrections to squark decay widths produce explicit
non-decoupling terms of the sort $\log \mg$.

To include the logarithmic terms in the effective descriptions, we have
extracted from the one-loop result the $\log\mg$ terms. We have
expanded the full one-loop result in the limit
$\mg\gg m_{\tilde{q}}$, and we have used the reduction factors
from Ref.\cite{Devaraj:1997es} to obtain scalar quantities. 
The results for top-squark decays into charginos are:
\begin{eqnarray}
\Delta \Amit&=&F_{-ai}^{(t)}+\delta
\Amit=-\frac{\al}{2\pi}\Amit\log\frac{\mg^{2}}{\mu_0^{2}}\ \ ,\nonumber\\
\Delta \Apit&=&F_{+ai}^{(t)}+\delta  \Apit=\left\lbrace \lt\Rtt\Vt^*\frac{\al}{\pi}+\frac{\al}{2\pi}\Apit\right\rbrace \log \frac{\mg^{2}}{\mu_0^{2}}\ \ ,
\label{eq:logterm1loop}
\end{eqnarray}
here $\Delta A_{+,-}$ are the full one-loop corrections to the form
factors,  $F_{+,-}$ are the contributions from the corresponding one-loop
diagrams, $\delta A_{+,-}$ the counterterm contributions, and $A_{+,-}$ are the
tree-level couplings defined in~(\ref{V1Apm}). 
$\mu_0$ is the scale appearing in the dimensional reduction of
  the one-loop UV-divergent integrals, which appears when applying the
  procedure from Ref.\cite{Devaraj:1997es}, and can be though as a
  renormalization scale.
However, these expressions
do not give any hint at the origin of the $\log\mg$ terms, or how can
they be computed. 
To understand those terms a renormalization group analysis is in order.

\section{Renormalization Group Analysis}
\label{sec:RGE}
In order to extract the exact dependence on the renormalization scale,
we make a renormalization group analysis, which will allow us to compute
the
logarithmic terms in $\mg$. To compute those terms, we construct an
effective theory below the gluino mass scale, which contains only
squarks, quarks, charginos, neutralinos and gluons in the light sector
of the theory, and integrate out the gluino contributions. We find out
the renormalization group equations (RGE) of the gaugino and higgsino
couplings, and perform the matching with the full MSSM couplings at the
gluino mass scale $\mg$. 
In the present computation we will consider only logarithmic RGE effects,
and neglect the possible threshold effects at the gluino mass scale.
Since the effective theory does not contain
gluinos, only the contributions from the gluon have to be taken into
account. We will present only the computation for the top-squark decay
into charginos, the other couplings follow from the present computation.

\subsection{Gluon contribution to $\Apit$}
The coupling $\Apit$ is the sum of two terms: a gaugino coupling, and a
higgsino coupling, as seen in eq.~(\ref{V1Apm}). To shorten up the
expressions, we will introduce the shortcuts $\hp$, $G$, which represent
the higgsino and gaugino part respectively:
\begin{equation}
\Apit=\hp+G \ \ ; \ \ \hp=-\lt\Rtt\Vt^* \ \ , \ \ G=\Rot\Vo^*\ \ .
\label{eq:defHpG}
\end{equation}

The total gluon contribution to the divergent part of the vertex form factors and the wave function renormalization constants is:
\begin{equation}\label{deltacoef}
(Coef \Delta)_{+}=\left(\frac{\al}{3\pi}\Apit+\frac{1}{2}\frac{\al}{3\pi}\Apit\right)=\frac{\al}{2\pi}\Apit
\ \ ,
\end{equation}
where the first term comes from the vertex form factor's divergent part and the
second from the fermion and sfermion wave function renormalization constants.
Then, the gluon contribution to the $\beta$ function is
\begin{equation}\label{beta1}
\beta_{+,g}=-2(Coef \Delta)_{+}=-\frac{\al}{\pi}\Apit=-\frac{\al}{\pi}\left(\hp+G\right)\ \ .
\end{equation}
Therefore, the renormalization group equation is
\begin{equation}\label{rge}
\frac{d\Apit(t)}{dt}=-\frac{\al(t)}{\pi}\Apit(t)\ \ ,
\end{equation}
where $t=\log Q$, $Q$ being the renormalization
scale. To solve that equation, we make use of the standard RGE for the
QCD coupling constant (see e.g.~\cite{Buras:1998raa})
\begin{equation}\label{buras}
\frac{d\al(t)}{dt}=-\frac{1}{2\pi}\beta_{0}\al^2(t)\ \ ,
\end{equation} 
$\beta_0$ being the standard QCD $\beta$-function:
$$
\beta_0=\frac{11 N_c-2N_f}{3}-\frac{ N_{\sfr}}{6}-2 N_{\tilde{g}}\  \ ,
$$
where $N_c=3$ is the number of colors, $N_f$, $N_{\sfr}$, and
$N_{\tilde{g}}$ are the number of quarks, squarks and gluino that have a
mass below the scale $Q$ at which we compute the $\beta_0$
function\footnote{The RGE evolution is performed by steps,
    taking into account the change in the $\beta_0$ function as the
    scale $Q$ crosses  thresholds of colored particles.}. 
Inserting 
expression~(\ref{buras}) in (\ref{rge}) we obtain
\begin{equation}
\frac{d\Apit(t)}{\Apit(t)}=\frac{2}{\beta_{0}}\frac{d\al(t)}{\al(t)}\ \ .
\end{equation} 
Solving the equation we obtain
\begin{equation}
\frac{\Apit(t)}{\Apit(t_{0})}=\left(\frac{\al(t)}{\al(t_{0})}\right)^\frac{2}{\beta_{0}}\ \ .
\label{eq:runningAplus1}
\end{equation} 
Finally, using the QCD running coupling constant 
\begin{equation}
\frac{\al(Q)}{\al(Q_{0})}=1-\beta_{0}\frac{\al(Q)}{2\pi}\log\frac{Q}{Q_{0}}\ \ ,
\label{eq:alphasQCD}
\end{equation}
the running of the right-handed vertex coupling constant is,
approximated to ${\cal O}{(\al)}$:
\begin{equation}
\Apit(Q)\simeq \Apit(Q_ {0})\left(1-\frac{\al(Q)}{\pi}\log\frac{Q}{Q_{0}}\right)
\ \ .
\end{equation} 
The boundary conditions at $Q_{0}=\mg$ are 
\begin{equation}
\Apit(\mg)=\hp(m_{q}(\mg))+G(\mg)\ \ ,
\end{equation} 
then, the running coupling constant is 
\begin{equation}
\Apit(Q)\simeq\left(\hp(m_{q}(\mg))+G(\mg)\right)\left(1-\frac{\al(Q)}{\pi}\log\frac{Q}{\mg}\right)\ \ .
\label{eq:runningAplus2}
\end{equation} 
Now, we have the squark-chargino  running coupling constant as a
function of the gauge and Higgs boson couplings at the gluino mass
scale, but we want to express it as a function of couplings at the
renormalization scale $Q$. Note that the gauge part of the coupling
$G$~(\ref{eq:defHpG}) only contains EW gauge couplings, and they do not
receive one-loop running contributions from the QCD sector, therefore it is a constant
term. 
\begin{equation}
G(\mg)=G(Q)\equiv G\ \ .
\label{eq:runningG}
\end{equation}
The higgsino coupling $\hp$~(\ref{eq:defHpG}), on the other hand, has a dependence on the quark
Yukawa coupling (or mass), which does run due to QCD corrections,
according to the RGE,
\begin{equation}
m(Q)=m(Q_{0})\left(\frac{\al(Q)}{\al(Q_{0})}\right)^\frac{4}{\beta_{0}}\ \ ,
\label{eq:runningmq}
\end{equation} 
by inserting these expression into eq.~(\ref{eq:runningAplus1}) we can
obtain
\begin{eqnarray}
  \Apit(Q)&=&\Apit(\mg)\left(\frac{\al(Q)}{\al(\mg)}\right)^\frac{2}{\beta_{0}}\nonumber\\
  &=&\hp(m_q(\mg))
  \left(\frac{\al(Q)}{\al(\mg)}\right)^\frac{2}{\beta_{0}}+G\left(\frac{\al(Q)}{\al(\mg)}\right)^\frac{2}{\beta_{0}}\nonumber\\
  &=&\hp(m_q(Q))
  \left(\frac{\al(Q)}{\al(\mg)}\right)^\frac{-2}{\beta_{0}}+G\left(\frac{\al(Q)}{\al(\mg)}\right)^\frac{2}{\beta_{0}}\nonumber\\
&\simeq&
\hp(m_q(Q))\left(1+\frac{\al(Q)}{\pi}\log\frac{Q}{\mg}\right)+G\left(1-\frac{\al(Q)}{\pi}\log\frac{Q}{\mg}\right)
  \label{eq:runningAplus3}  
\end{eqnarray}
where, in the last line, we have made the ${\cal O}(\al)$
approximation. 
Note that the expressions for the higgsino and gaugino couplings
  are different. Actually, if we write the higgsino and gaugino couplings
at the scale $Q$ as a function of the couplings at the  scale $\mg$ they
have the same form (\ref{eq:runningAplus2}), since they have the same
RGE (\ref{eq:runningAplus1}). The
difference appears when we write the higgsino/gaugino couplings at the
scale $Q$ as a function of the gauge/Higgs couplings at the same
scale~(\ref{eq:runningAplus3}), due to the different running of the
gauge~(\ref{eq:runningG}) and Higgs-boson (\ref{eq:runningmq}) couplings
between the scales $\mg$ and $Q$. 
The last line in eq.~(\ref{eq:runningAplus3}) agrees with the
$\log\mg/\mu_0$ term of the fixed order one-loop expression
in~(\ref{eq:logterm1loop}).

\subsection{Gluon contribution to $\Amit$}
The gluon contribution to the divergent part of the left-handed
couplings $\Amit$, eq.~(\ref{V1Apm}), is the same that in the previous case.
\begin{equation}\label{deltacoef2}
(Coef \Delta)_{-}=\left(\frac{\al}{3\pi}\Amit+\frac{1}{2}\frac{\al}{3\pi}\Amit\right)=\frac{\al}{2\pi}\Amit\ \ ,
\end{equation}
But now, the coupling only contains a Yukawa like coupling
\begin{equation}\label{beta2}
\beta_{-,g}=-2(Coef \Delta)_{-}=-\frac{\al}{\pi}\Amit=-\frac{\al}{\pi}\hm\ \ .
\end{equation}
The renormalization group equation is
\begin{equation}\label{rge2}
\frac{d\hm(t)}{dt}=-\frac{\al(t)}{\pi}\hm(t)\ \ ,
\end{equation}
which has as a solution
\begin{equation}
\hm(Q) = 
\hm(m_{q}(\mg))\left(\frac{\al(Q)}{\al(\mg)}\right)^{\frac{2}{\beta_{0}}}\ \ .
\end{equation} 
Following the same steps as in the the previous section, we obtain
\begin{eqnarray} 
\hm(Q)
&=&\hm(m_{q}(\mg))\left(\frac{\al(Q)}{\al(\mg)}\right)^{\frac{2}{\beta_{0}}}
=\hm(m_{q}(Q))\left(\frac{\al(Q)}{\al(\mg)}\right)^{\frac{-2}{\beta_{0}}}\nonumber\\
&\simeq&\hm(m_{q}(Q))\left(1+\frac{\al(Q)}{\pi}\log\frac{Q}{\mg}\right)\ \ ,
\label{eq:runningHm}
\end{eqnarray}
where in the last line we have made the ${\cal O}(\al)$
approximation. This expression coincides with the higgsino running
coupling constant for $\Apit$~(\ref{eq:runningAplus3}), 
and it also agrees with the
$\log\mg/\mu_0$ term of the fixed order one-loop expression
in~(\ref{eq:logterm1loop}). 

\subsection{Renormalization Group summary}
The renormalization group running of the coupling constant, can be
summarized as follows: we can use effective gaugino and higgsino
couplings given by,
\begin{eqnarray}
g^{eff.}(Q)&=&g\,\left(\frac{\al(Q)}{\al(\mg)}\right)^\frac{2}{\beta_{0}}\simeq g\left( 1-\frac{\al(Q)}{\pi}\log  \frac{Q}{\mg}\right)\ \ , \nonumber\\
\tilde{\lambda}_{b,t}^{eff.}(Q)&=&\lambda_{b,t}^{eff.}(Q)\,\left(\frac{\al(Q)}{\al(\mg)}\right)^\frac{-2}{\beta_{0}}
\simeq\lambda_{b,t}^{eff.}(Q)\left( 1+\frac{\al(Q)}{\pi}\log \frac{Q}{\mg}\right)\ \ ,
\label{eq:logterms}
\end{eqnarray} 
where $\lambda^{eff.}(Q)$ are the effective Yukawa couplings
  defined in~(\ref{eq:hbhbteff1}).
We define then an effective partial decay width, computed by replacing
$g$ and $\lambda$ in the tree-level expression (\ref{eq:treleevelgamma})
by the expressions of eq.~(\ref{eq:logterms}):
$$
\Gamma^{eff.}=\Gamma^{tree}(g^{eff.},\tilde{\lambda}^{eff.}(Q))\ \ ,
$$
and we define \textit{improved} and \textit{remainder} widths and
corrections ($\Gamma^{imp.}$, $\delta^{imp.}$, $\delta^{rem.}$), in the
same fashion as the \textit{Yukawa-effective}, \textit{Yukawa-improved}
and \textit{Yukawa-remainder} of eqs.~(\ref{improved}),
(\ref{eq:defremainder}), (\ref{eq:defimproved}). 

The origin of the $\log\mg$ terms is a consequence of the
SUSY-breaking, and, technically, can be seen in different ways,
depending on the approximation used to make the computation. In the
one-loop fixed order computation they appear because of the
cancellation of UV-divergences between the gluon and gluino loops
(as explained above), but in the effective theory point of view, they
appear because of the different running of the gauge/Higgs boson and
gaugino/higgsino couplings. In a fully SUSY theory the gaugino
(higgsino) couplings are equal to the gauge (Higgs) boson couplings, and
they have the same RGE, but in a theory with broken SUSY (as the present
one) these couplings are no longer the same, they have different RGE,
and the difference between them is a measure of the SUSY-breaking
($\log\mg$). 

\begin{figure}[tbp]
\centering
\begin{tabular}{cc}
\includegraphics*[width=6cm]{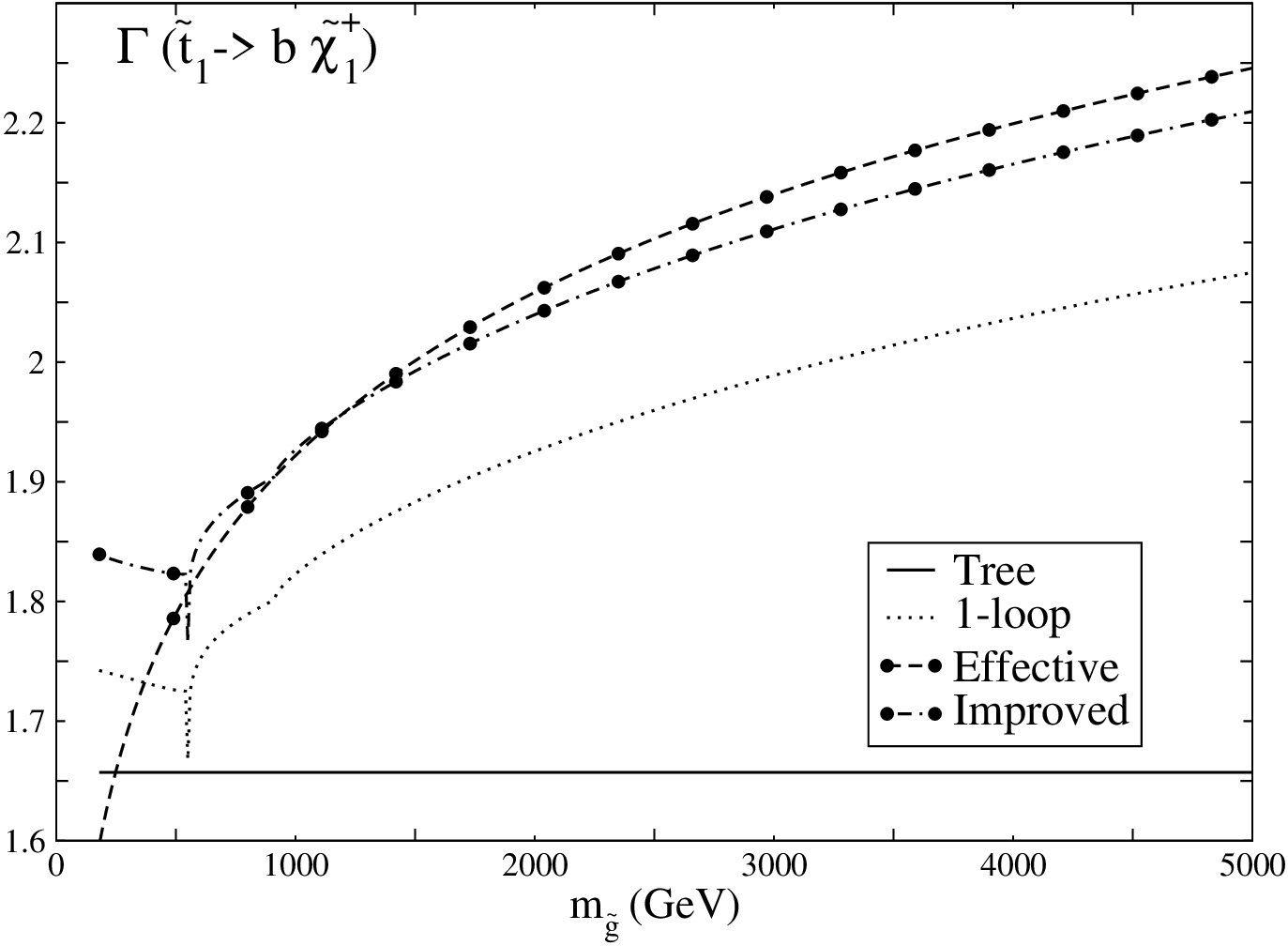}&
\includegraphics*[width=6cm]{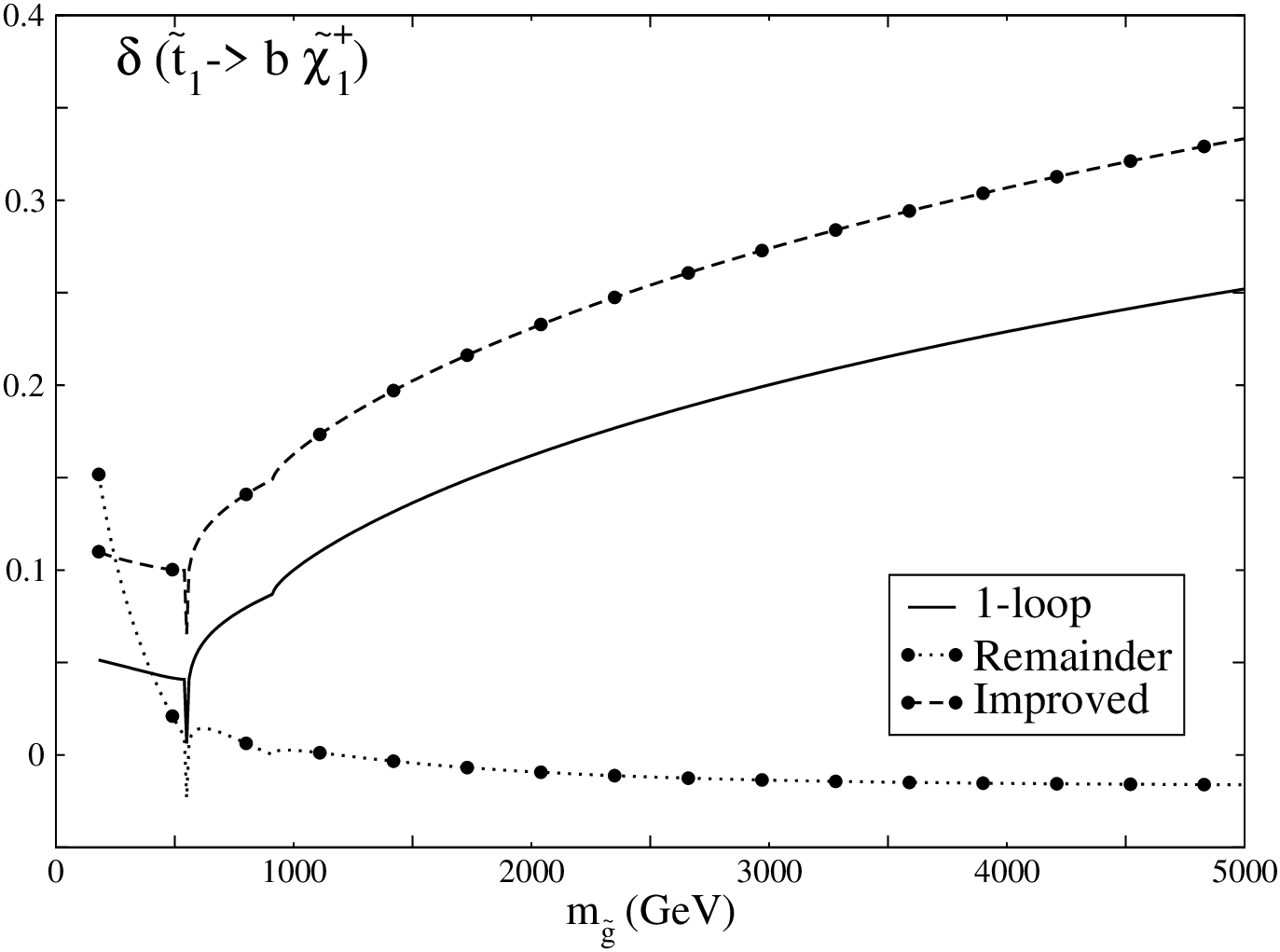}
\\
(a)&(b)
\end{tabular}
\caption{\textbf{a)} Partial decay widths and \textbf{b)} relative
  corrections  of the top-squark decay into the lightest chargino, as a
  function of the gluino mass $\mg$. Shown are different
  approximations to the computation. The \textit{effective},
  \textit{improved} and \textit{remainder}
  computations include the $\log\mg/Q$ terms.
  The input parameters are given in eq.~(\ref{eq:SUSYparams}).}
\label{fig:stopdeccormgl_log1}
\end{figure}

Fig.~\ref{fig:stopdeccormgl_log1} shows the partial decay width of the
top-squark into the lightest chargino, where we include the $\log\mg$ terms of
eq.~(\ref{eq:logterms}) in the effective description. The input
parameters are given in eq.~(\ref{eq:SUSYparams}). Now the effective
description follows the logarithmic behaviour of the full one-loop
corrections. Moreover, we have checked the validity of our result by
comparing the full one-loop corrections in the limit $\mg\gg M$ with the
one-loop expansion of the effective description, which contains only
$\log\mg/\mu_0$ terms\footnote{The comparison is performed by comparing
the slope of the different computations in the plots as a function
of $\log \frac{\mg}{\mu_0}$.}. The results agree with the previous
results of Ref.\cite{Hikasa:1995bw,Djouadi:1996wt} in the limit $\mg\gg
m_{\tilde{q}}$.
Our results go beyond the ones of Ref.\cite{Hikasa:1995bw}, by
  including the Yukawa terms, including explicitly all the chargino/neutra\-lino
  and squark mixing and couplings, showing the exact dependence from the
  renormalization group, and performing (see below) the numerical
  comparison with the fixed order computation.

Fig.~\ref{fig:stopdeccormgl_log1}b shows the new value of the
\textit{remainder} contributions. At low values of the gluino mass, the
\textit{remainder} is still large, because $\mg\simeq\msq$ and the
logarithmic approximation does not make sense, but for $\mg\geq 1\TeV$
the remainder contributions stay below the $2\%$ level.

\section{Numerical analysis}
\label{sec:numerical}
Following the computation and setup of the previous sections, we perform
a complete numerical analysis. We will concentrate on third generation
squark decays (top-squarks and bottom-squarks), as they have the
richest phenomenology, since only on them the higgsino couplings are
large enough. The input parameters that we use are given in
eq.~(\ref{eq:SUSYparams}), and the resulting spectrum is discussed in
section~\ref{sec:numsetup}. We have chosen a large value for the
gluino mass to enhance the effects of the logarithmic terms.

\begin{figure}
\centering
\rotatebox{90}{\resizebox{20cm}{!}{\begin{tabular}{cccc}
\includegraphics*[width=12.5cm]{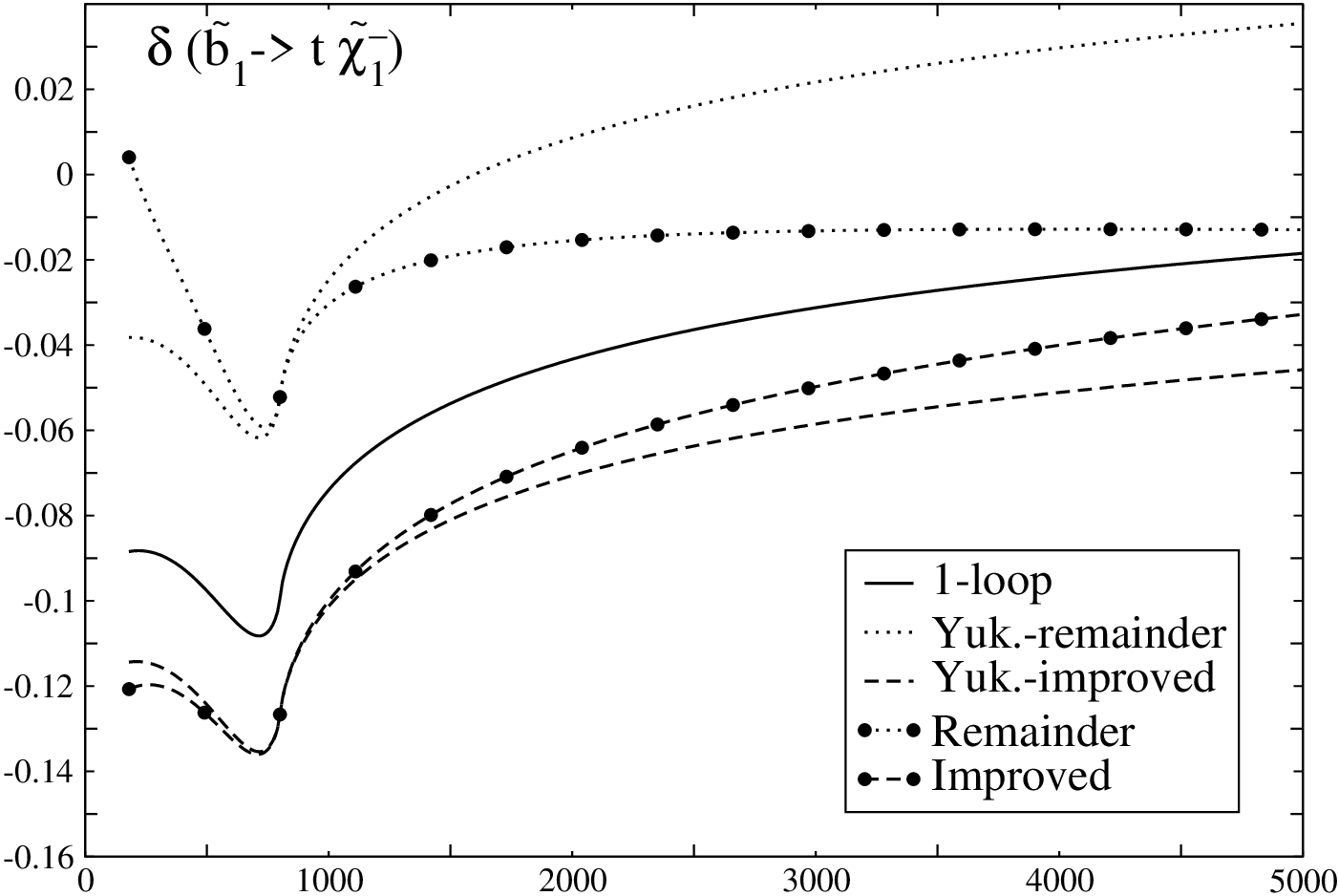}&
\includegraphics*[width=12.5cm]{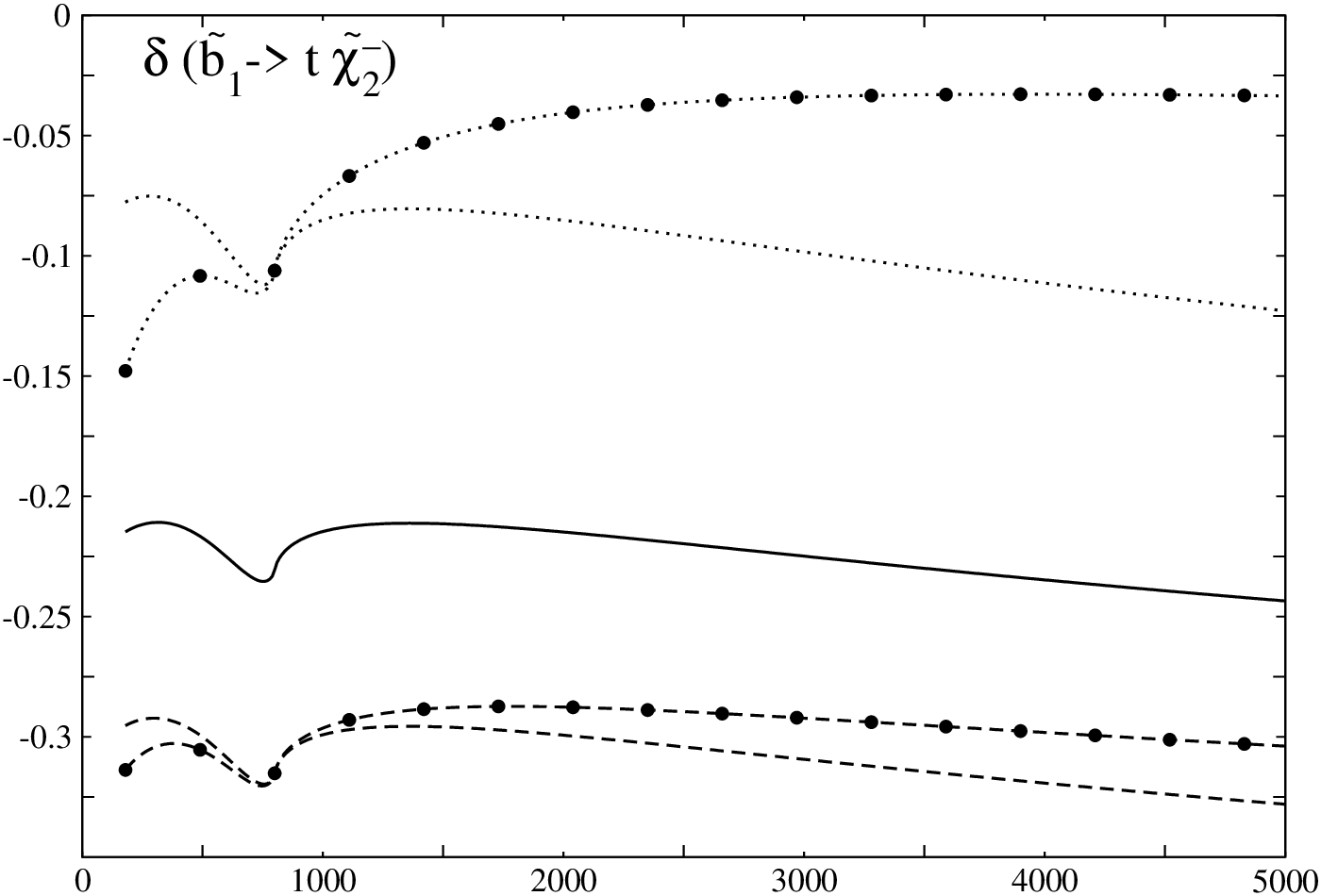}&
\includegraphics*[width=12.5cm]{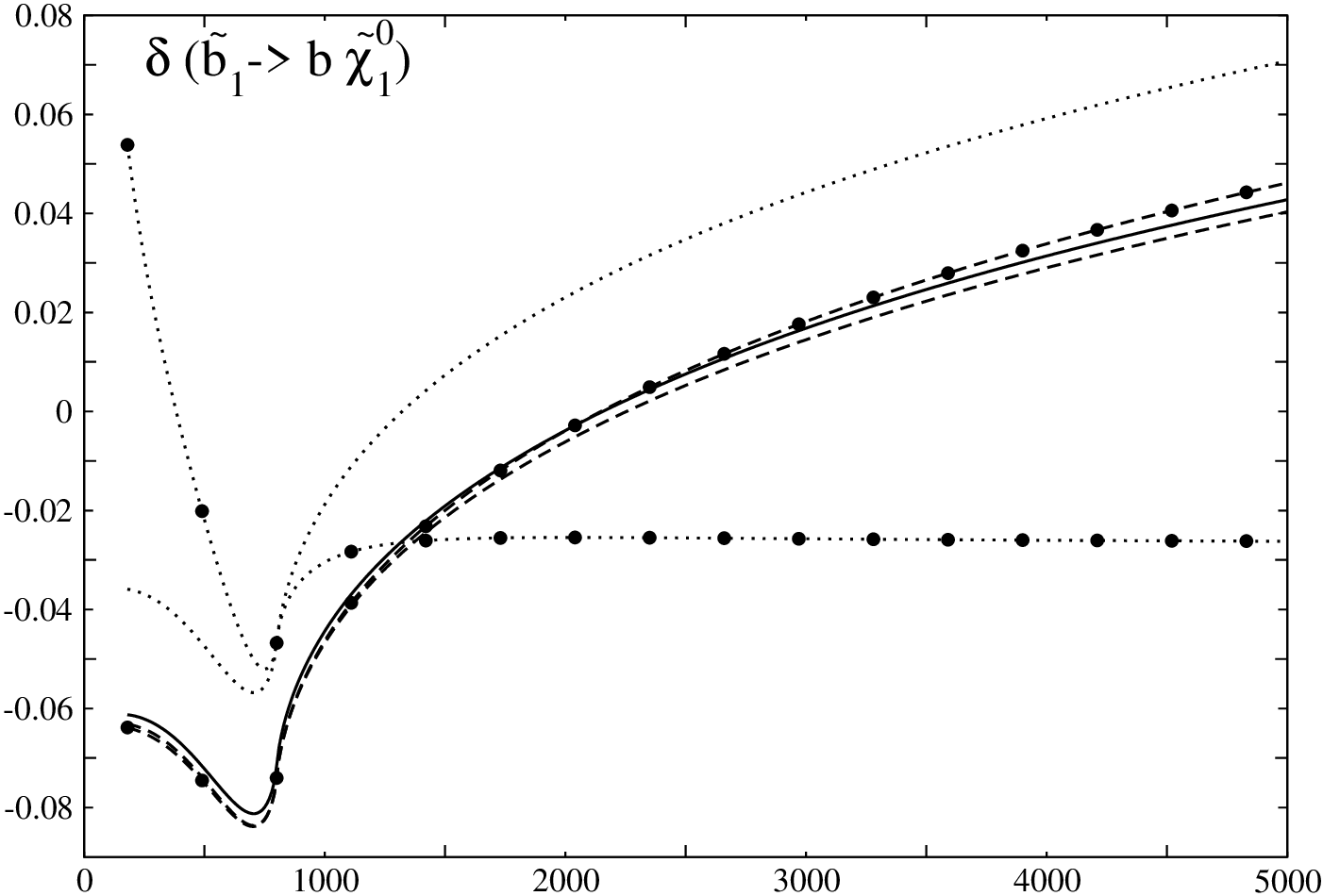}&
\includegraphics*[width=12.5cm]{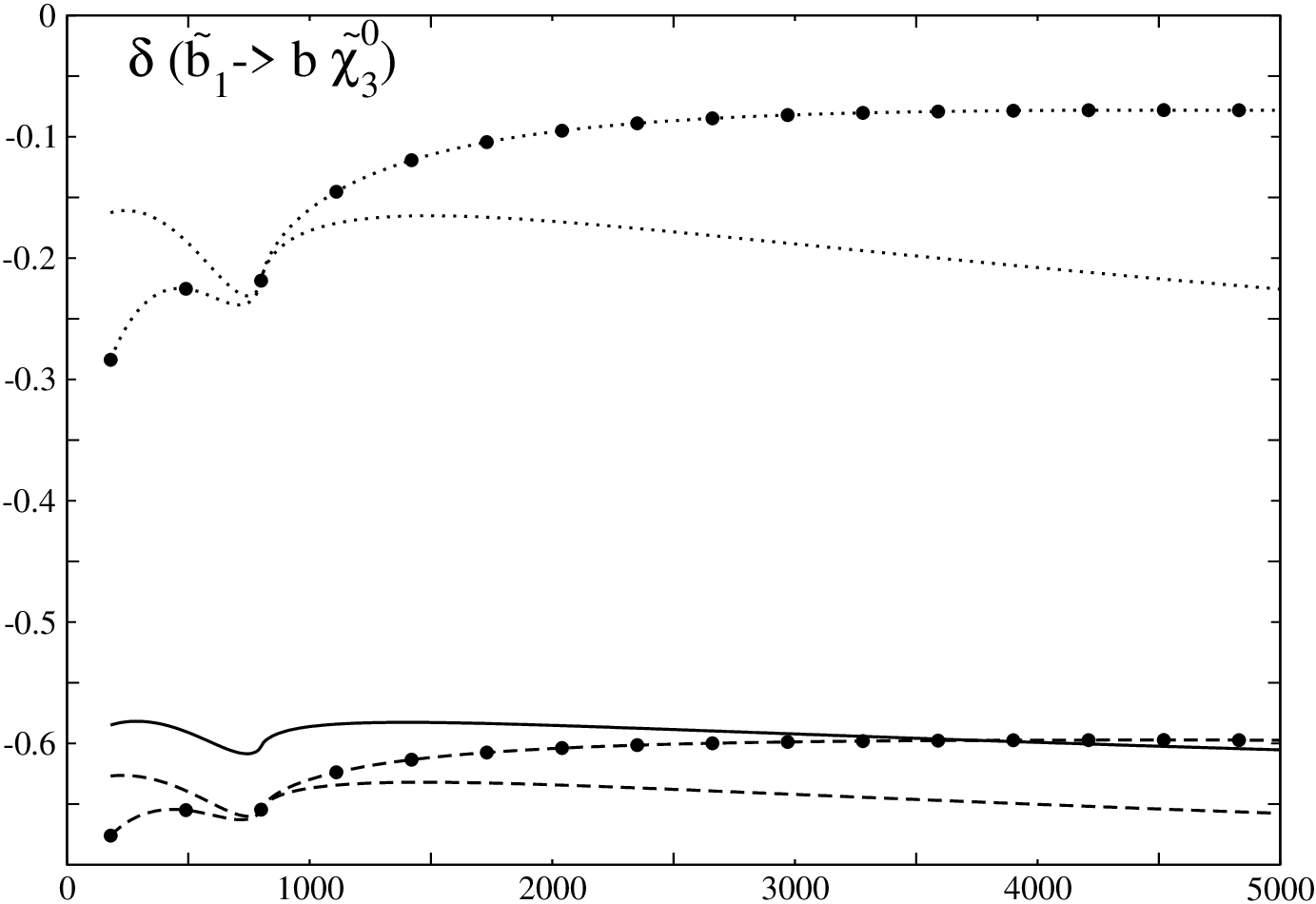}\\
\includegraphics*[width=12.5cm]{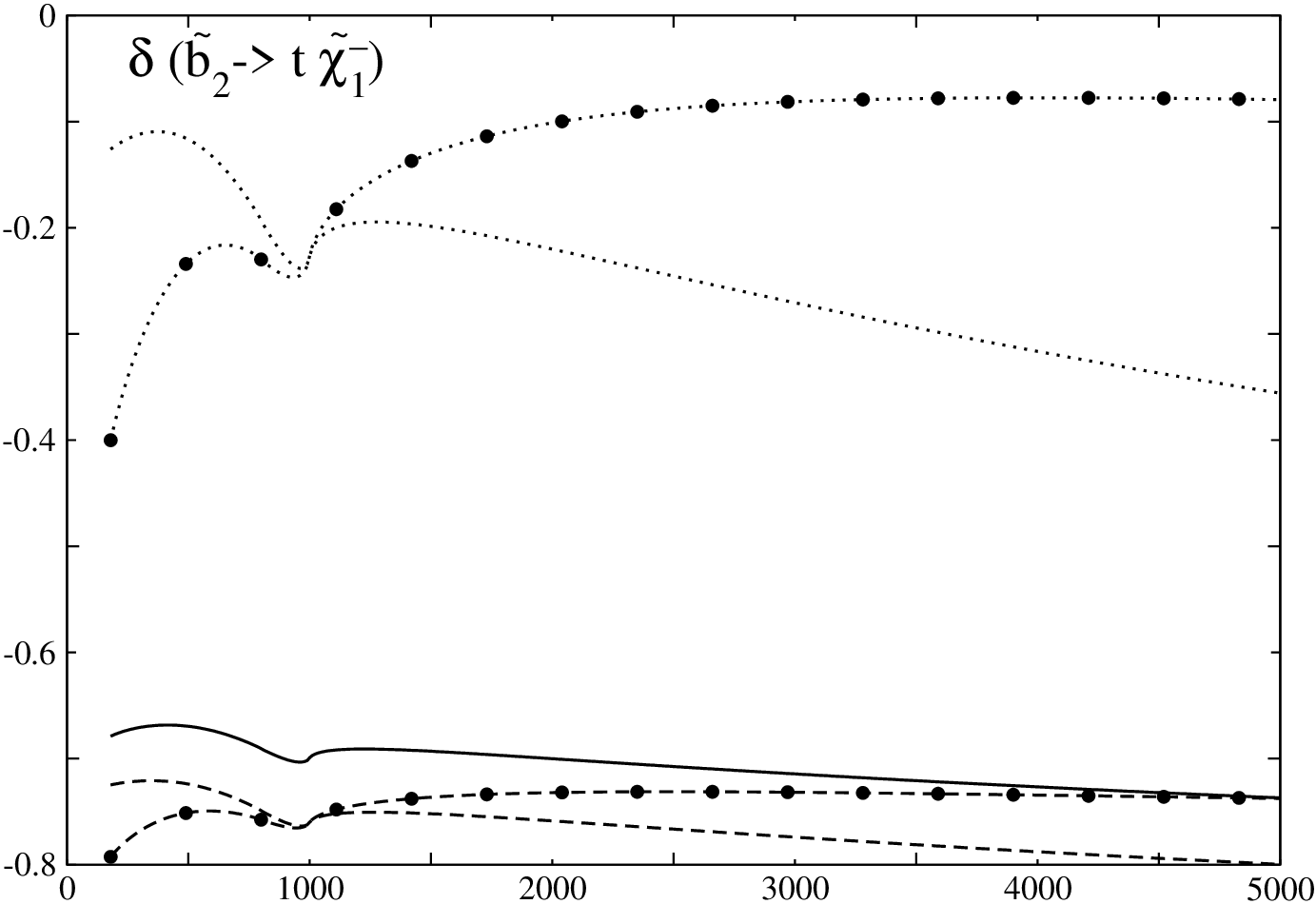}&
\includegraphics*[width=12.5cm]{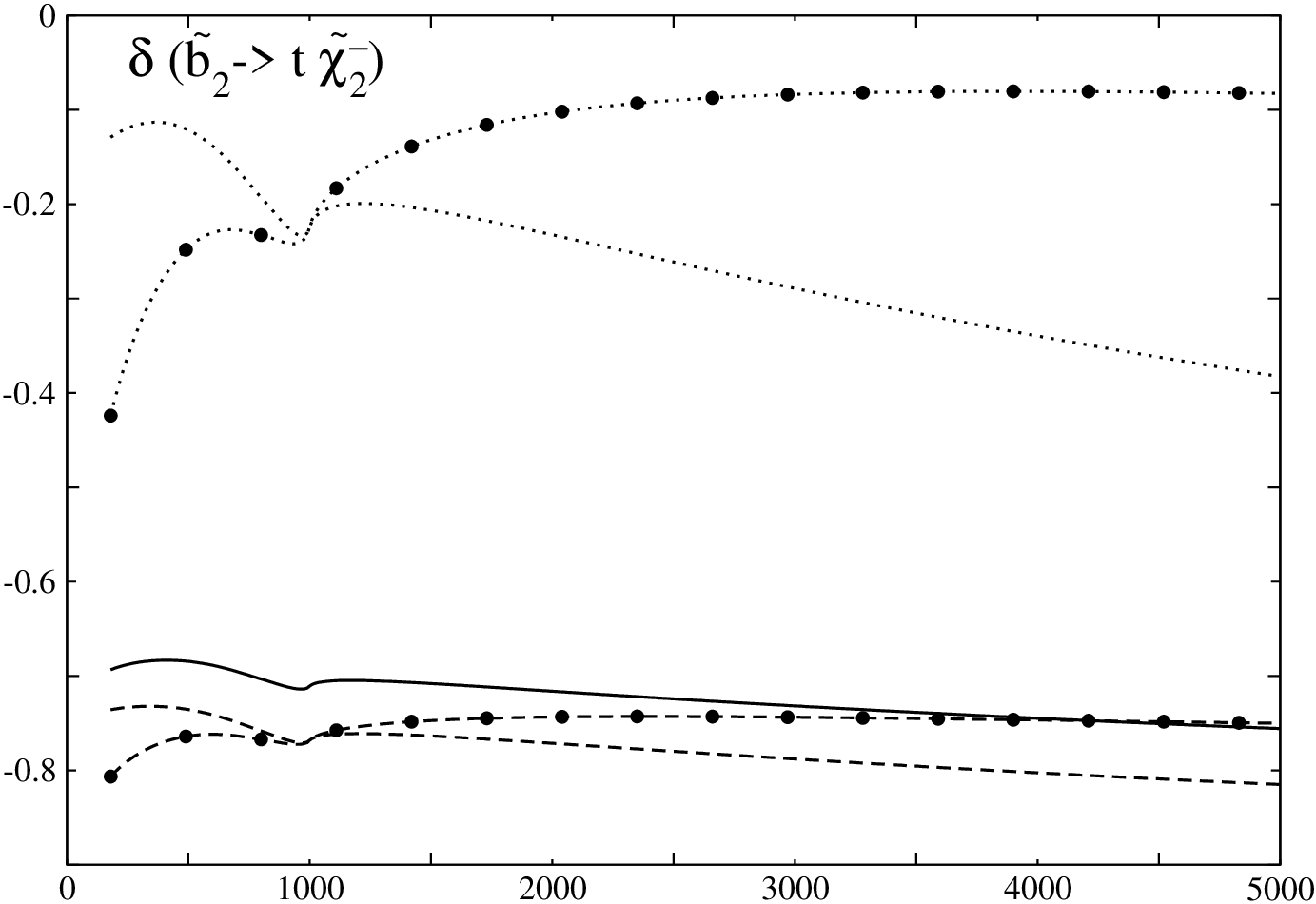}&
\includegraphics*[width=12.5cm]{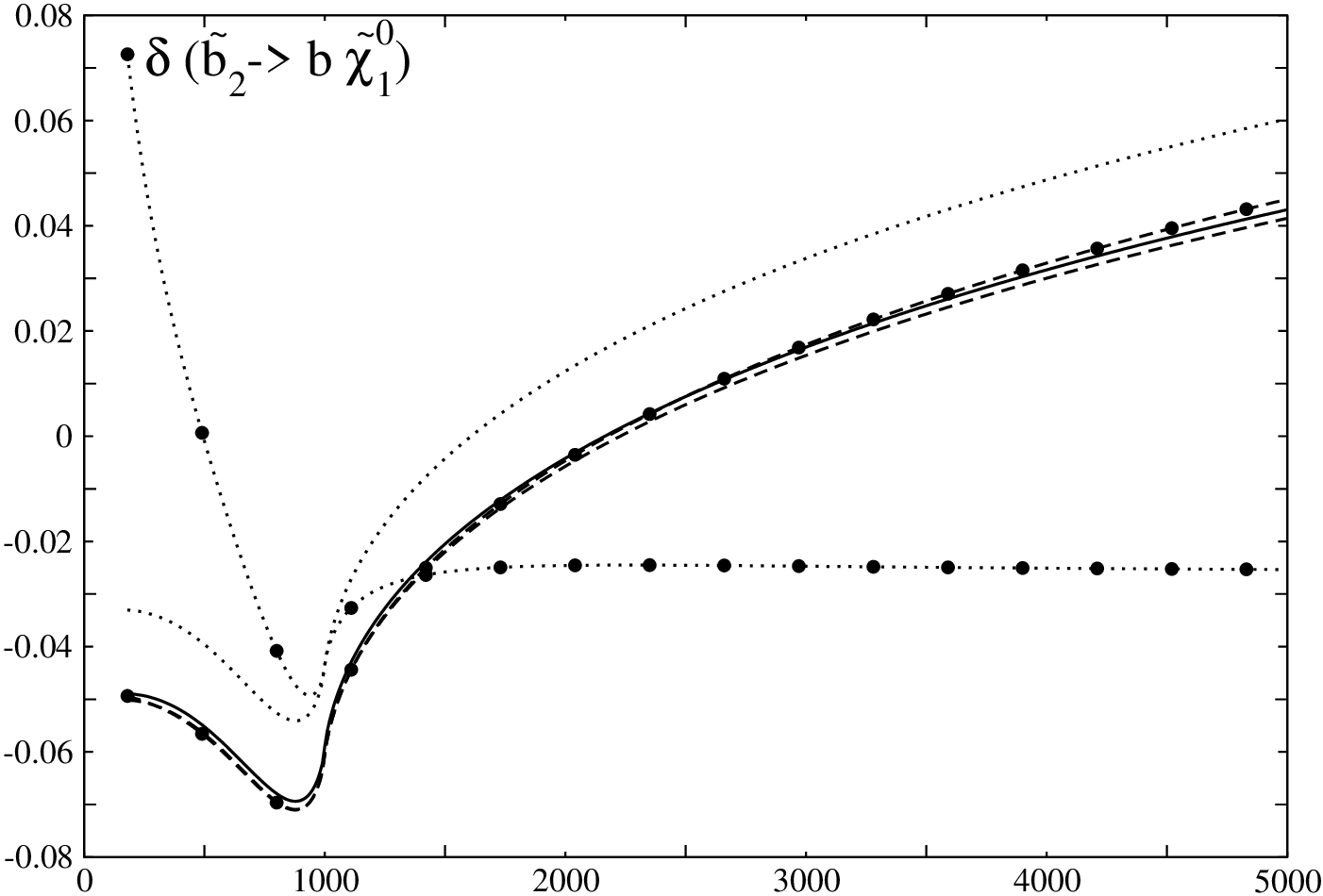}&
\includegraphics*[width=12.5cm]{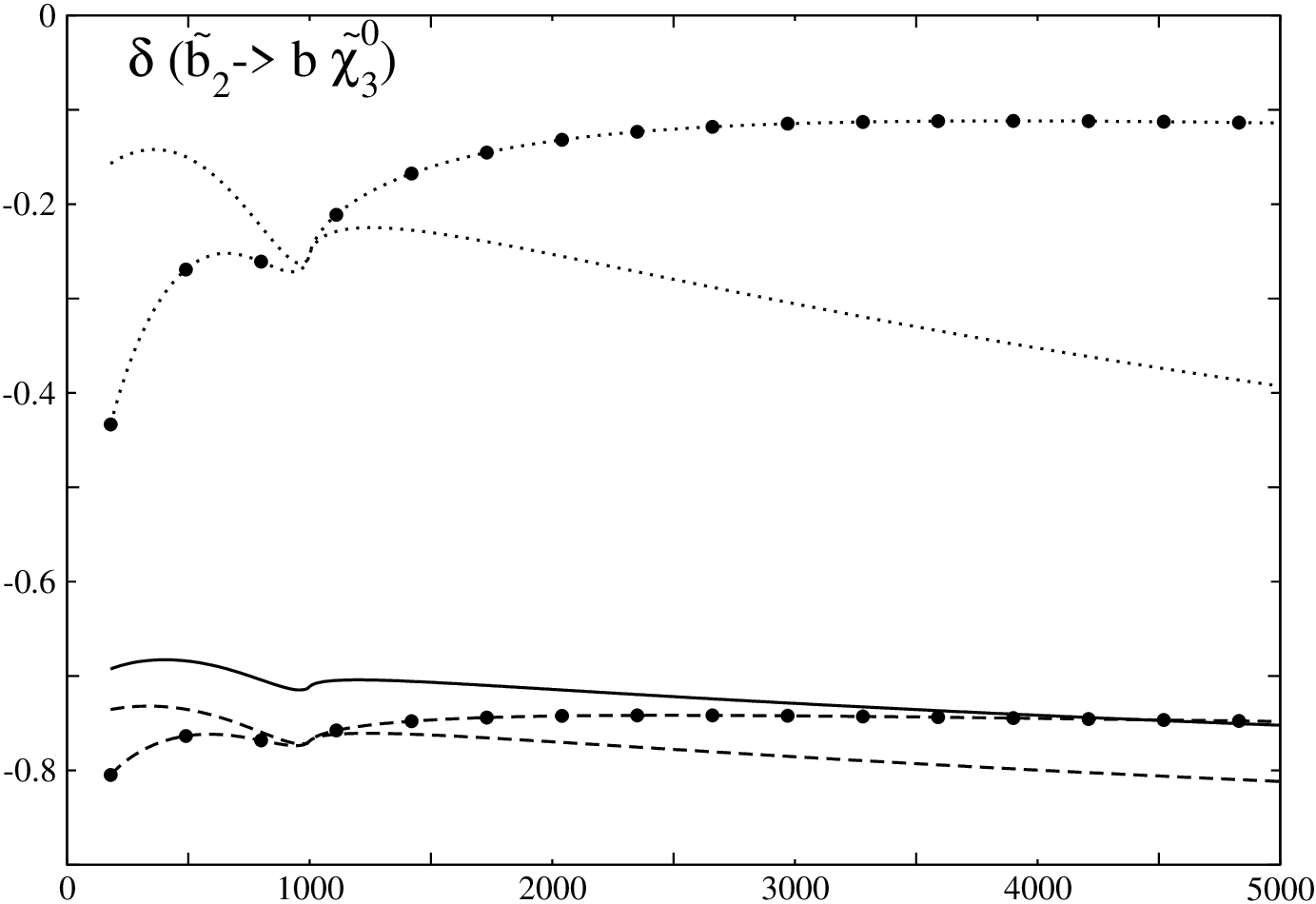}\\
\includegraphics*[width=12.5cm]{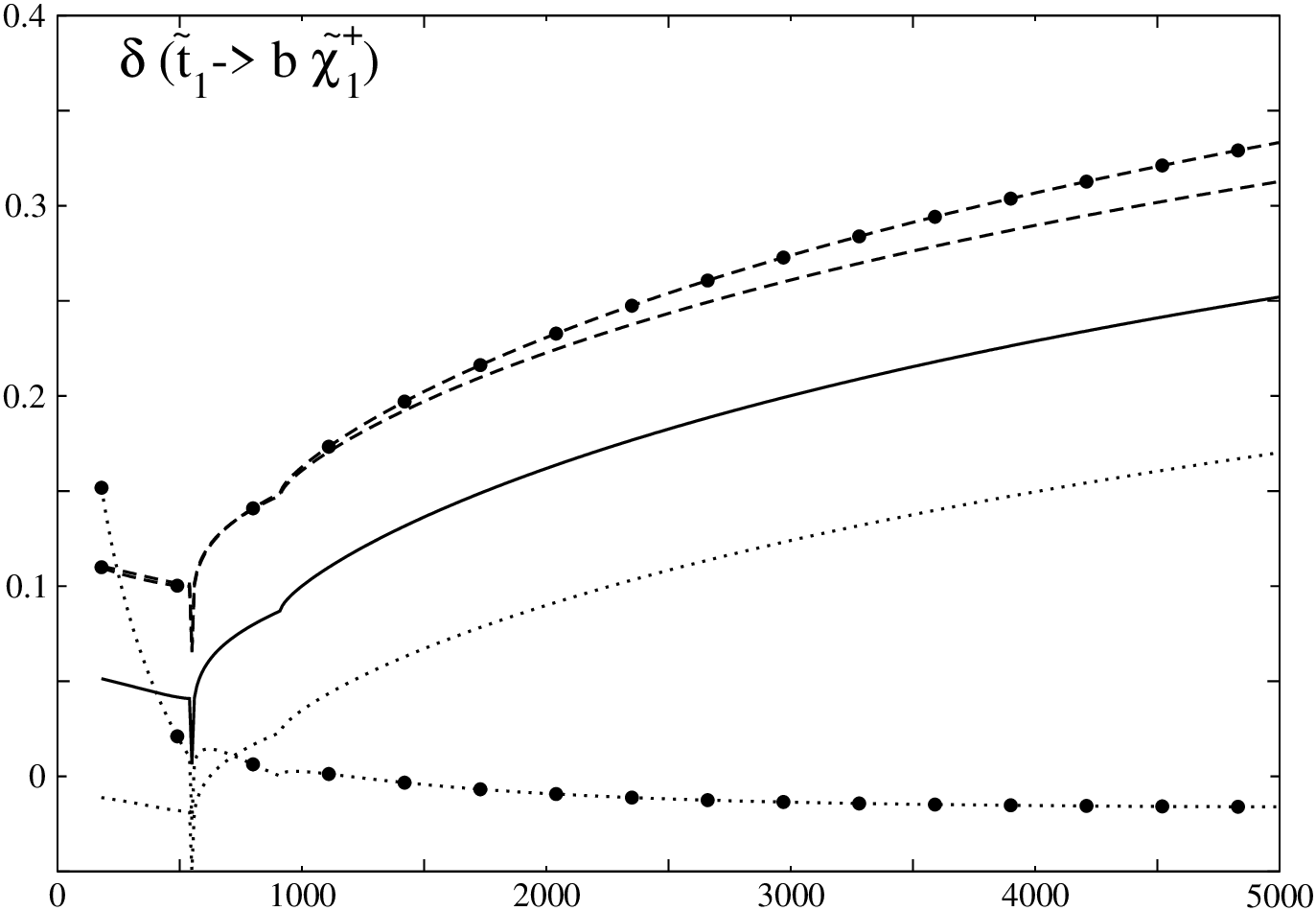}&
\includegraphics*[width=12.5cm]{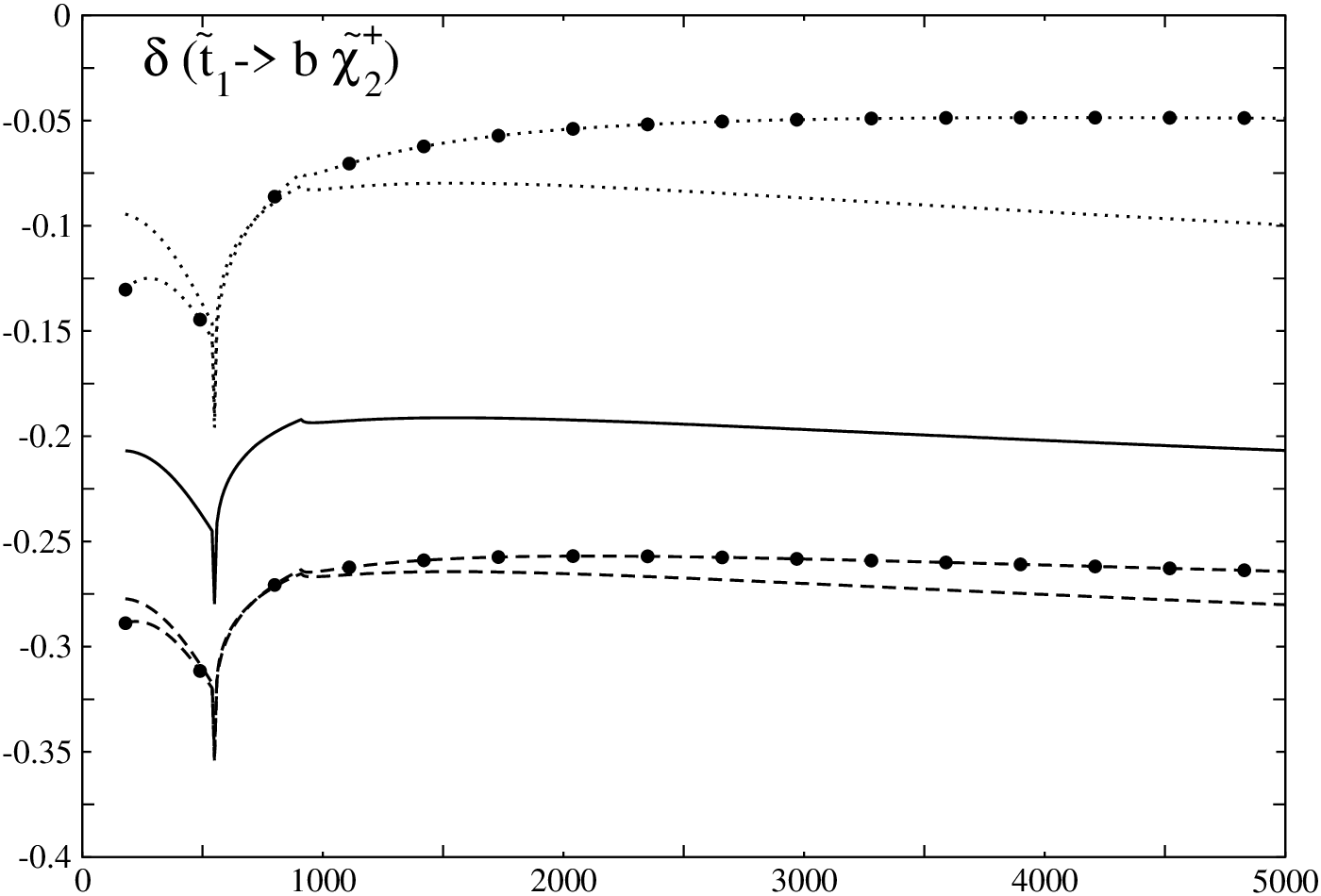}&
\includegraphics*[width=12.5cm]{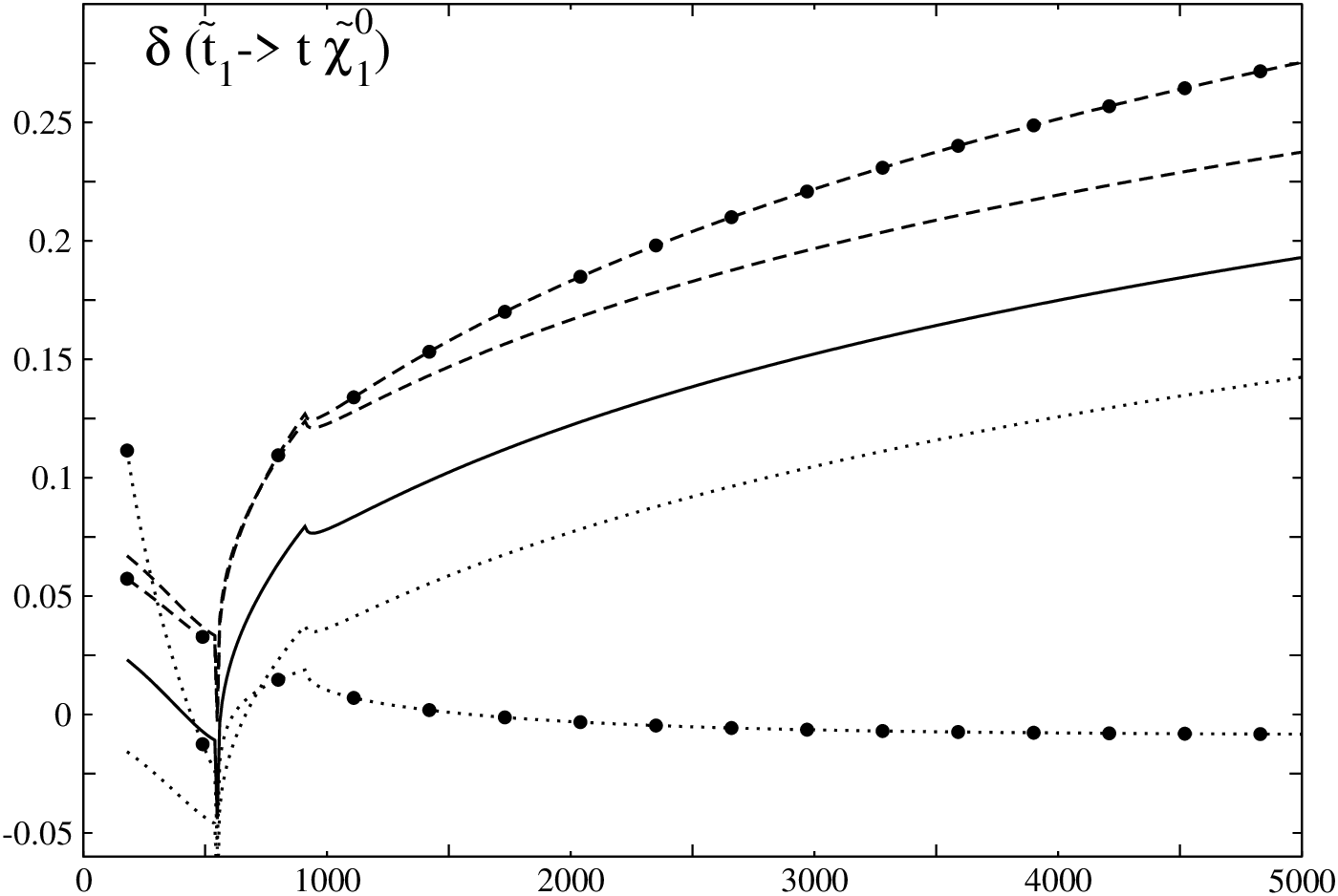}&
\includegraphics*[width=12.5cm]{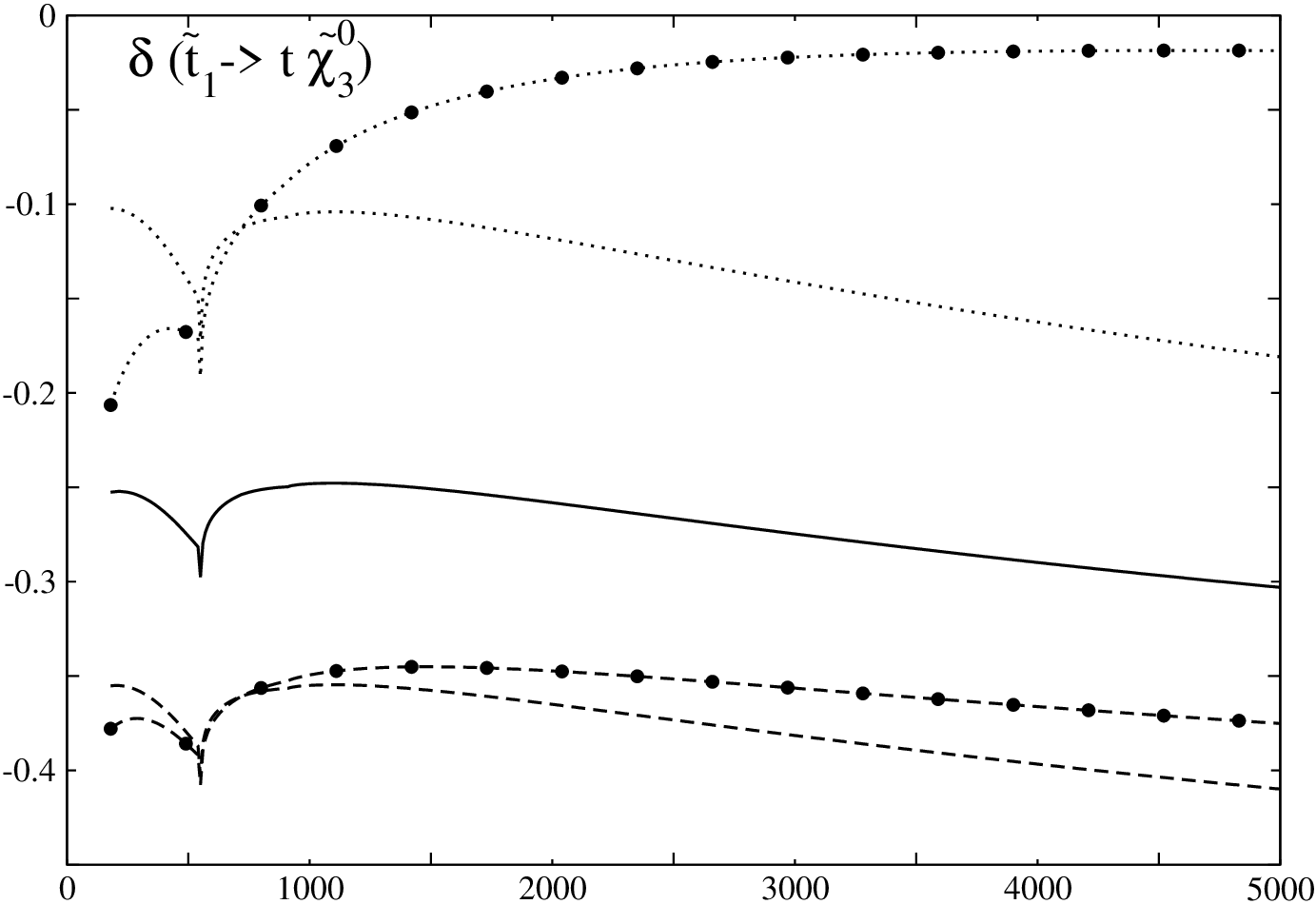}\\
\includegraphics*[width=12.5cm]{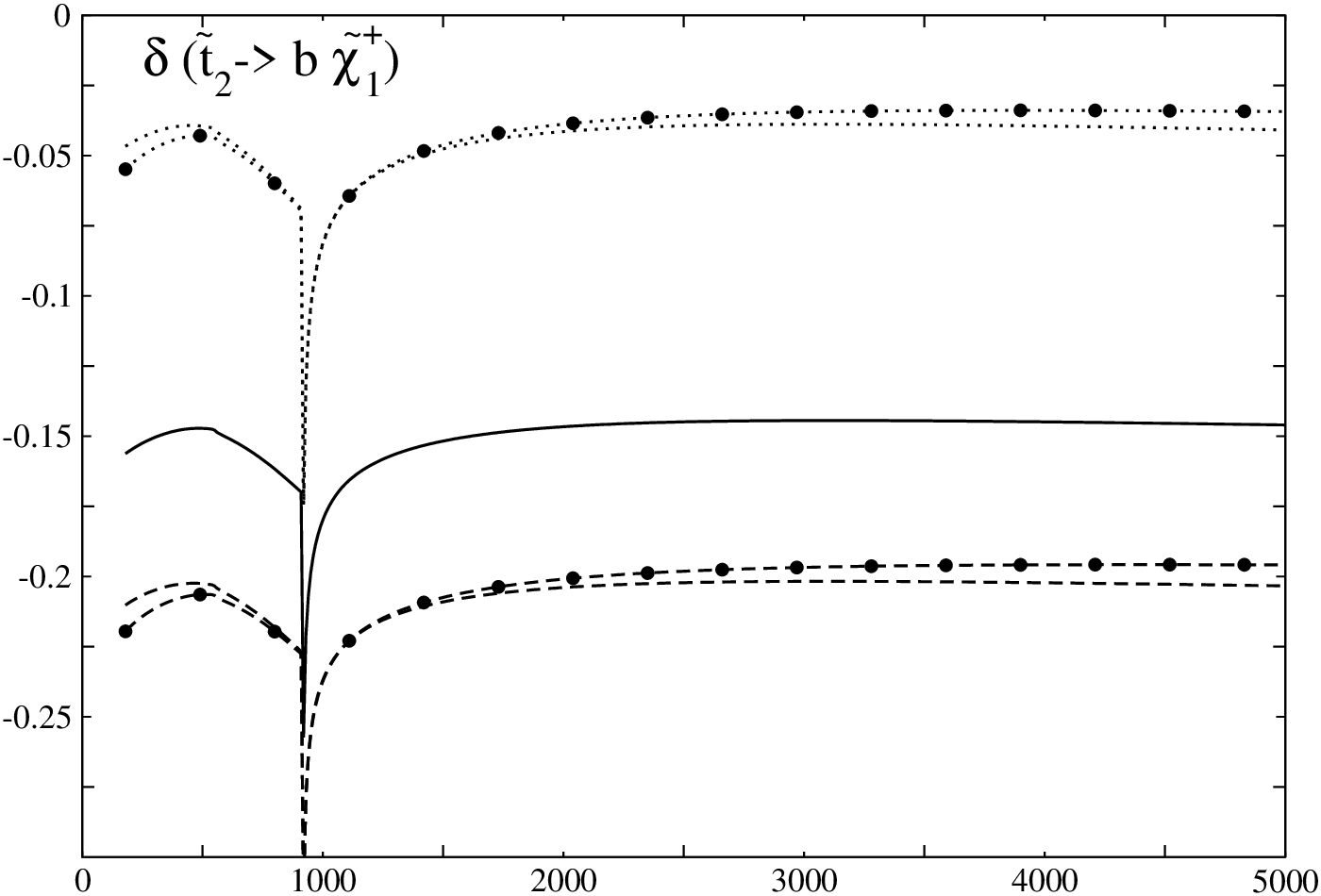}&
\includegraphics*[width=12.5cm]{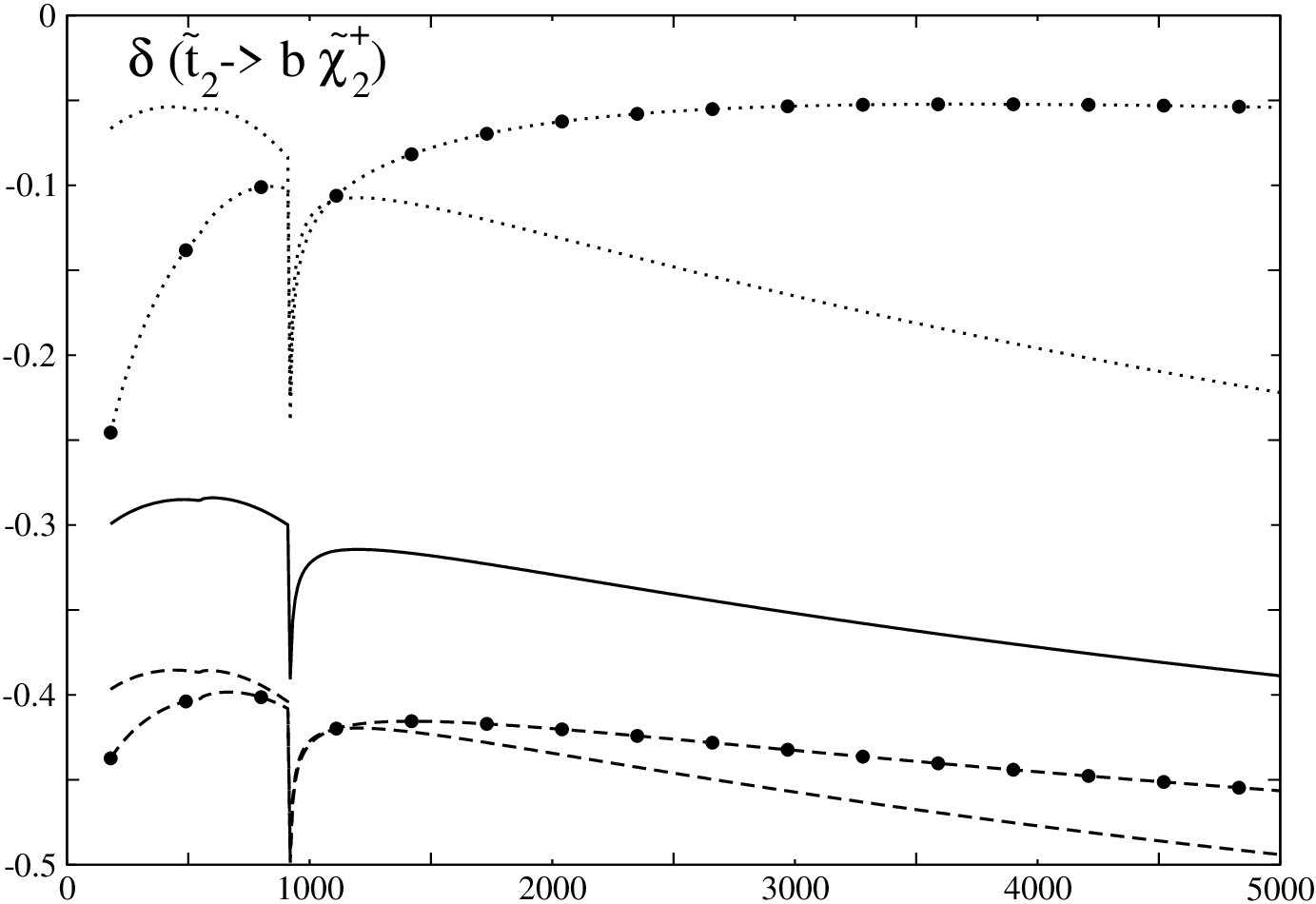}&
\includegraphics*[width=12.5cm]{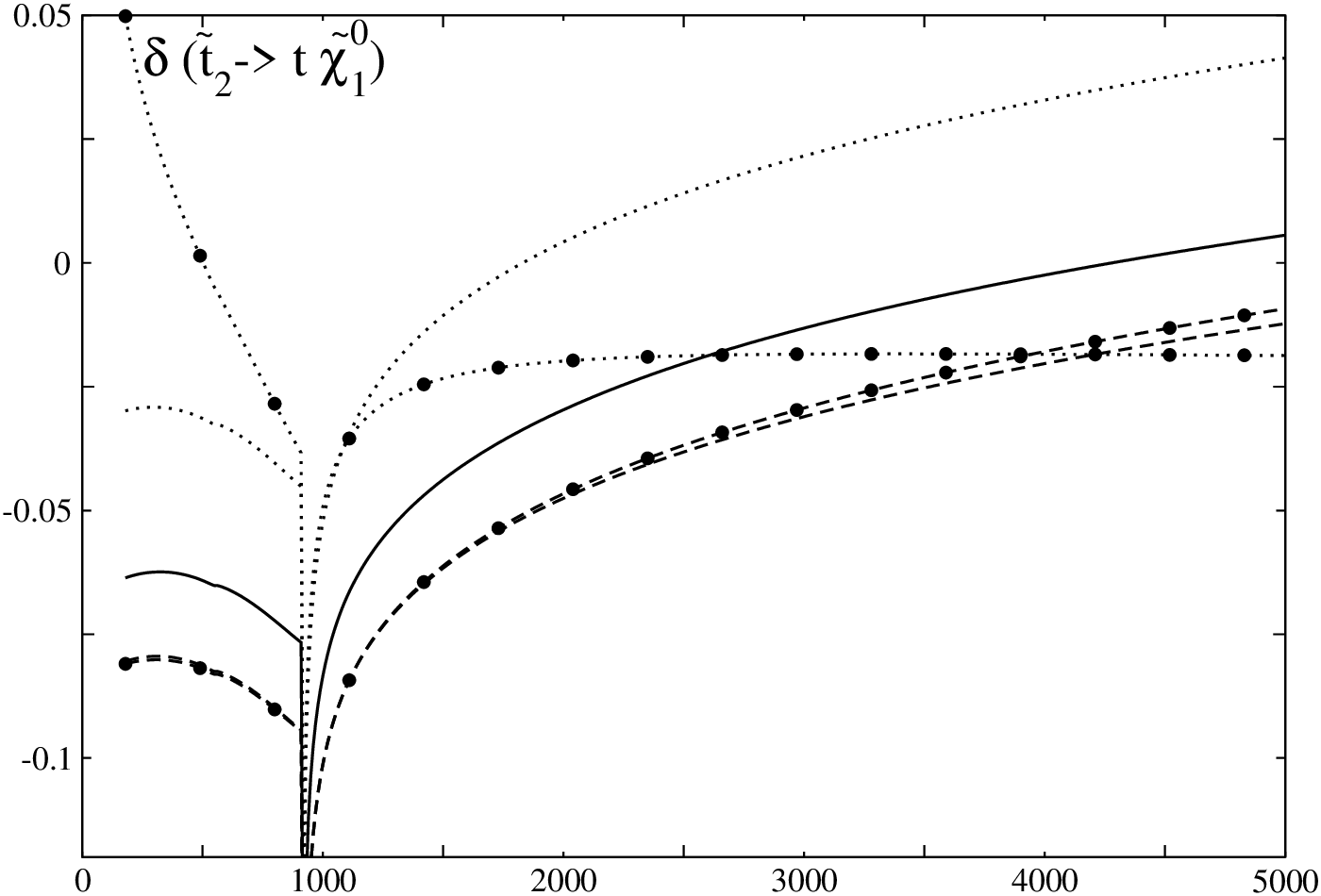}&
\includegraphics*[width=12.5cm]{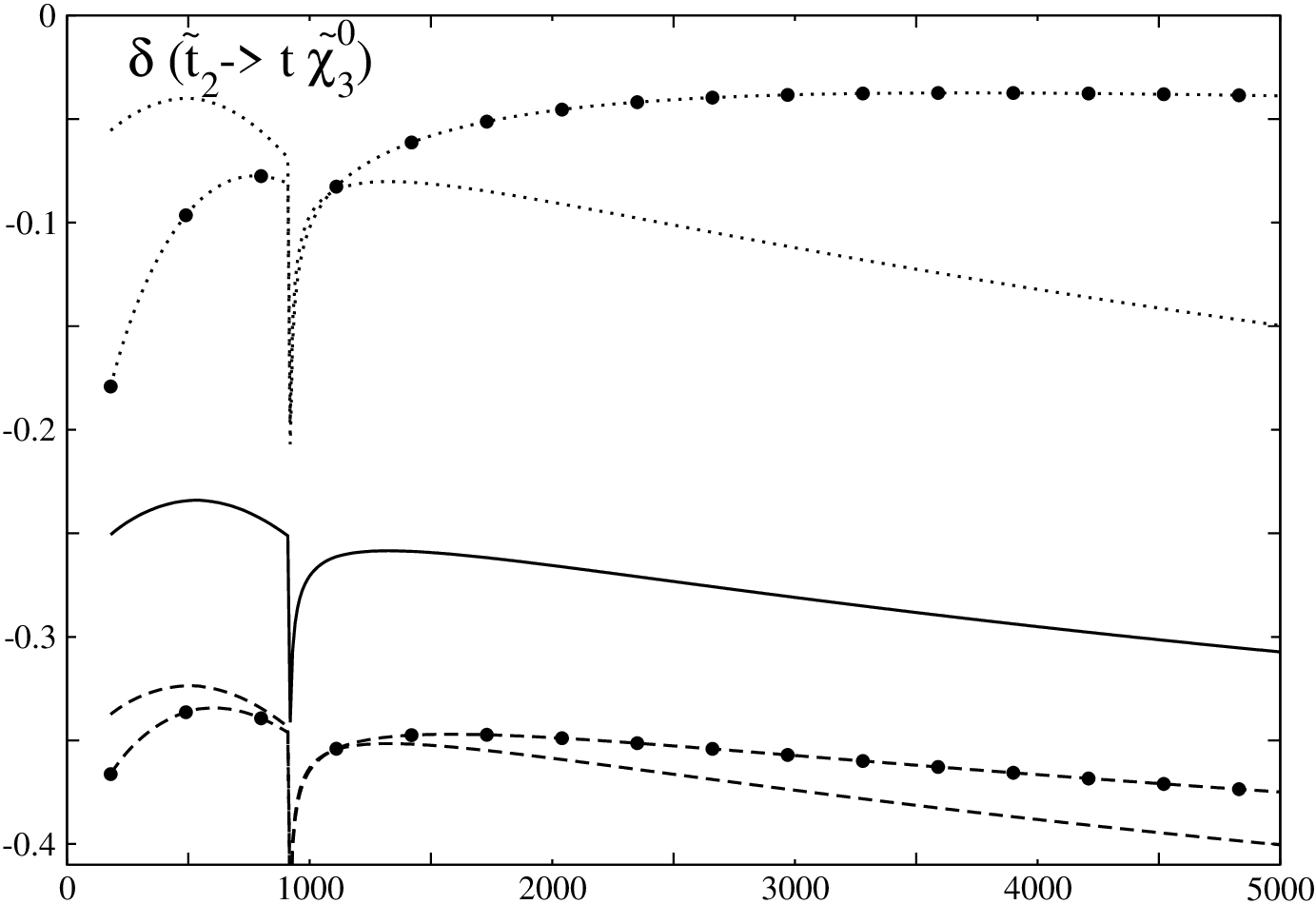}\\
\multicolumn{4}{c}{\huge $\mg\ ({\rm GeV})$}
\end{tabular}}}
\caption{Relative corrections of the
  squark decays into charginos and neutralinos, as a function of $\mg$. 
  We show the results  
  including only the effective Yukawa couplings,
  eq.~(\ref{eq:hbhbteff1}), and results including the
  $\log\mg/Q$ terms, eq.~(\ref{eq:logterms}) (lines with circles).
  The input parameters are given in eq.~(\ref{eq:SUSYparams}).}
\label{fig:mgl}
\end{figure}

As a first example we show in Fig.~\ref{fig:mgl} the 
relative corrections  
to the squark (stop and sbottom) partial decay widths
into charginos and neutralinos, as a function of the gluino mass
($\mg$). We show the predictions of the effective description including
only the effects of the effective Yukawa couplings~(\ref{eq:hbhbteff1})
(labelled \textit{Yuk.}),
and including also the $\log\mg/Q$ terms of eq.~(\ref{eq:logterms})
(lines marked with full circles). We show the partial decays
  into the two charginos and two neutralinos $\neut_1$, $\neut_3$, the
  results for the other neutralinos are similar to the ones shown. The
  second neutralino ($\neut_2$) is mostly a gaugino ($\tilde{w}$) and
  its results are very similar to $\cmin_1$, whereas the fourth
  neutralino ($\neut_4$) is mostly of higgsino-type ($\tilde{h}$) and
  its results are very similar to $\neut_3$.
In all plots we observe the same pattern: the
effective Yukawa couplings~(\ref{eq:hbhbteff1}) do not describe
correctly the variation with $\mg$, only after including the log-terms
of eq.~(\ref{eq:logterms}) does the effective description follow the shape
of the one-loop corrections. The \textit{remainder} corrections
$\delta^{Yuk.-rem.}$~(\ref{eq:defremainder}) -- those terms that are not
described by the effective couplings -- have a dependence on $\mg$
before including the log-terms, but after including them we see that in
all channels $\delta^{rem.}$ is essentially flat above 
$\mg\sim1500\GeV$, which means that they have absorbed the bulk of the
dependence on $\mg$. Moreover, after including the log-terms
$\delta^{rem.}$ is much smaller -- between a $-2\%$ and a $-5\%$ -- than
without them -- between a $5\%$ and a $-10\%$. The effects of the
log-terms are more visible in the gaugino-like channels, where the
Yukawa couplings play no role, and the bulk of the corrections
corresponds to the log-terms. In these channels the corrections change in
a range of a $10\%$ in the interval $\mg=[500,5000]\GeV$. In the
higgsino-like channels their importance is less apparent. On one side
the effective Yukawa couplings carry the bulk of the corrections (around
a $-30\%$), and on the other side the sign difference among the two
contributions (\ref{eq:logterms}) make them to partially compensate each
other. But also in this case the effective Yukawa couplings
alone~(\ref{eq:hbhbteff1}) do not describe correctly the
$\mg$-dependence, and there is a range of a $3\%$ variation in the
corrections in the studied $\mg$-interval. 
The spikes that are seen in some plots correspond to the
  threshold singularities for the opening of the squark decay into gluinos
  ($\tilde{q}\rightarrow\tilde{g}q$), we recall that in this work we are
  not interested in the region where the gluino decay channel is open
  ($\mg+m_q<\msq$), since in that region the strong decay is the leading
  one, and the chargino/neutra\-lino channels have a negligible branching
  ratio, and are phenomenologically irrelevant. Nevertheless we include the
  plots also in that region to show the trend of the corrections.

\begin{figure}[tbp]
\centering
\rotatebox{90}{\resizebox{20cm}{!}{\begin{tabular}{cccc}
\includegraphics*[width=12.5cm]{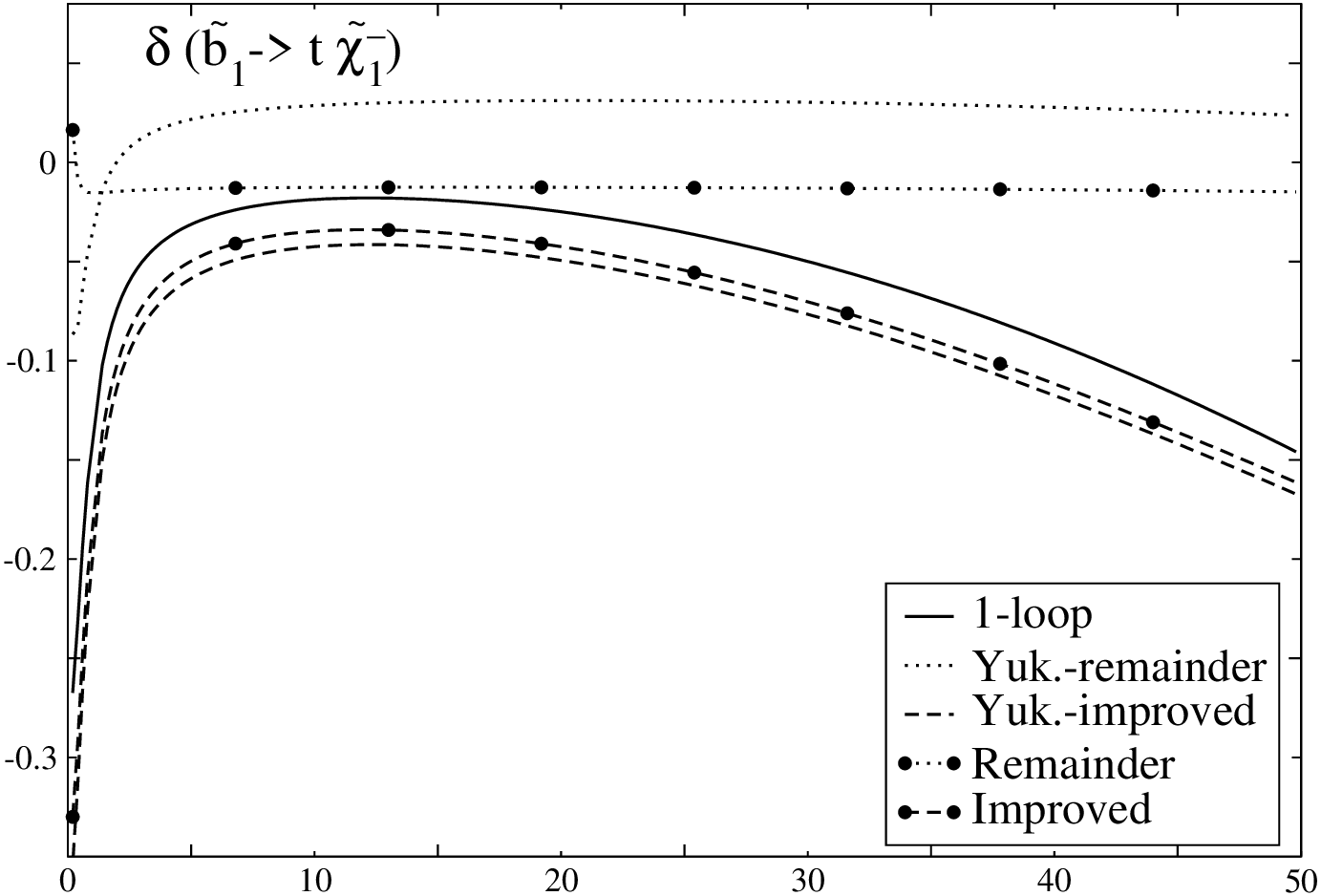}&
\includegraphics*[width=12.5cm]{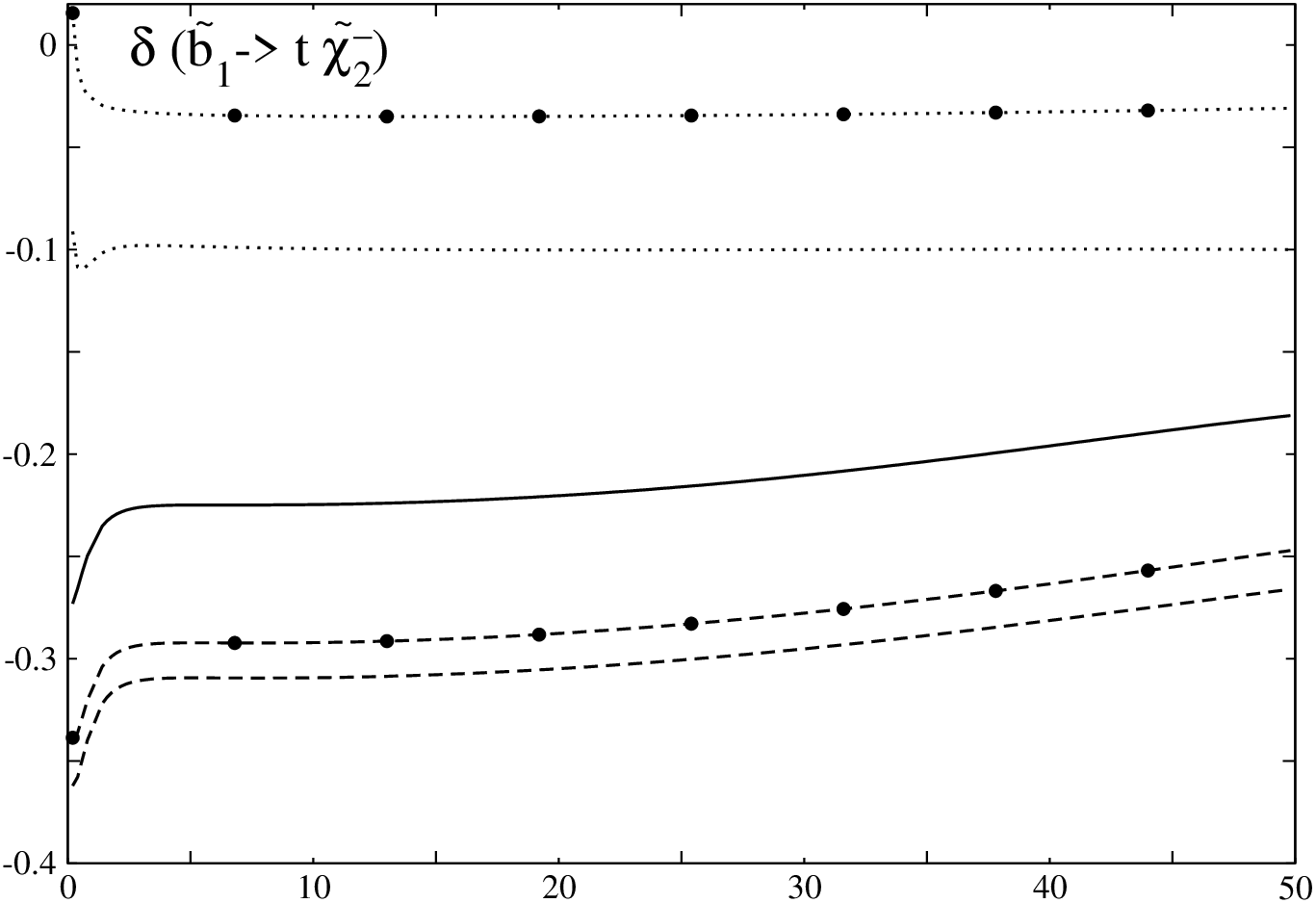}&
\includegraphics*[width=12.5cm]{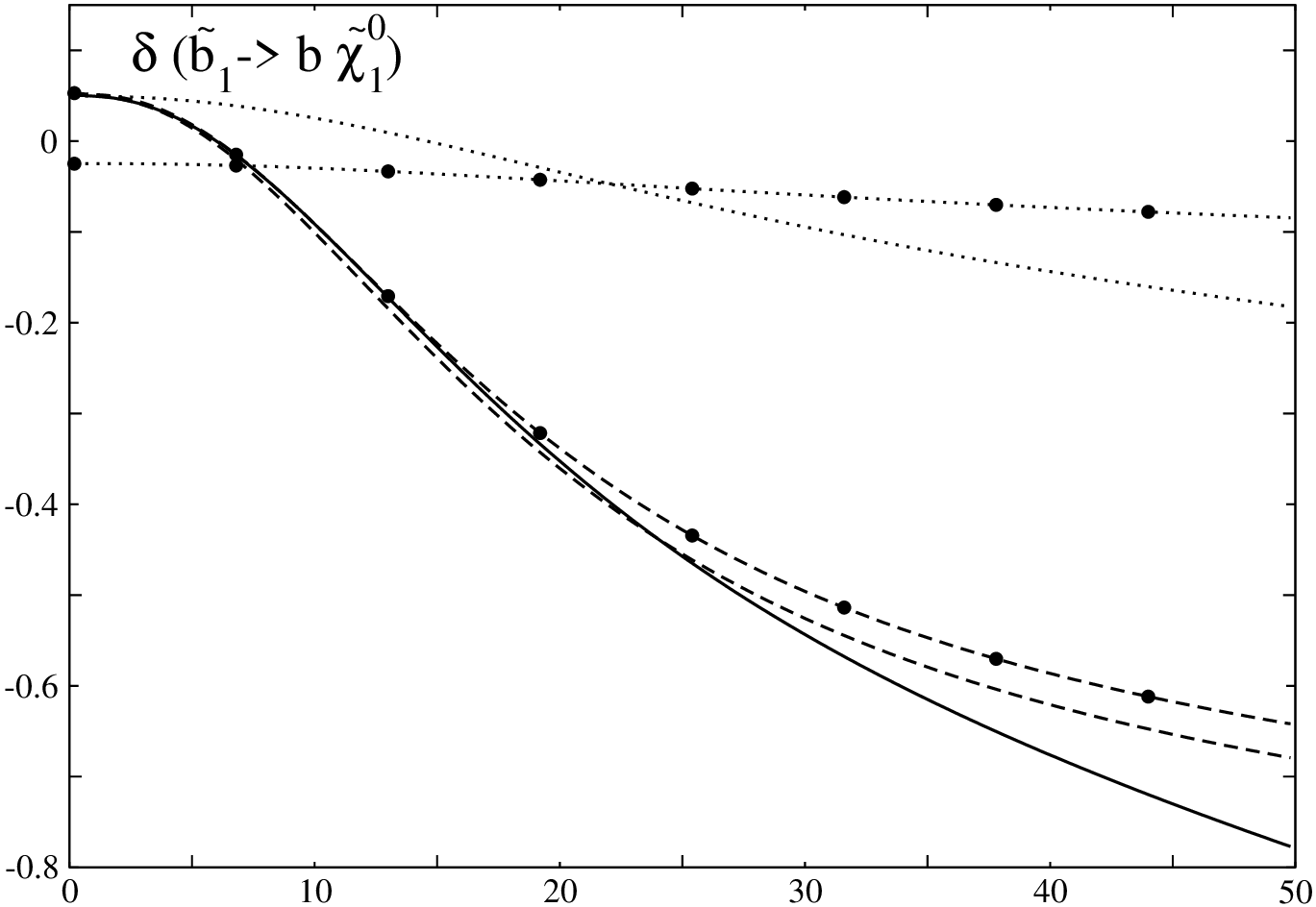}&
\includegraphics*[width=12.5cm]{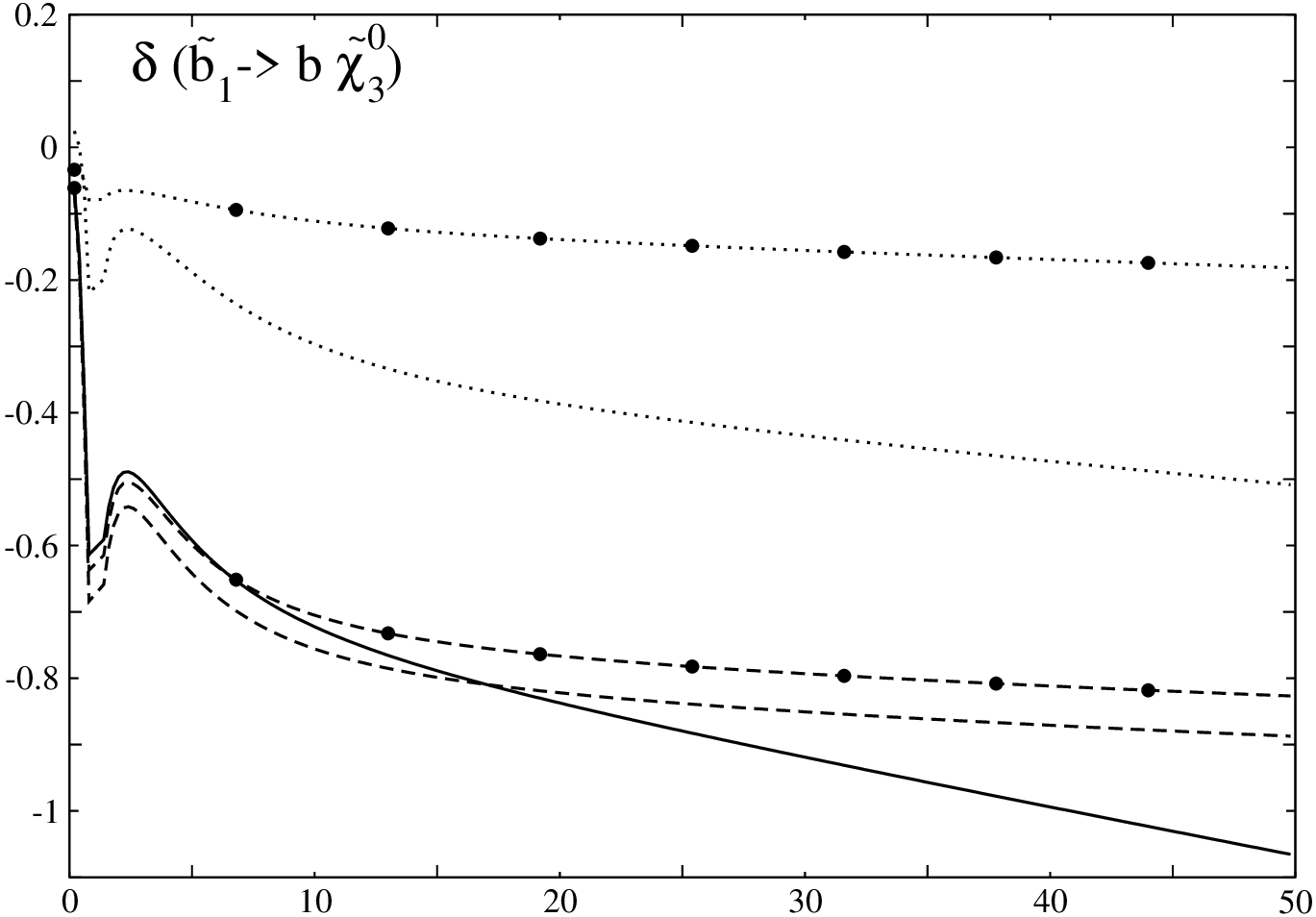}\\
\includegraphics*[width=12.5cm]{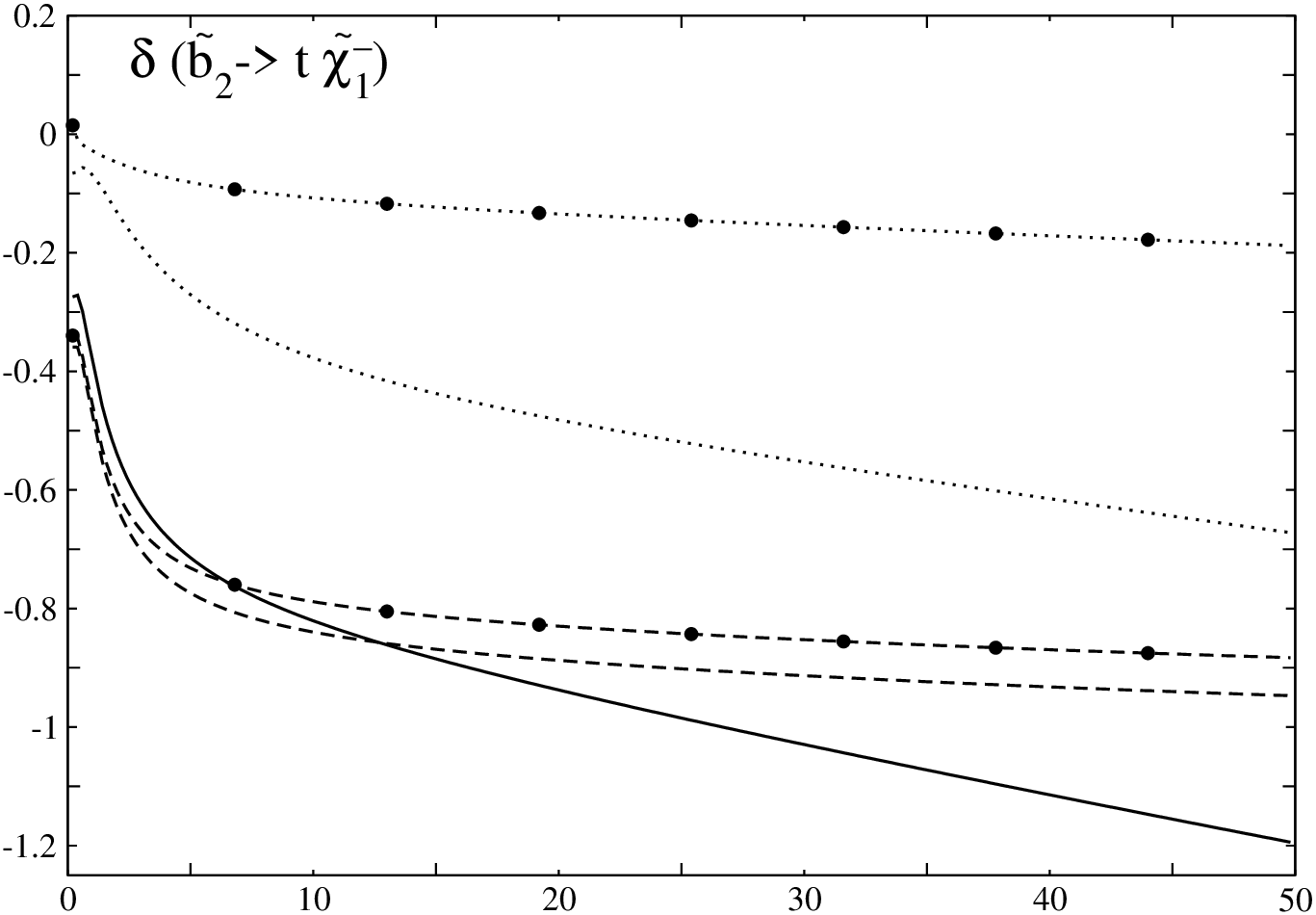}&
\includegraphics*[width=12.5cm]{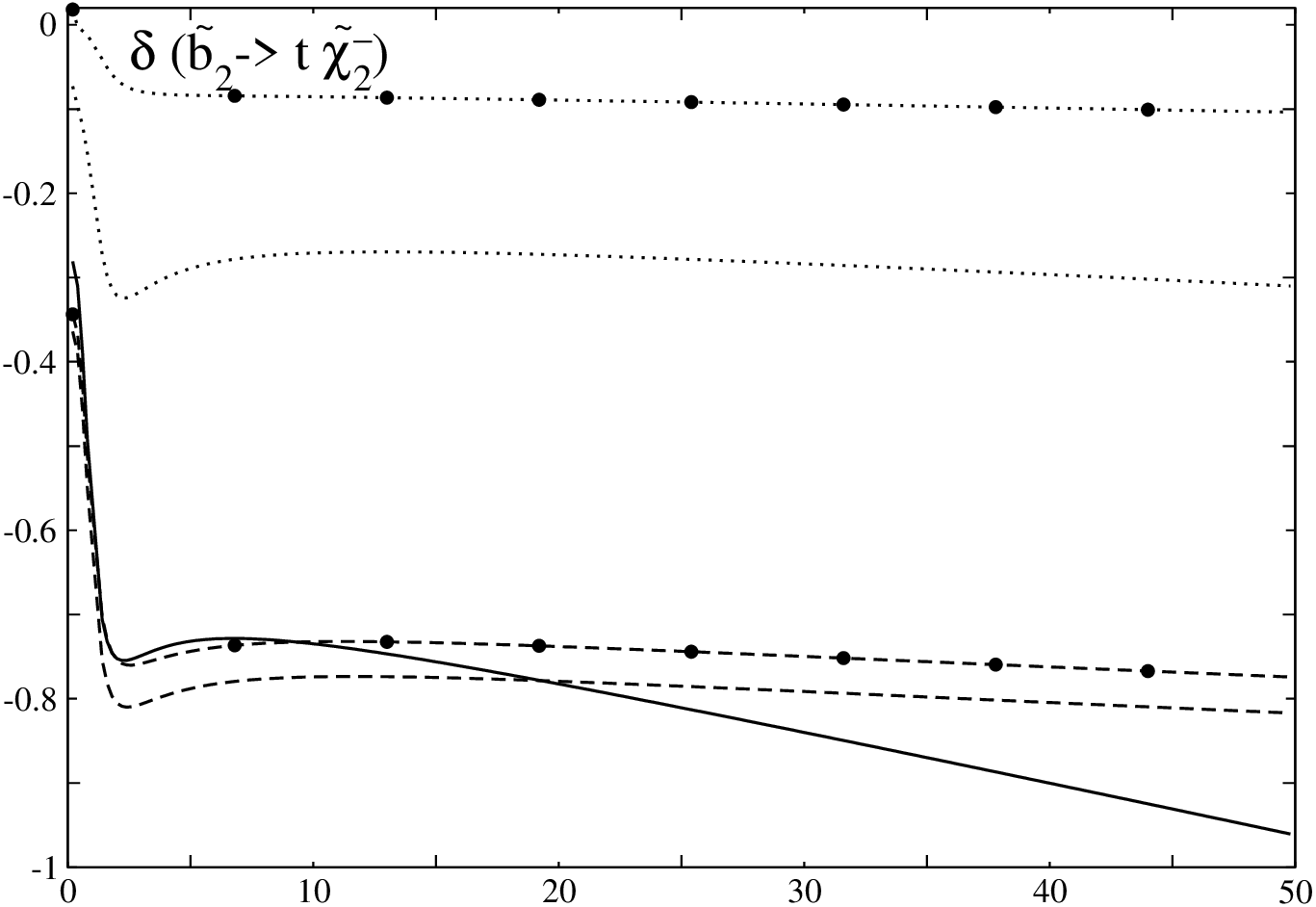}&
\includegraphics*[width=12.5cm]{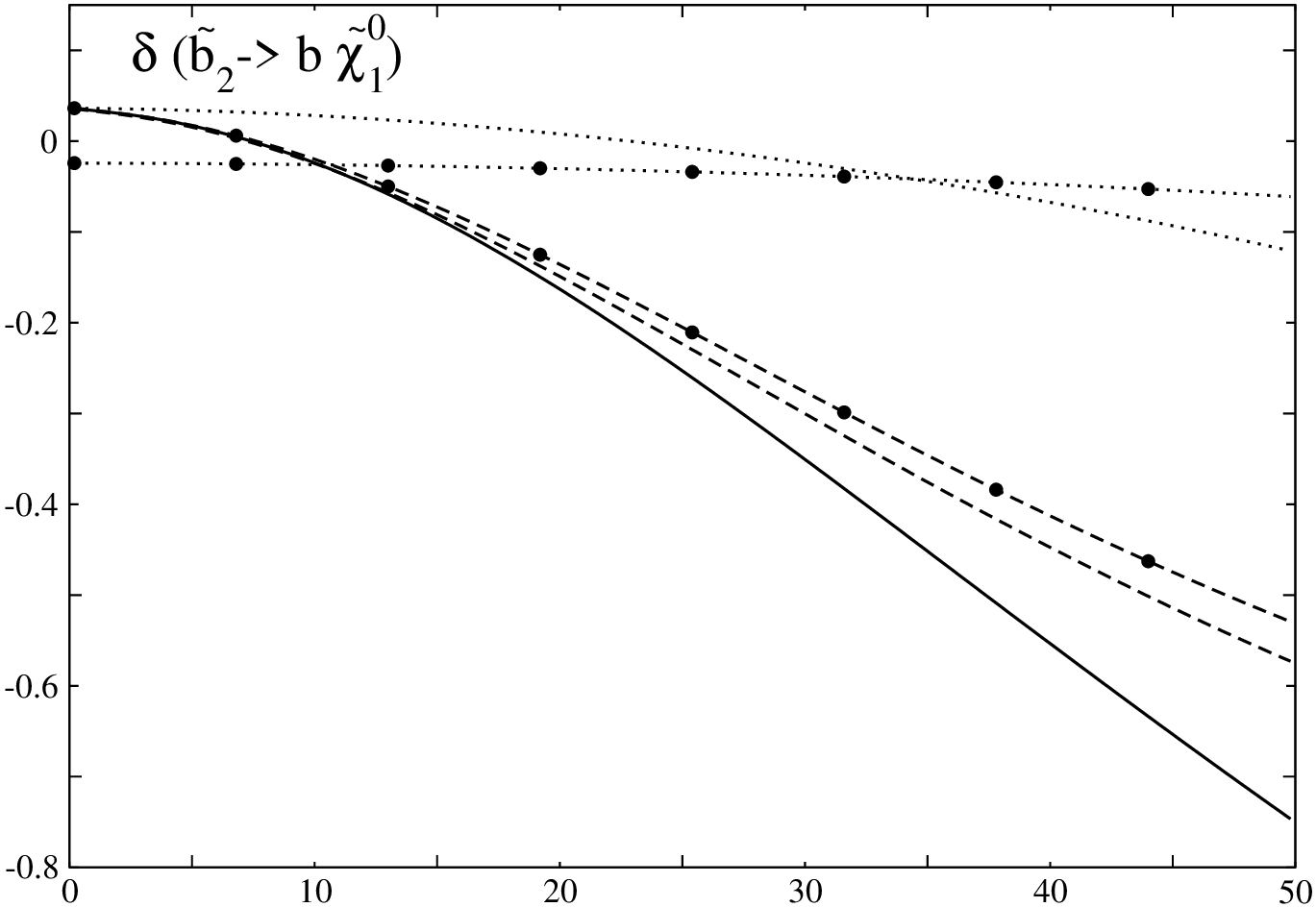}&
\includegraphics*[width=12.5cm]{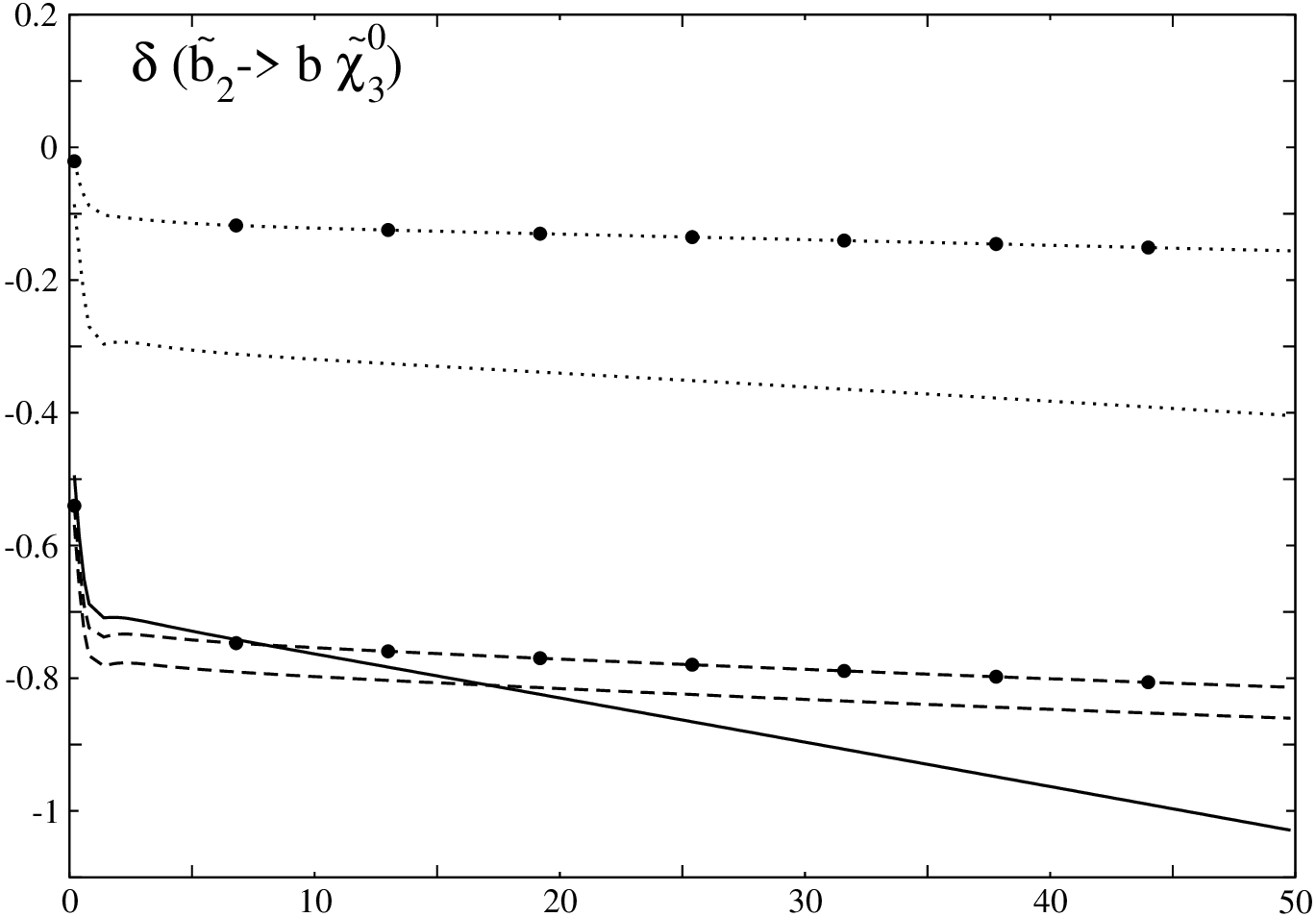}\\
\includegraphics*[width=12.5cm]{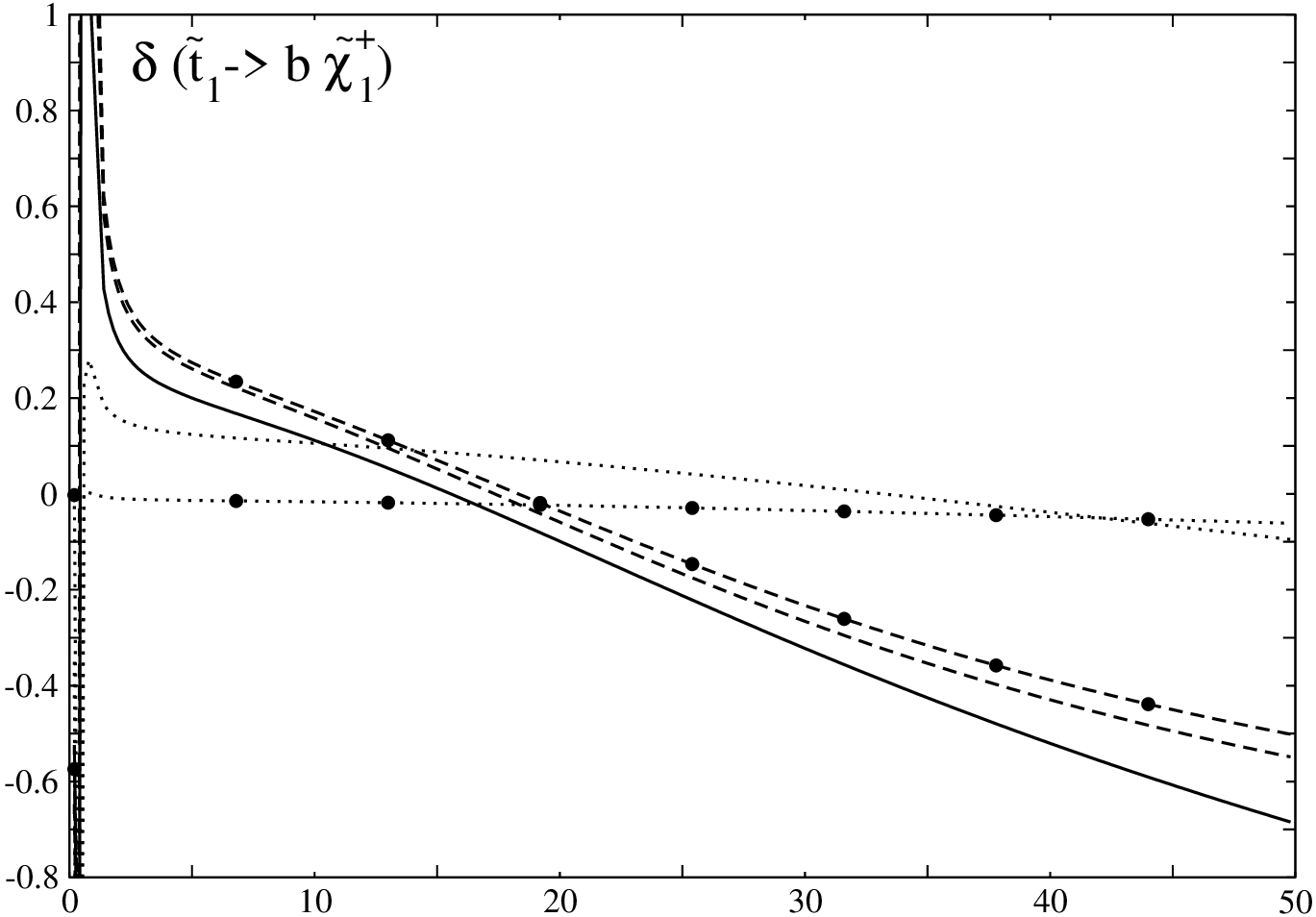}&
\includegraphics*[width=12.5cm]{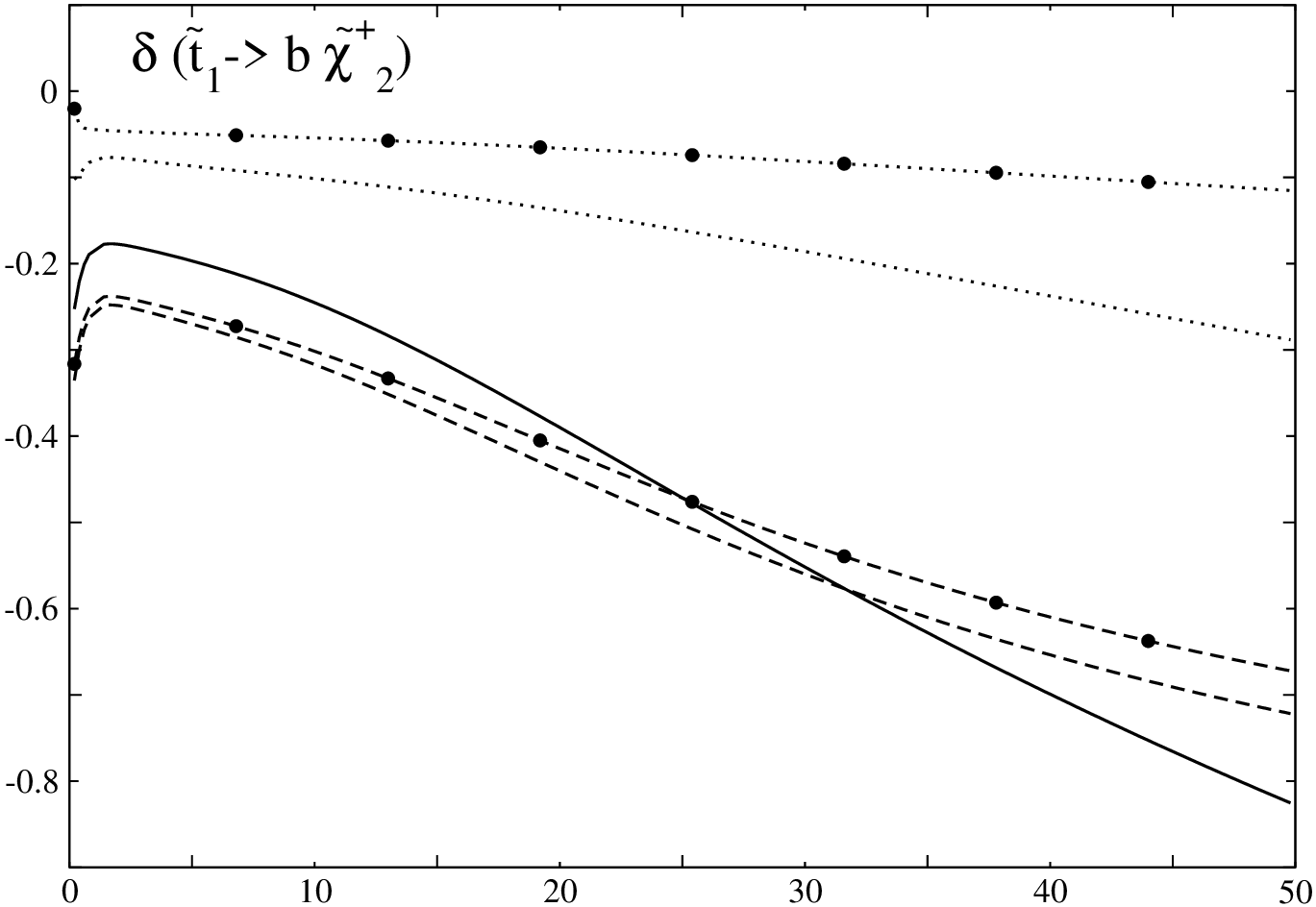}&
\includegraphics*[width=12.5cm]{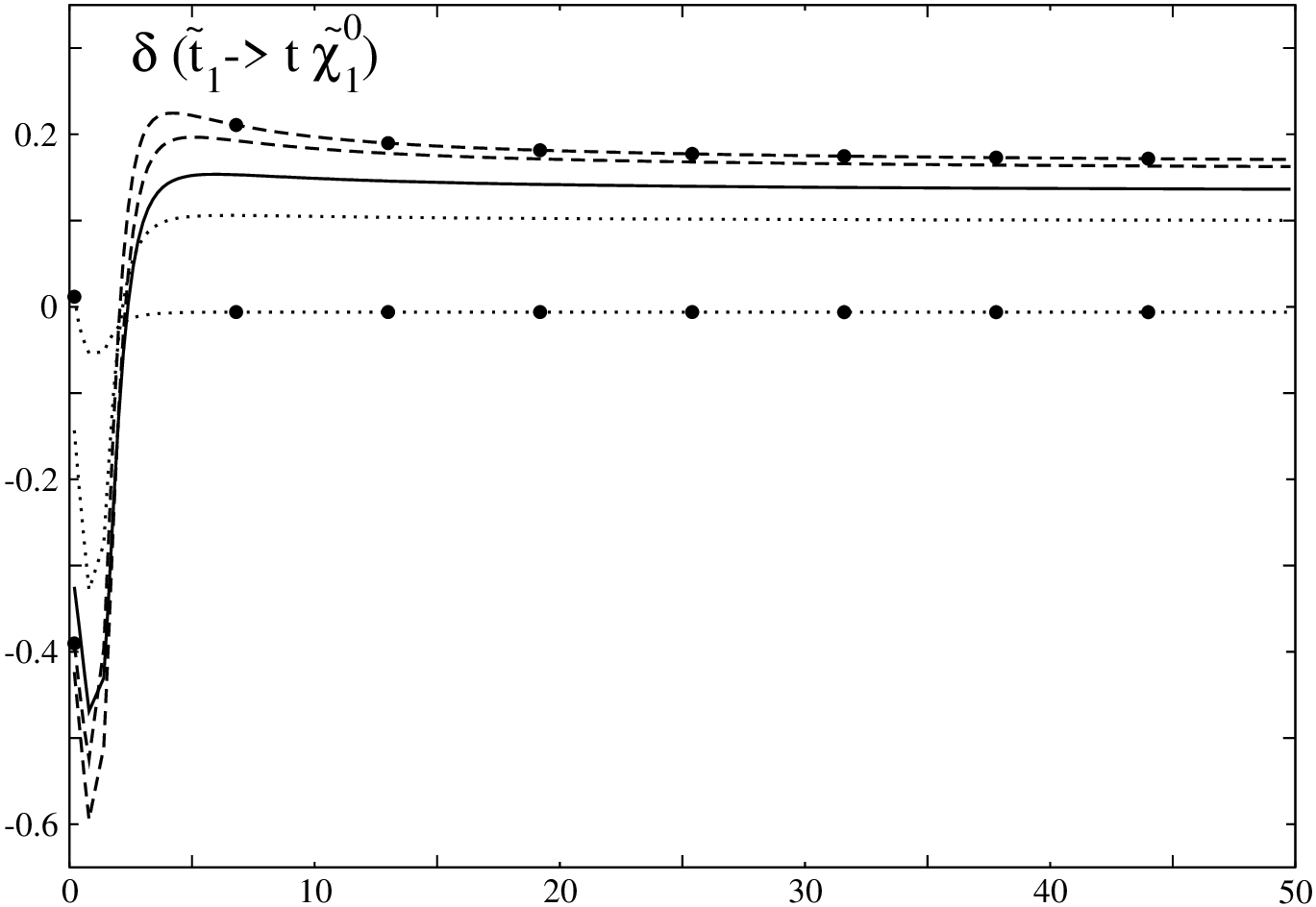}&
\includegraphics*[width=12.5cm]{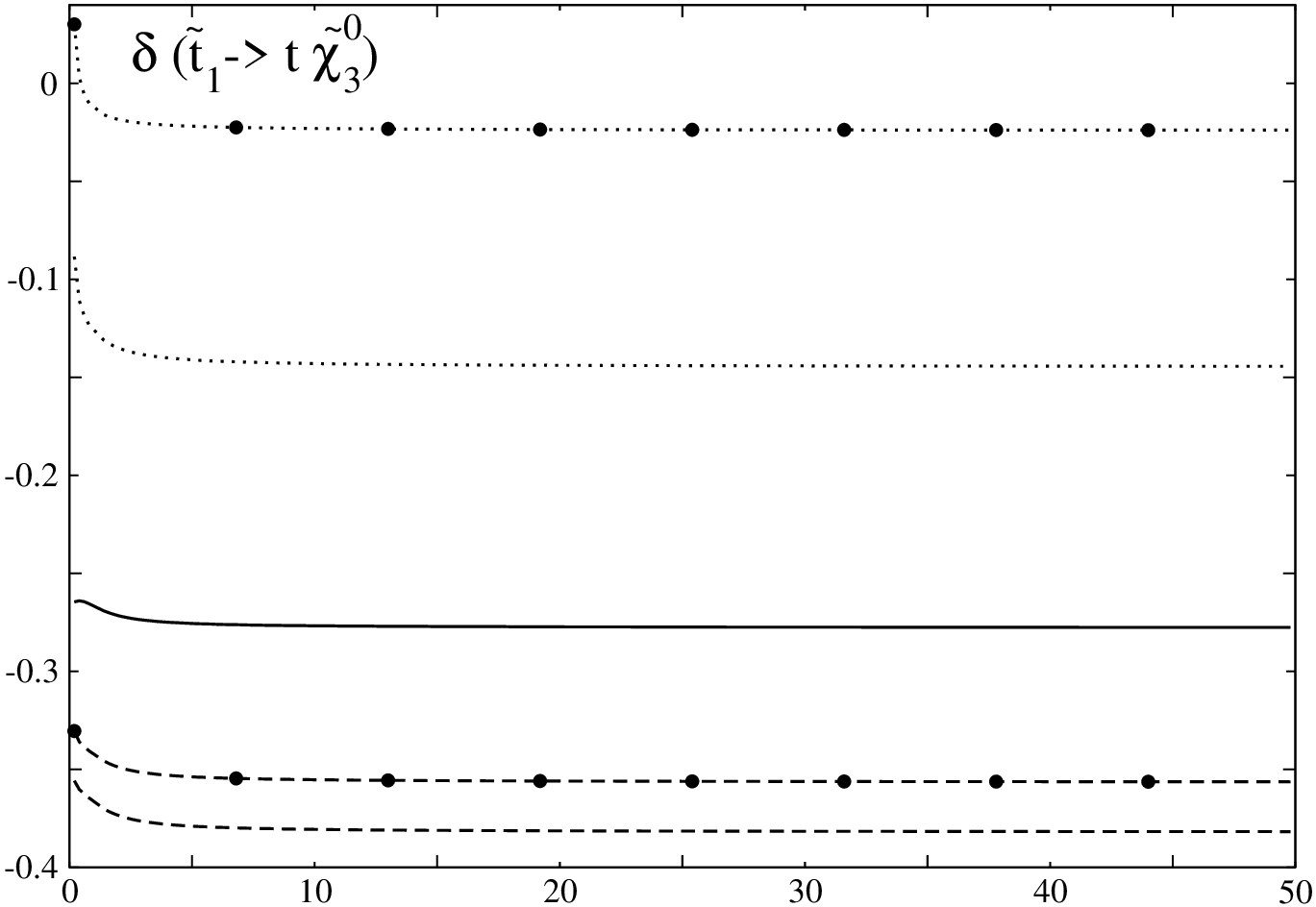}\\
\includegraphics*[width=12.5cm]{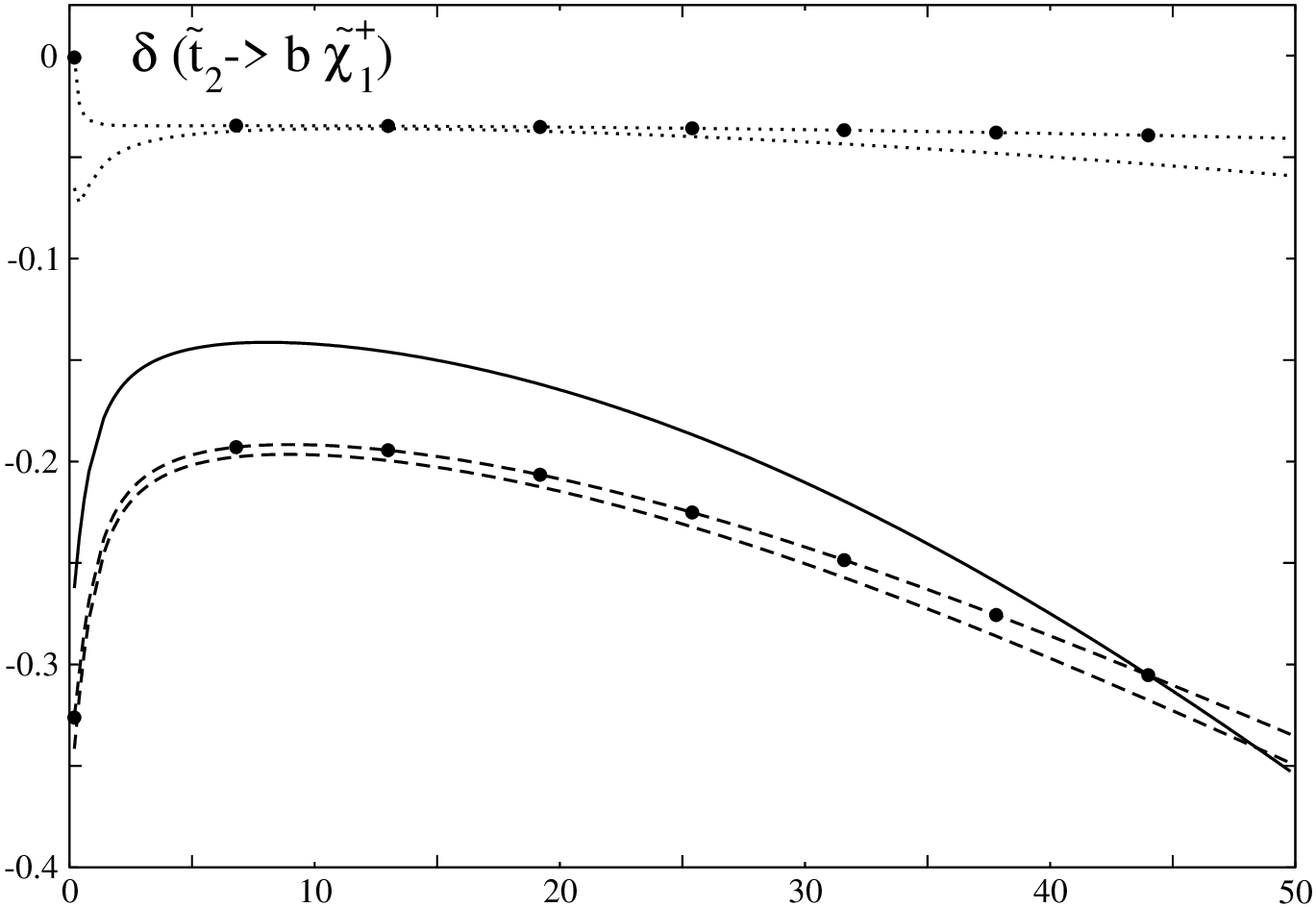}&
\includegraphics*[width=12.5cm]{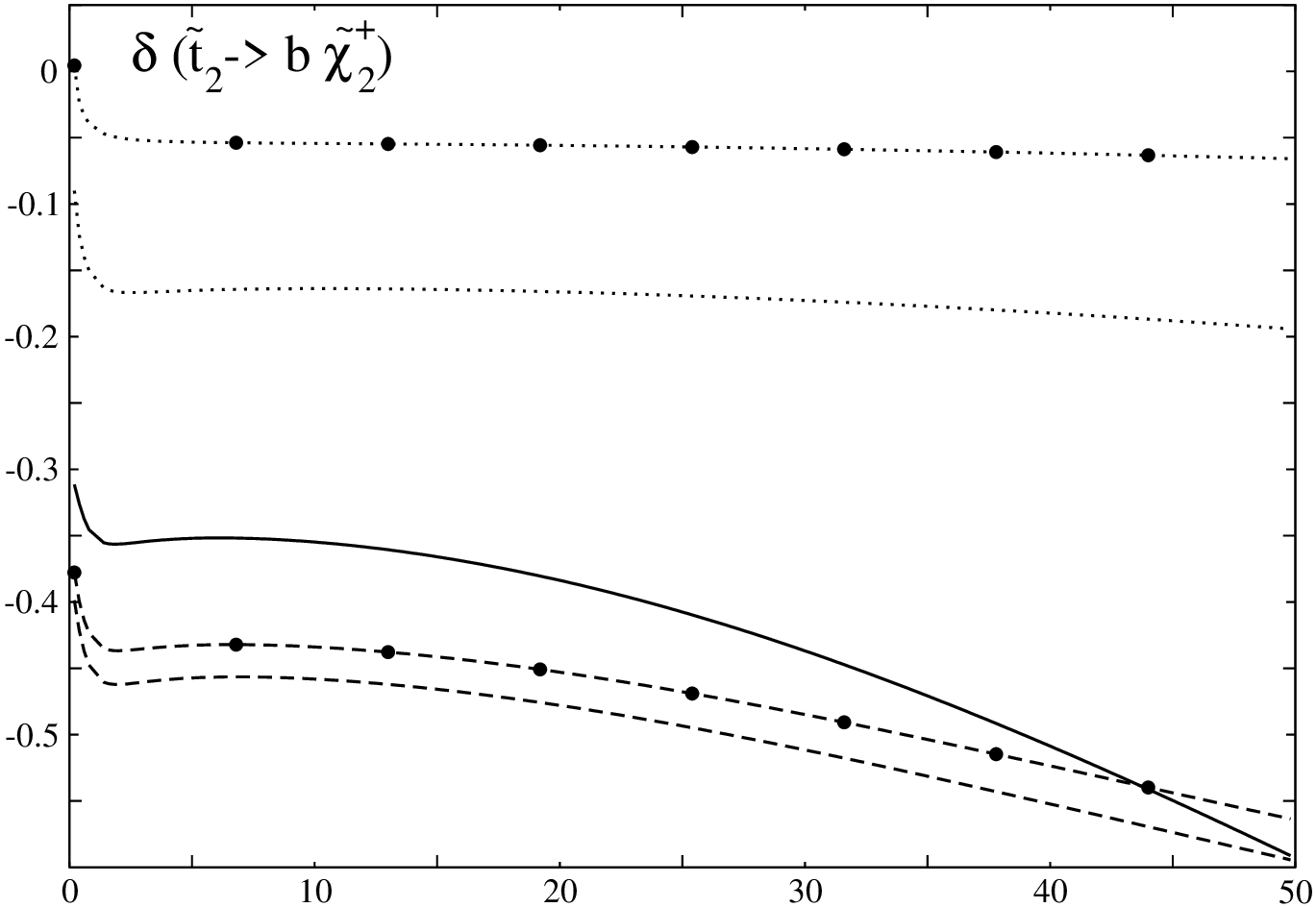}&
\includegraphics*[width=12.5cm]{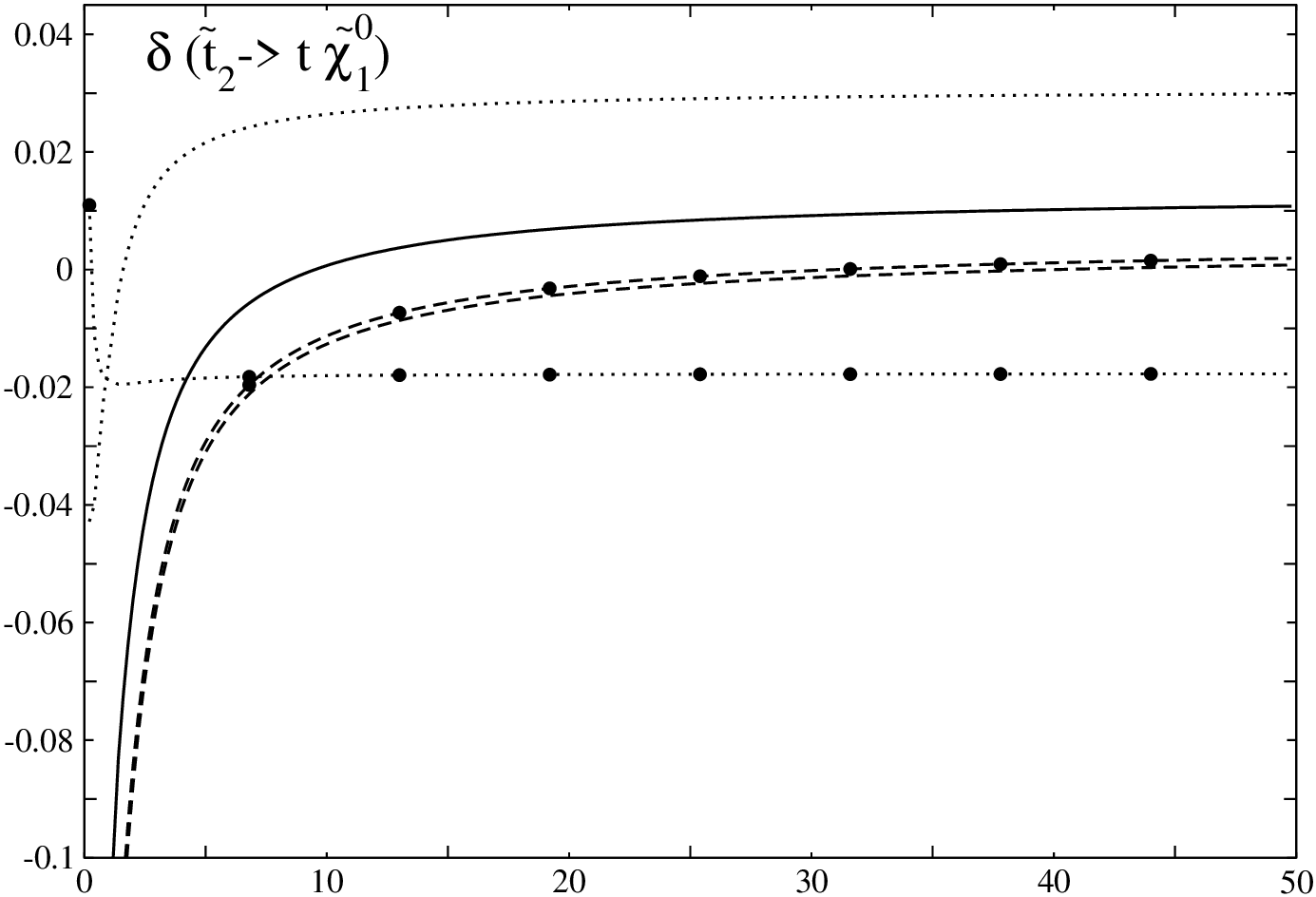}&
\includegraphics*[width=12.5cm]{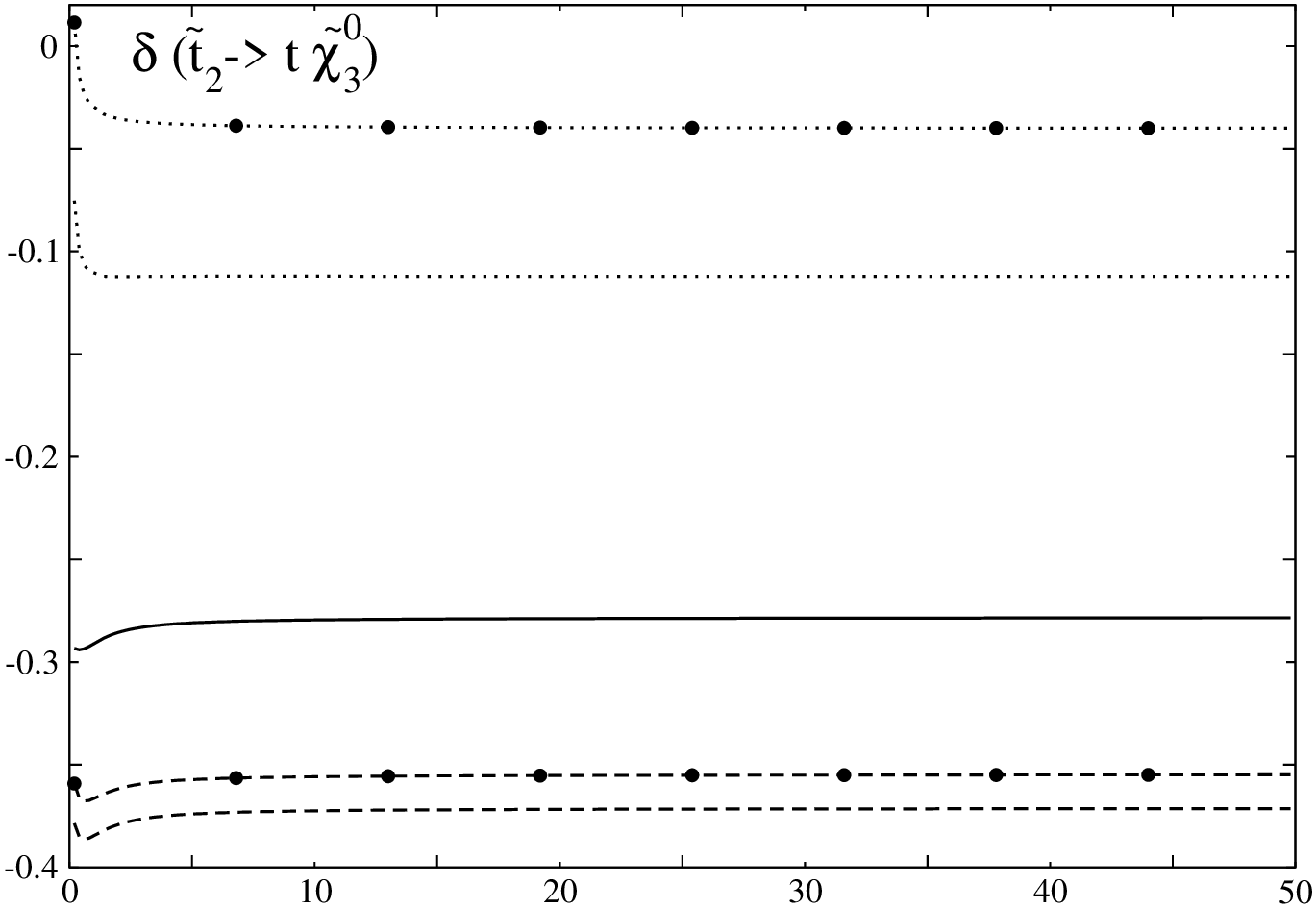}\\
\multicolumn{4}{c}{\huge $\tb$}
\end{tabular}}}
\caption{Relative corrections of the
  squark decays into charginos and neutralinos, as a function of $\tb$. 
  We show the results  
  including only the effective Yukawa couplings,
  eq.~(\ref{eq:hbhbteff1}), and results including the
  $\log\mg/Q$ terms, eq.~(\ref{eq:logterms}) (lines with circles).
  The input parameters are given in eq.~(\ref{eq:SUSYparams}).}
\label{fig:tb}
\end{figure}

Next, we show in Fig.~\ref{fig:tb} the evolution of the different
corrections as a function of $\tb$. 
Here we see large negative corrections,
growing with $\tb$. The origin of the negative corrections is twofold:
on one side the standard QCD running of the quark mass reduces
significantly the Yukawa coupling, and on the other, for positive values
of the higgsino mass parameter $\mu$ the contributions to 
$\Delta m_q$~(\ref{eq:deltamq}) are also positive, decreasing even more the
effective Yukawa couplings~(\ref{eq:hbhbteff1}). However, even after
taking into account these two sources of corrections, still there is a
large \textit{remainder} of relative corrections (up to $-67\%$ at $\tb=50$
for some channels) which
can not be accounted for. After including
the logarithmic terms, eq.~(\ref{eq:logterms}), the situation is quite
different. Now, the effective
description can reproduce quite well the one-loop results, and the
\textit{remainder} corrections (those that can not be described by the
effective couplings) are reduced at a level (in absolute
value) below the $20\%$. Table~\ref{tab:remainder}
  shows the value of the remainder corrections~(\ref{eq:defremainder})
  for all studied channels at $\tb=50$, including the full effective
  description, and including only the Yukawa corrections. We see that in
all channels the corrections are reduced significantly after including
the logarithmic terms.

\begin{table}
\centering\begin{tabular}{||c|c|c||c|c|c||c|c|c||}
\hline
Channel & $\delta^{Yuk.-rem.}$ & $\delta^{rem.}$ &
Channel & $\delta^{Yuk.-rem.}$ & $\delta^{rem.}$ \\ \hline
$\sbottom_1\to t\cmin_1$ & $2.4\%$ & $-1.5\%$&
$\stopp_1\to t\cplus_1$ & $-10\%$ & $-6\%$ \\\hline
$\sbottom_1\to t\cmin_2$ & $-10\%$ & $-3\%$&
$\stopp_1\to t\cplus_2$ & $-29\%$ & $-12\%$ \\\hline
$\sbottom_1\to b\neut_1$ & $-18\%$ & $-8\%$&
$\stopp_1\to t\neut_1$ & $10\%$ & $-0.6\%$ \\\hline
$\sbottom_1\to b\neut_3$ & $-51\%$ & $-18\%$&
$\stopp_1\to t\neut_3$ & $-14\%$ & $-2\%$ \\\hline
$\sbottom_2\to t\cmin_1$ & $-67\%$ & $-19\%$&
$\stopp_2\to t\cplus_1$ & $-6\%$ & $-4\%$ \\\hline
$\sbottom_2\to t\cmin_2$ & $-31\%$ & $-10\%$&
$\stopp_2\to t\cplus_2$ & $-19\%$ & $-7\%$ \\\hline
$\sbottom_2\to b\neut_1$ & $-12\%$ & $-6\%$&
$\stopp_2\to t\neut_1$ & $3\%$ & $-1.8\%$ \\\hline
$\sbottom_2\to b\neut_3$ & $-40\%$ & $-16\%$&
$\stopp_2\to t\neut_3$ & $-11\%$ & $-4\%$ \\\hline
\hline
\end{tabular}
\caption{Value of the \textit{remainder}
  corrections~(\ref{eq:defremainder}) for specific squark partial decay
  widths, including only the effective Yukawa couplings, 
  eq.~(\ref{eq:hbhbteff1}), and results including the
  $\log\mg/Q$ terms, eq.~(\ref{eq:logterms}), for $\tb=50$, and the rest
  of input parameters as given in eq.~(\ref{eq:SUSYparams}).}
\label{tab:remainder}
\end{table}

\begin{figure}
\centering
\rotatebox{90}{\resizebox{20cm}{!}{\begin{tabular}{cccc}
\includegraphics*[width=12.5cm]{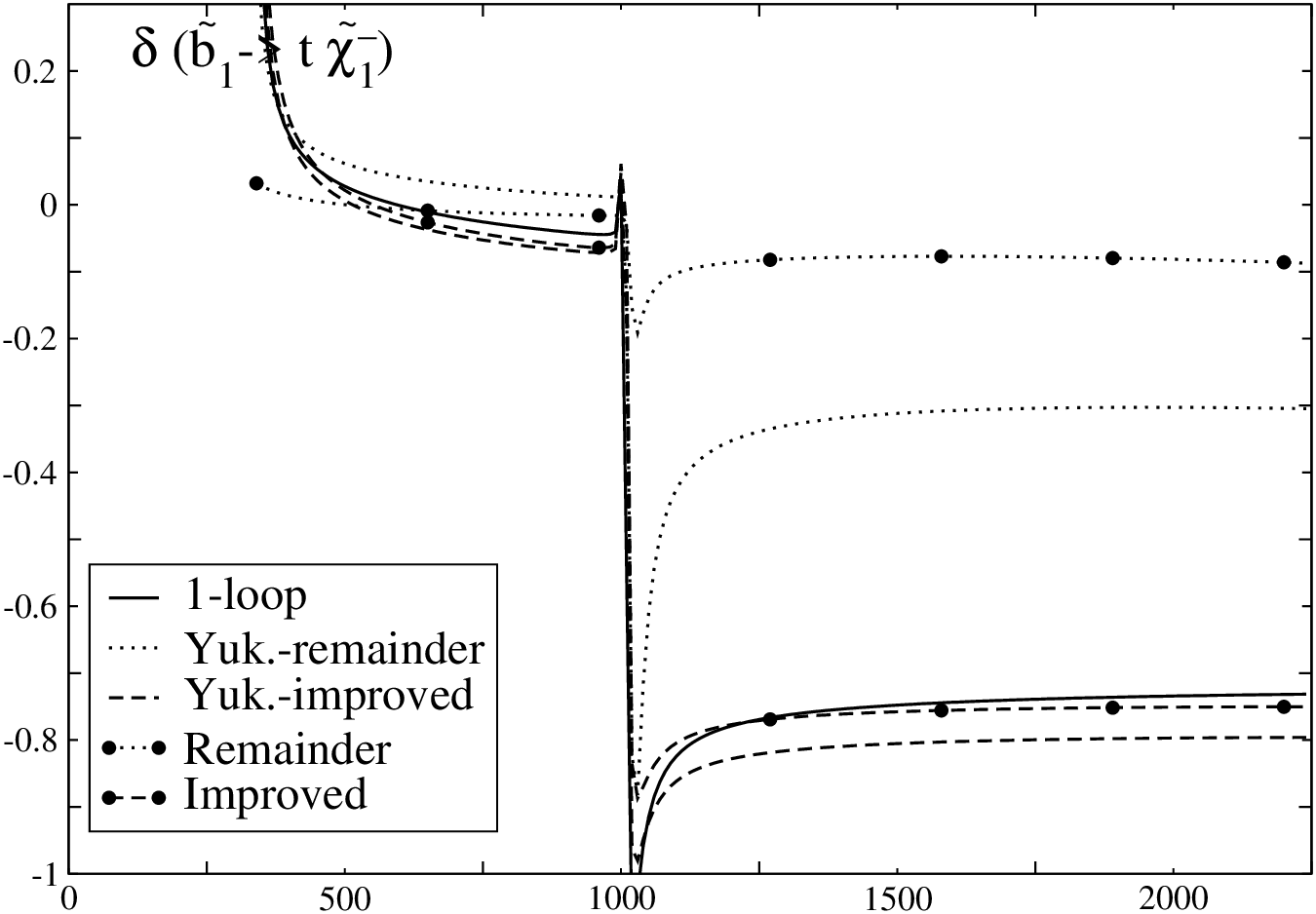}&
\includegraphics*[width=12.5cm]{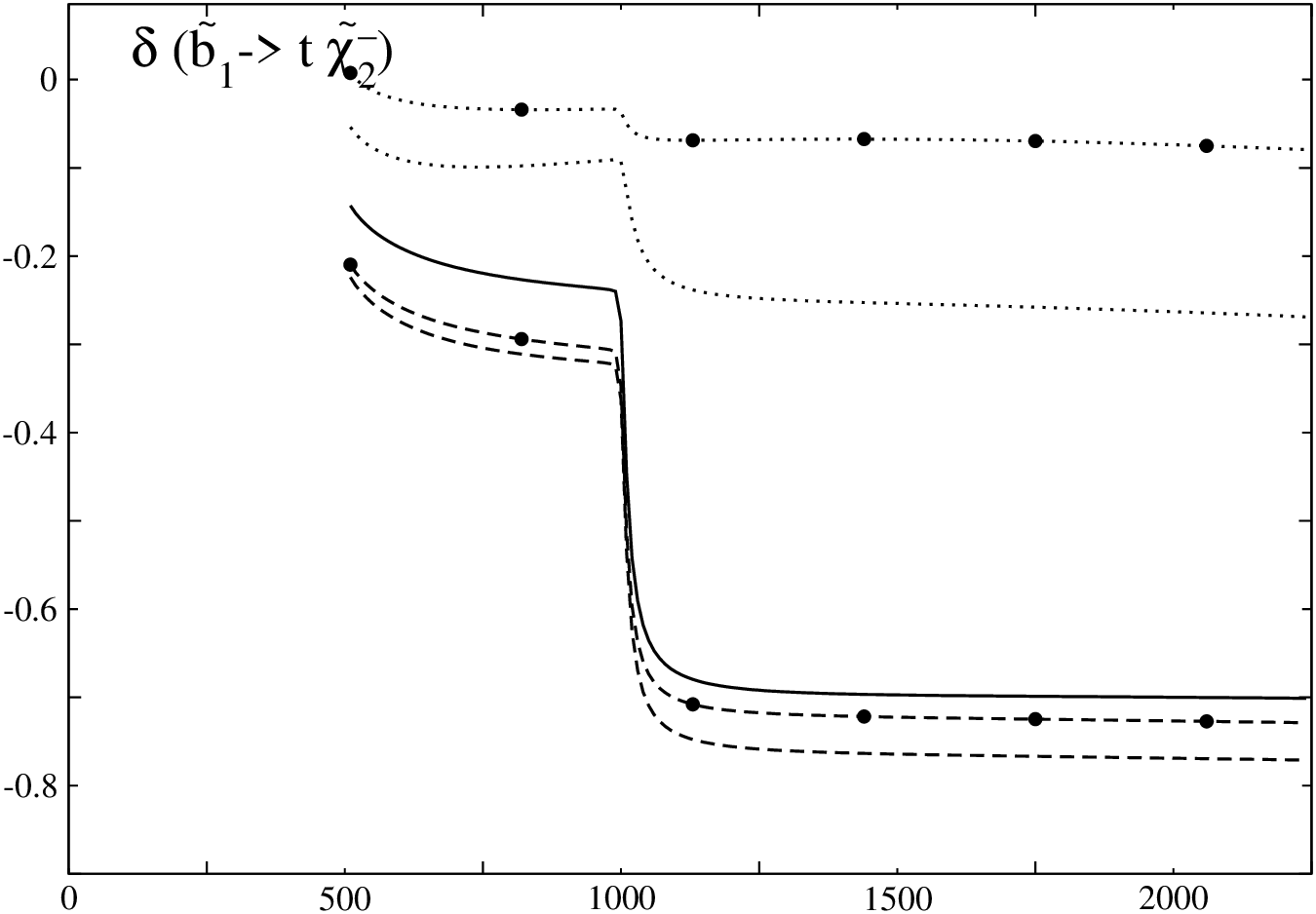}&
\includegraphics*[width=12.5cm]{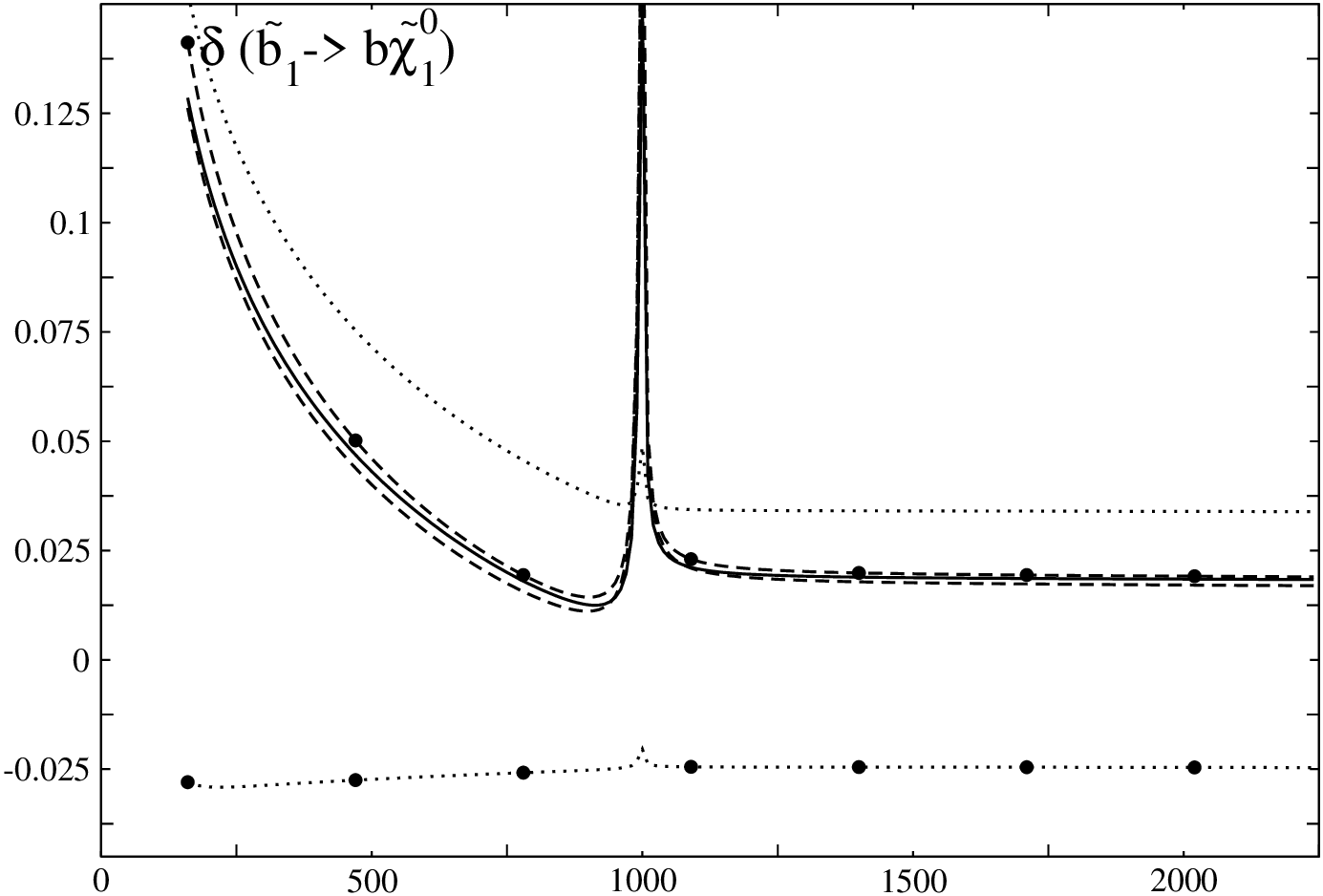}&
\includegraphics*[width=12.5cm]{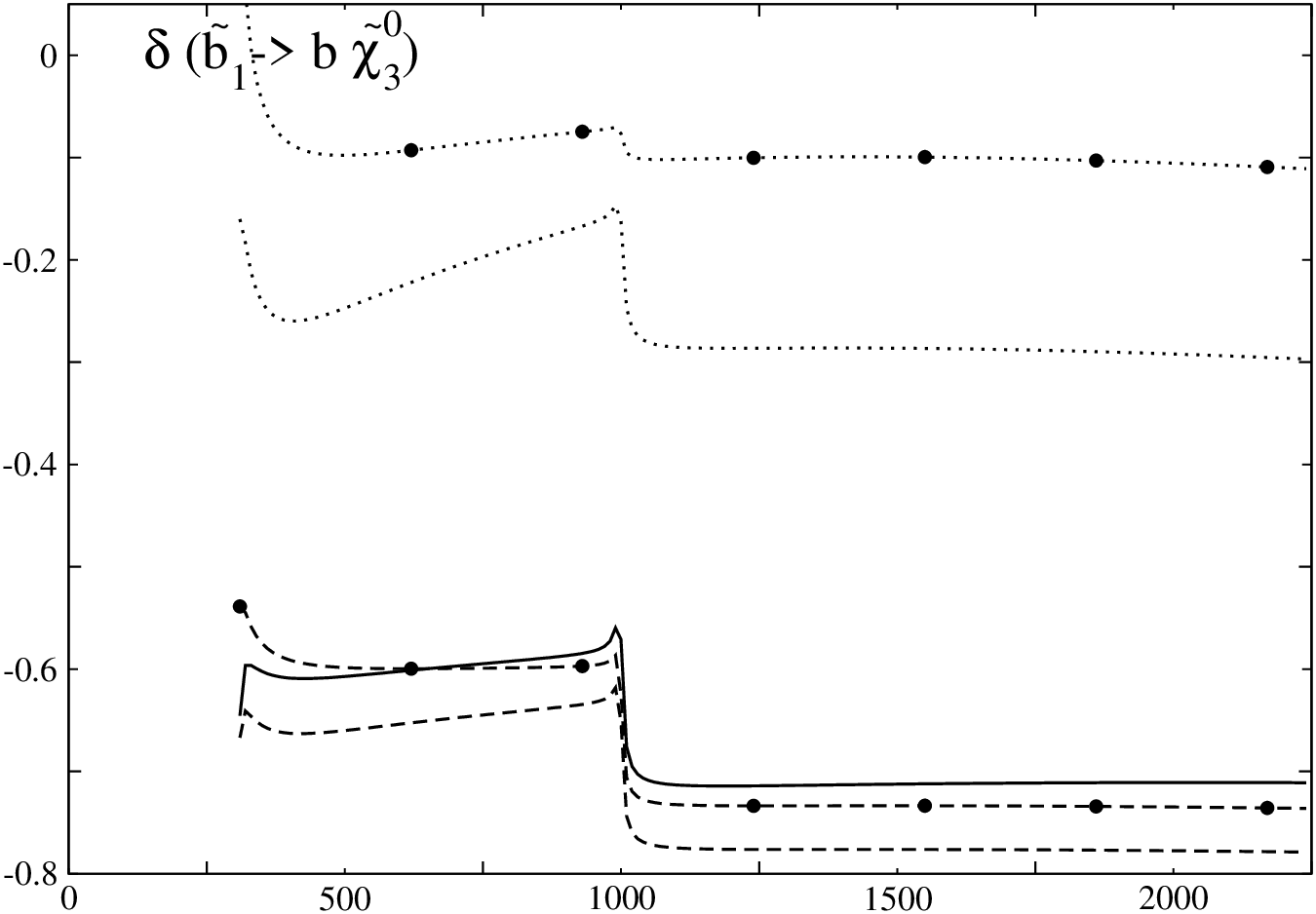}\\
\includegraphics*[width=12.5cm]{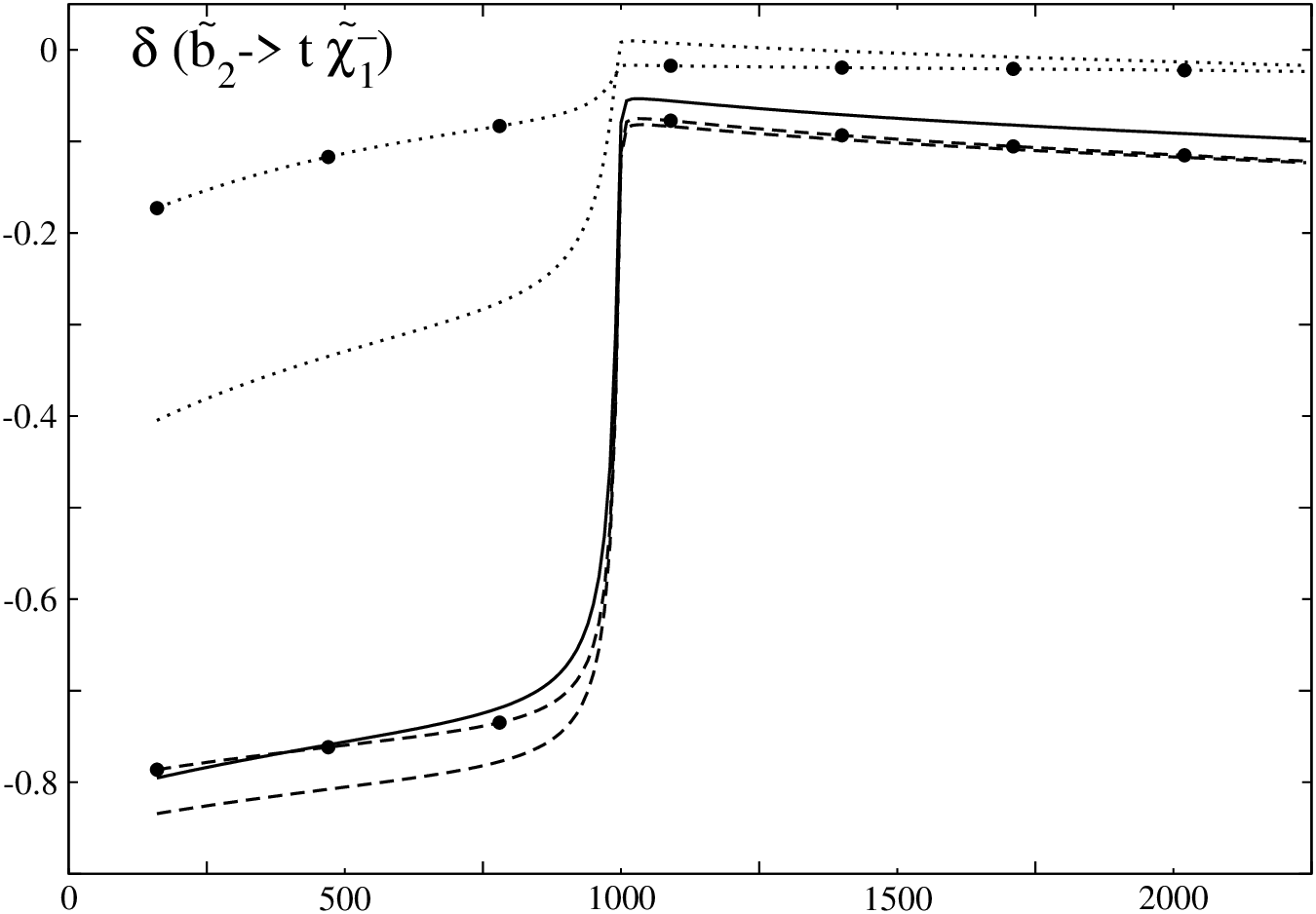}&
\includegraphics*[width=12.5cm]{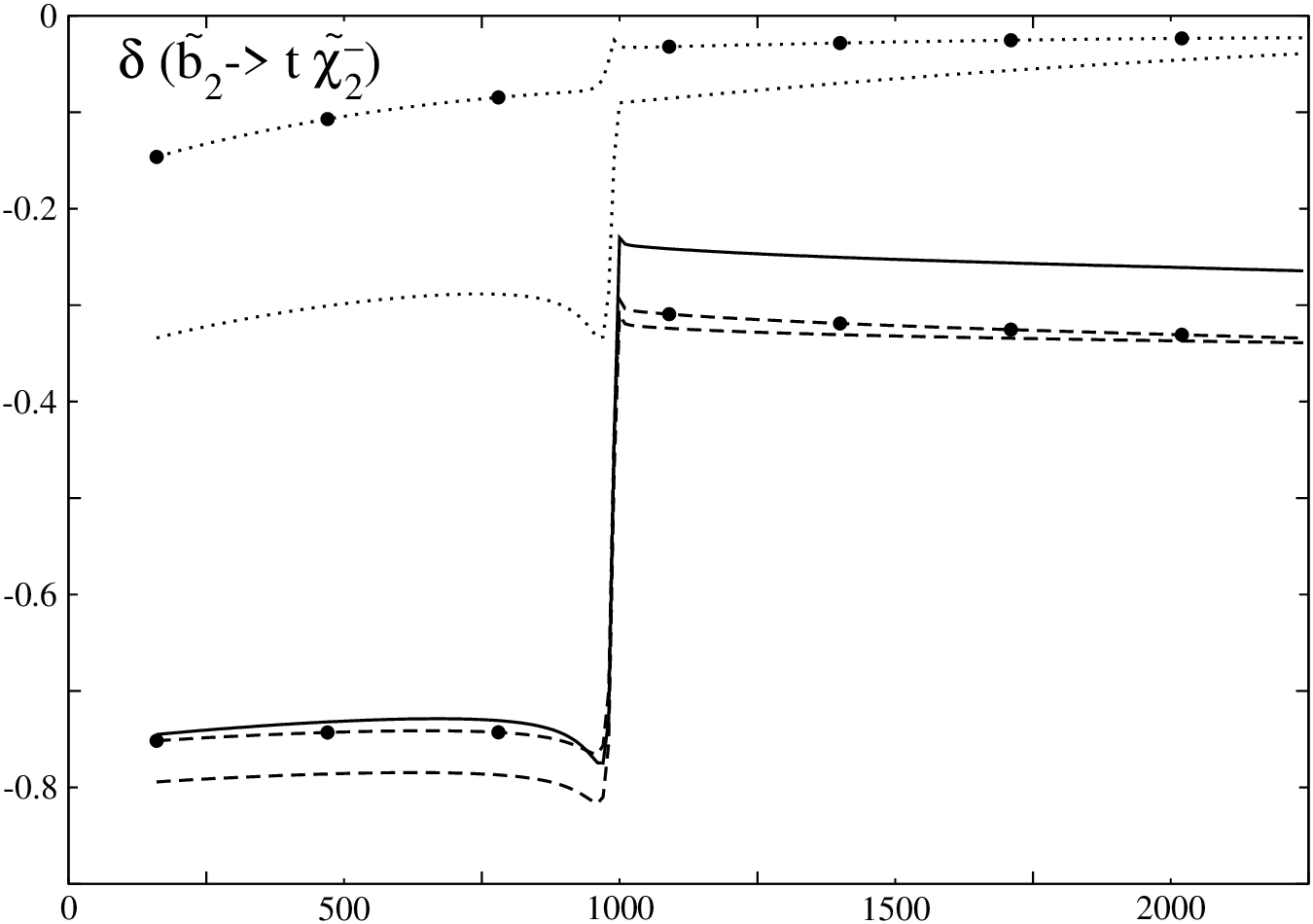}&
\includegraphics*[width=12.5cm]{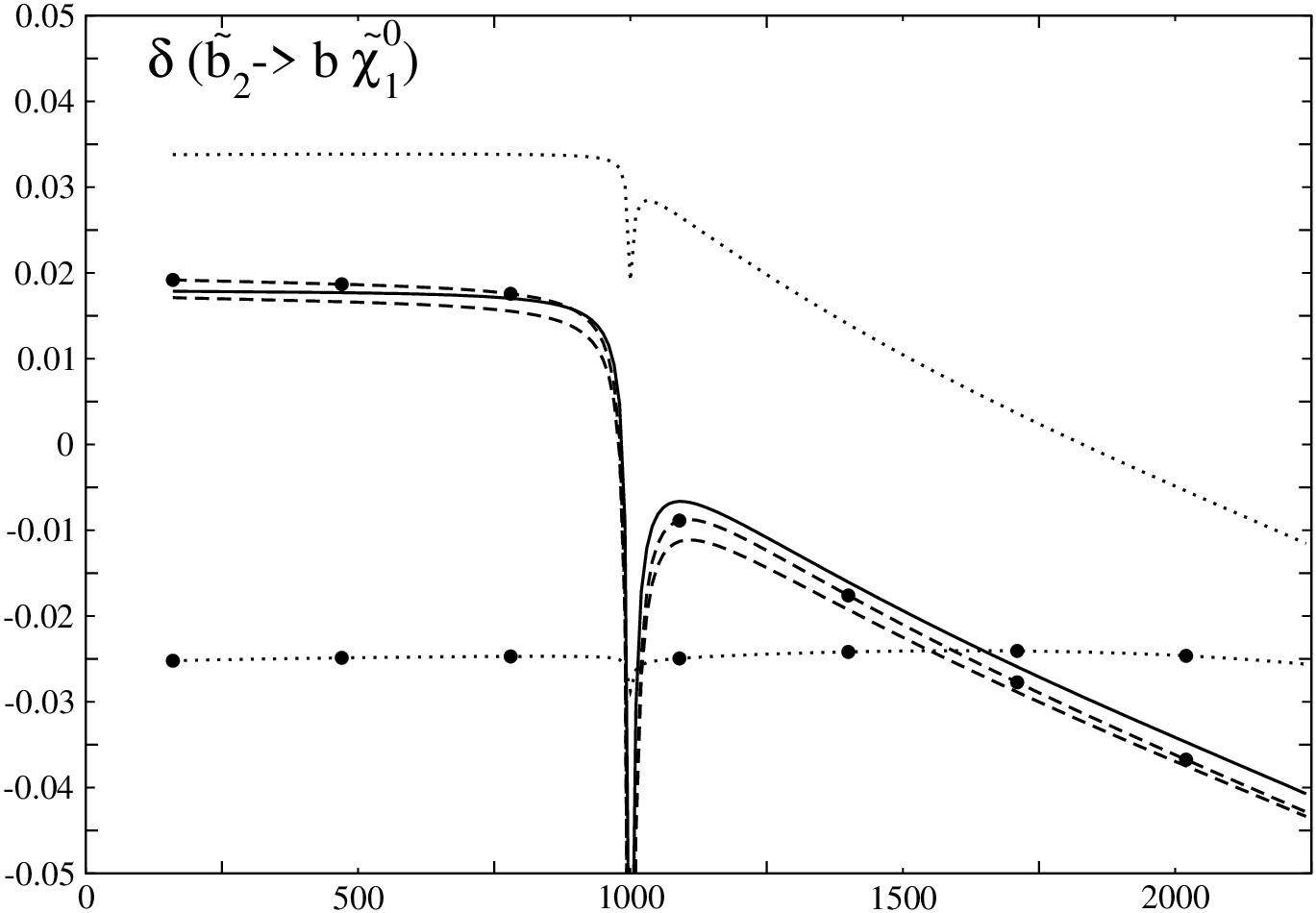}&
\includegraphics*[width=12.5cm]{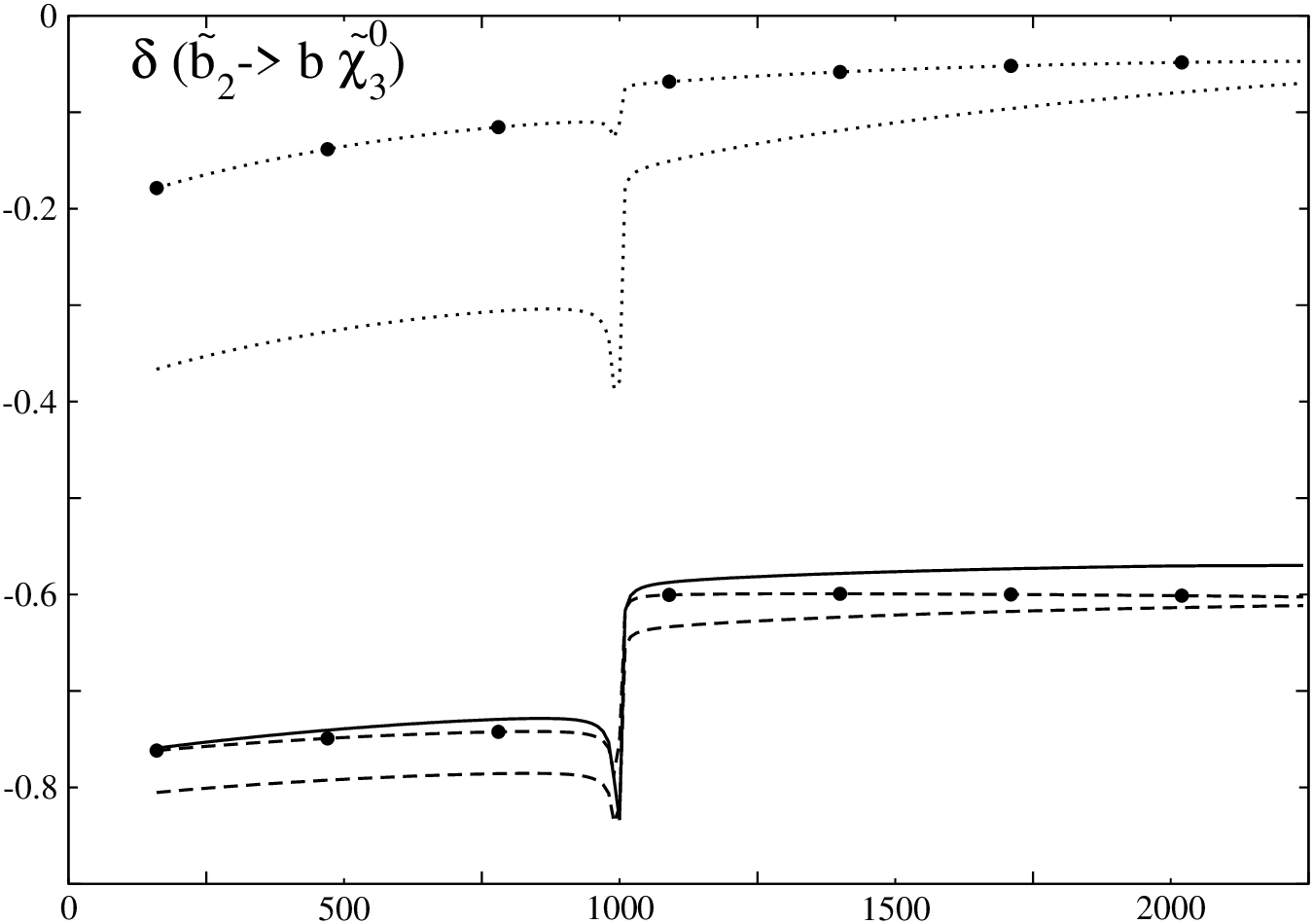}\\
\includegraphics*[width=12.5cm]{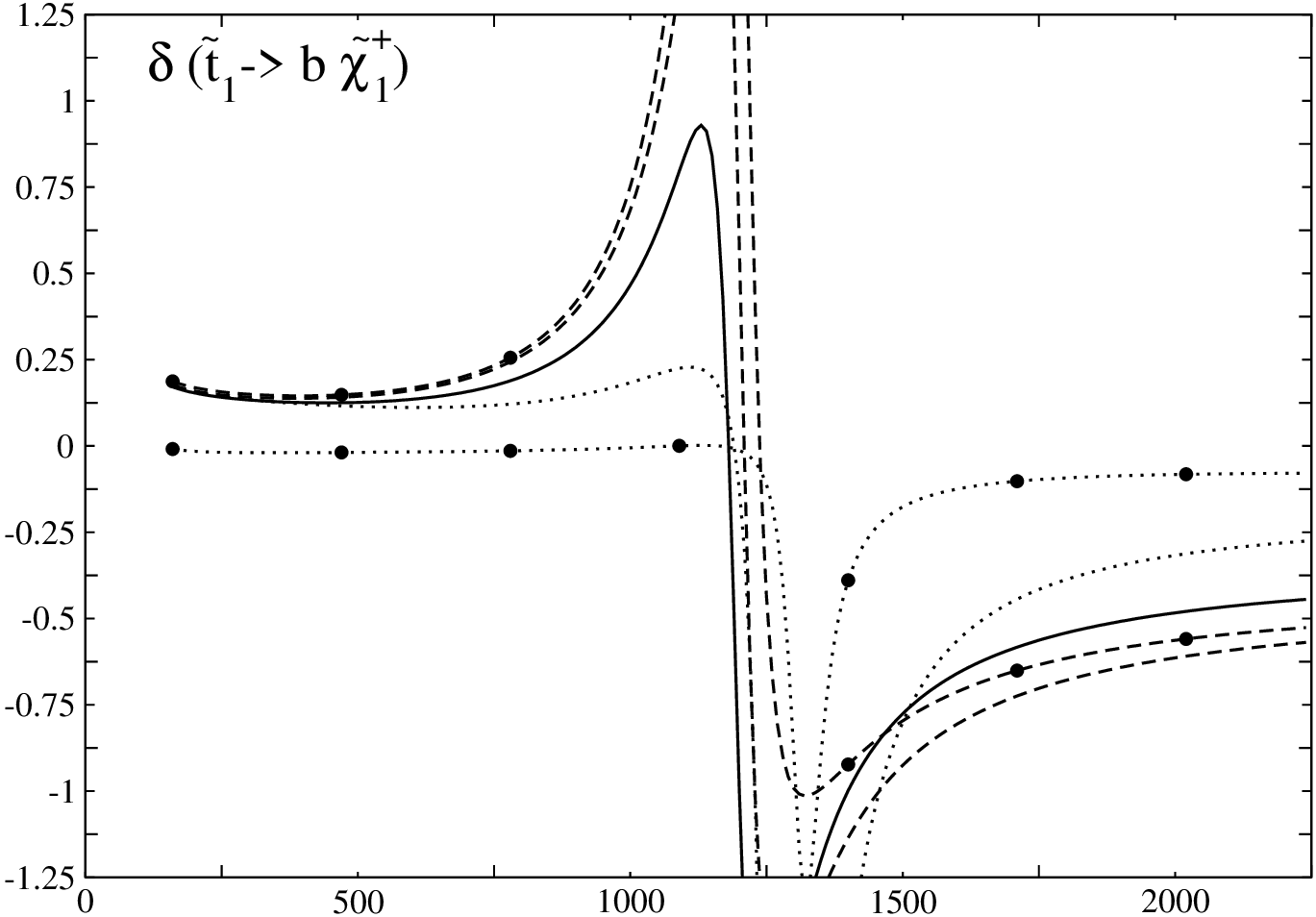}&
\includegraphics*[width=12.5cm]{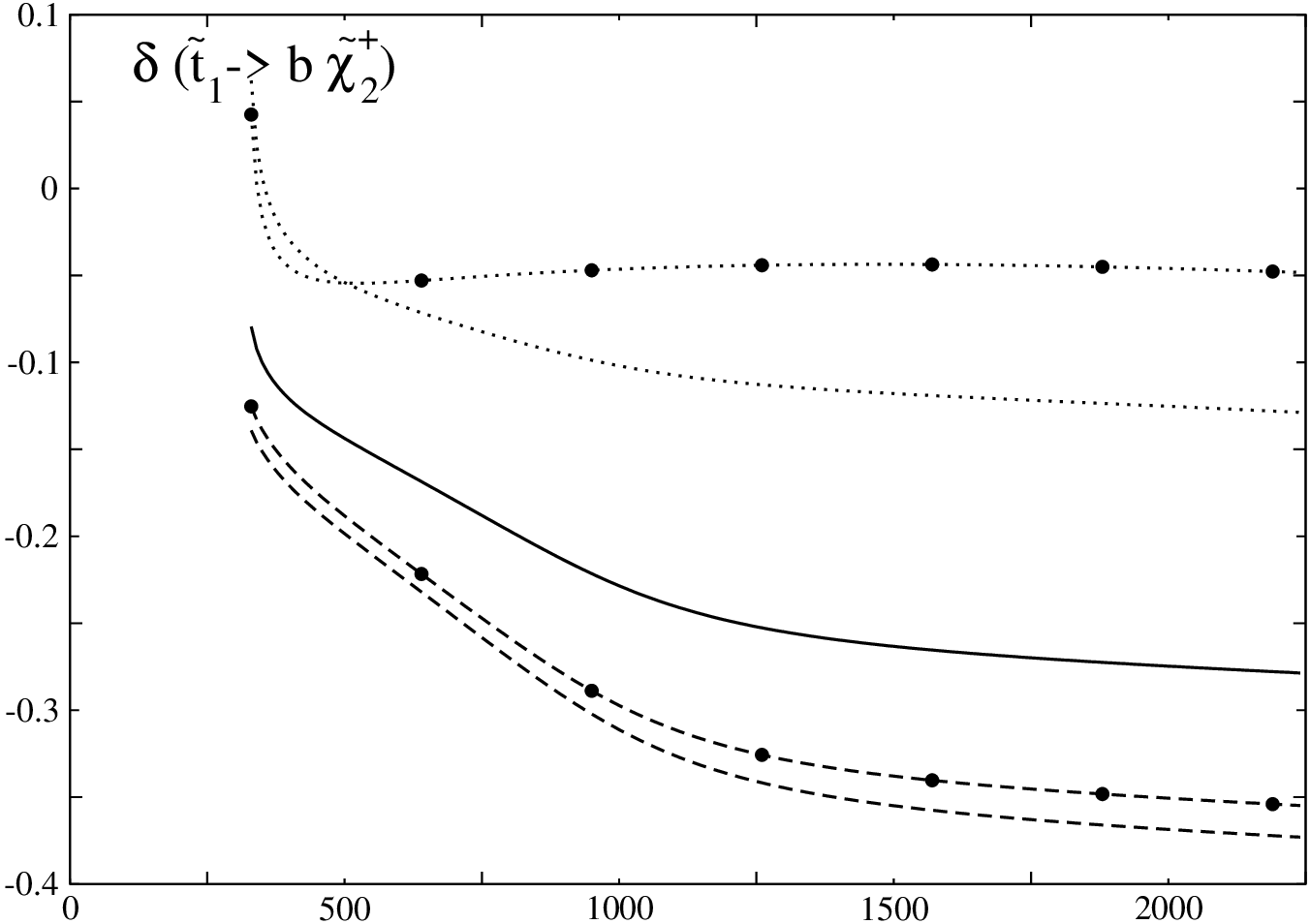}&
\includegraphics*[width=12.5cm]{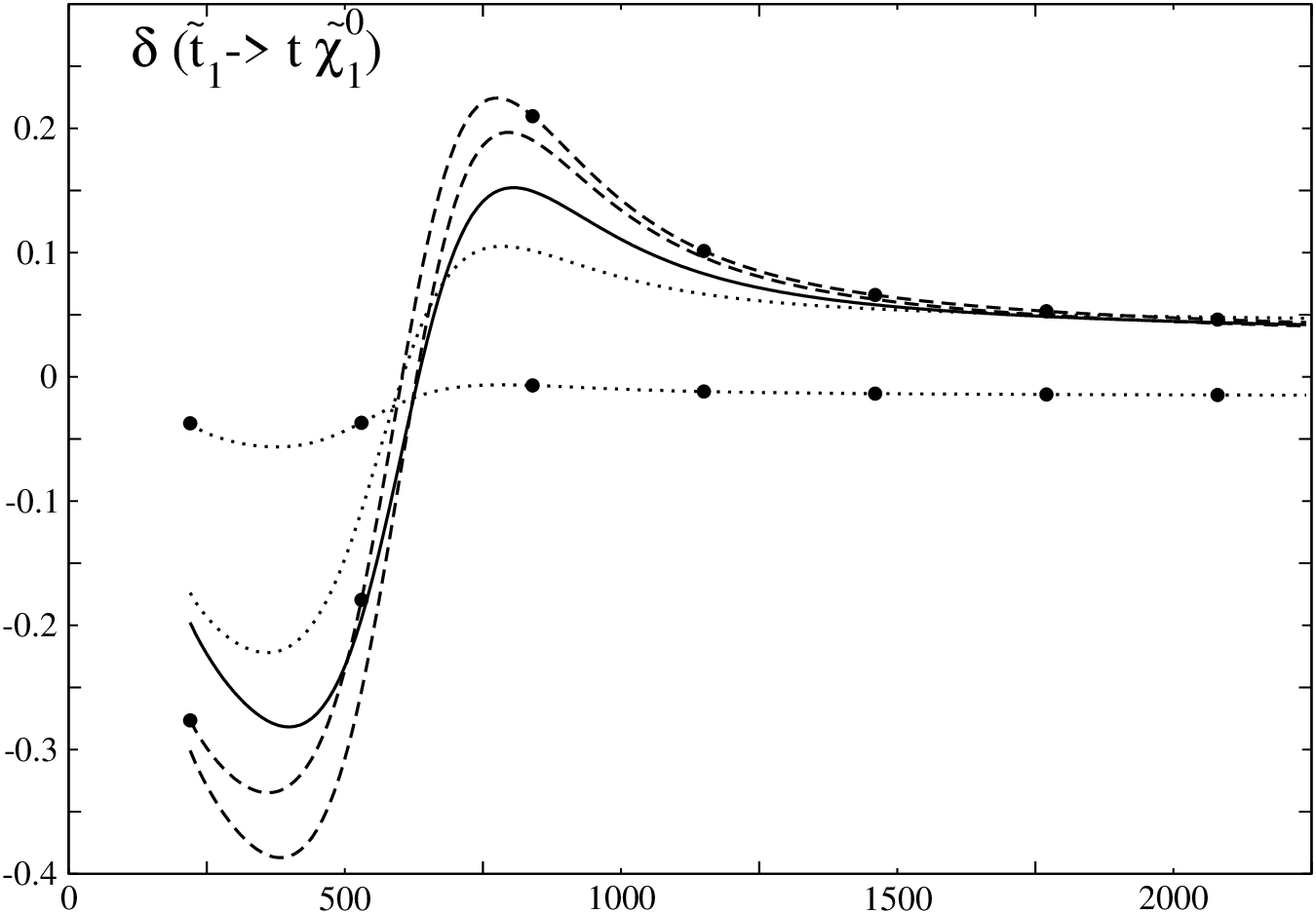}&
\includegraphics*[width=12.5cm]{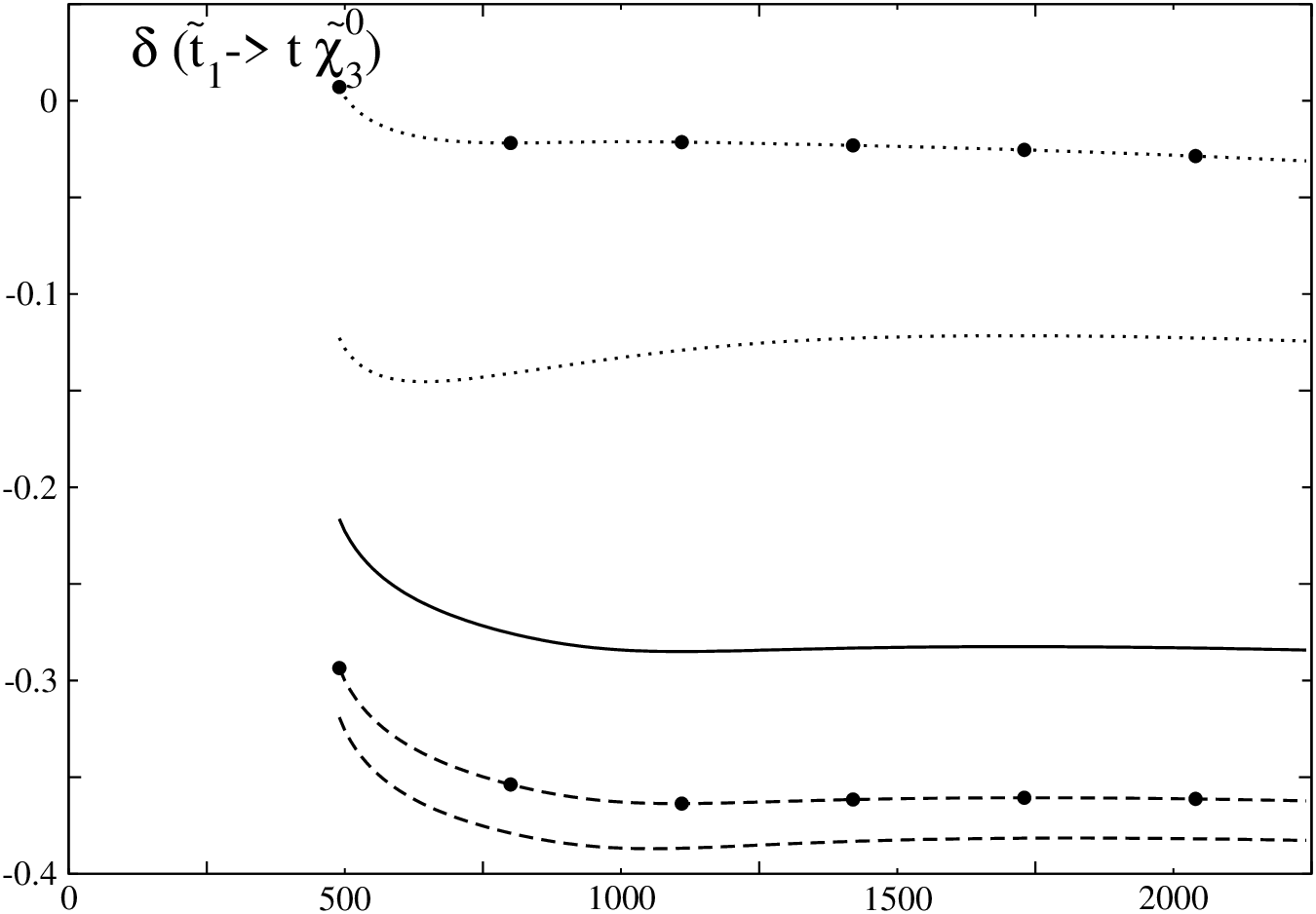}\\
\includegraphics*[width=12.5cm]{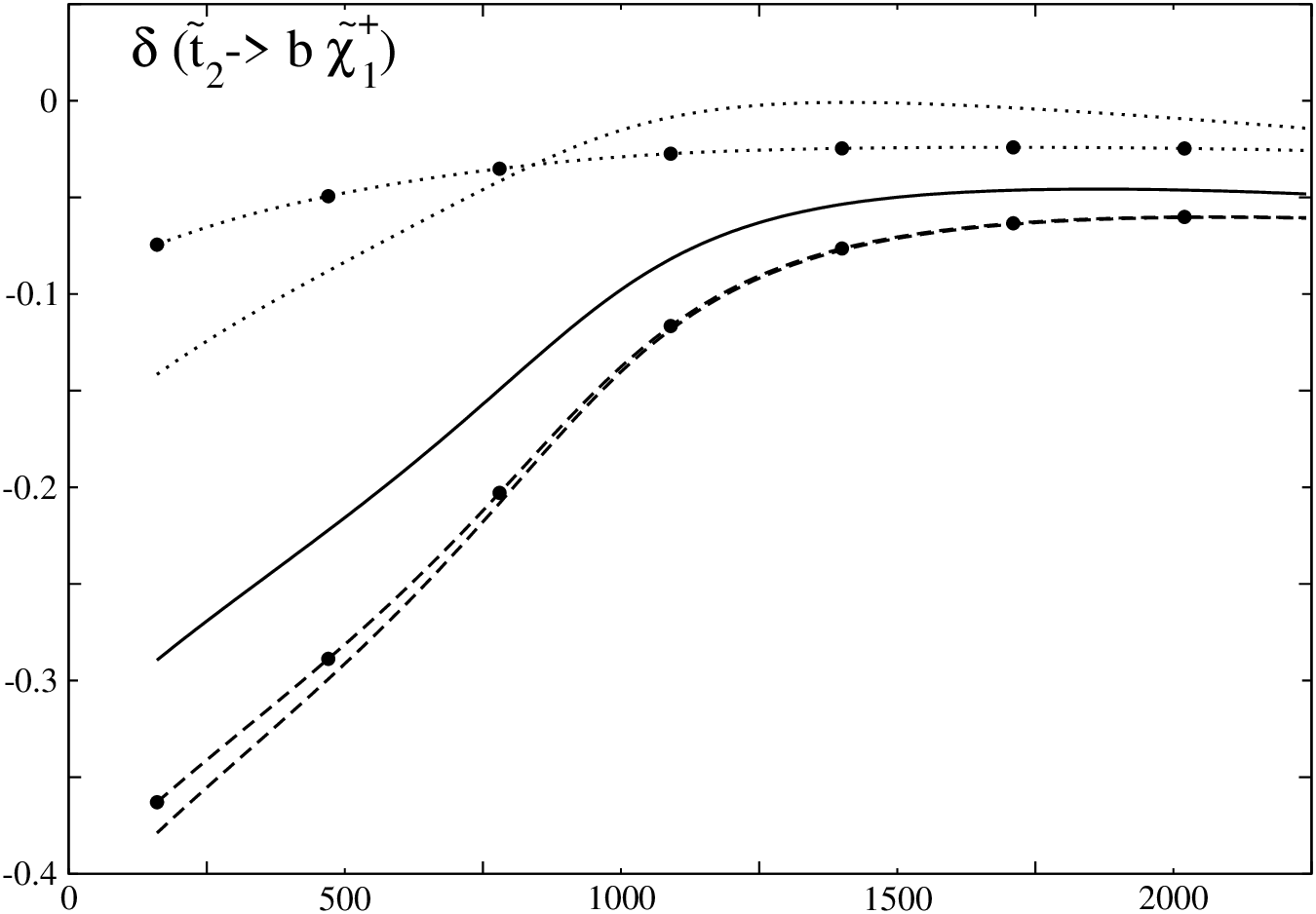}&
\includegraphics*[width=12.5cm]{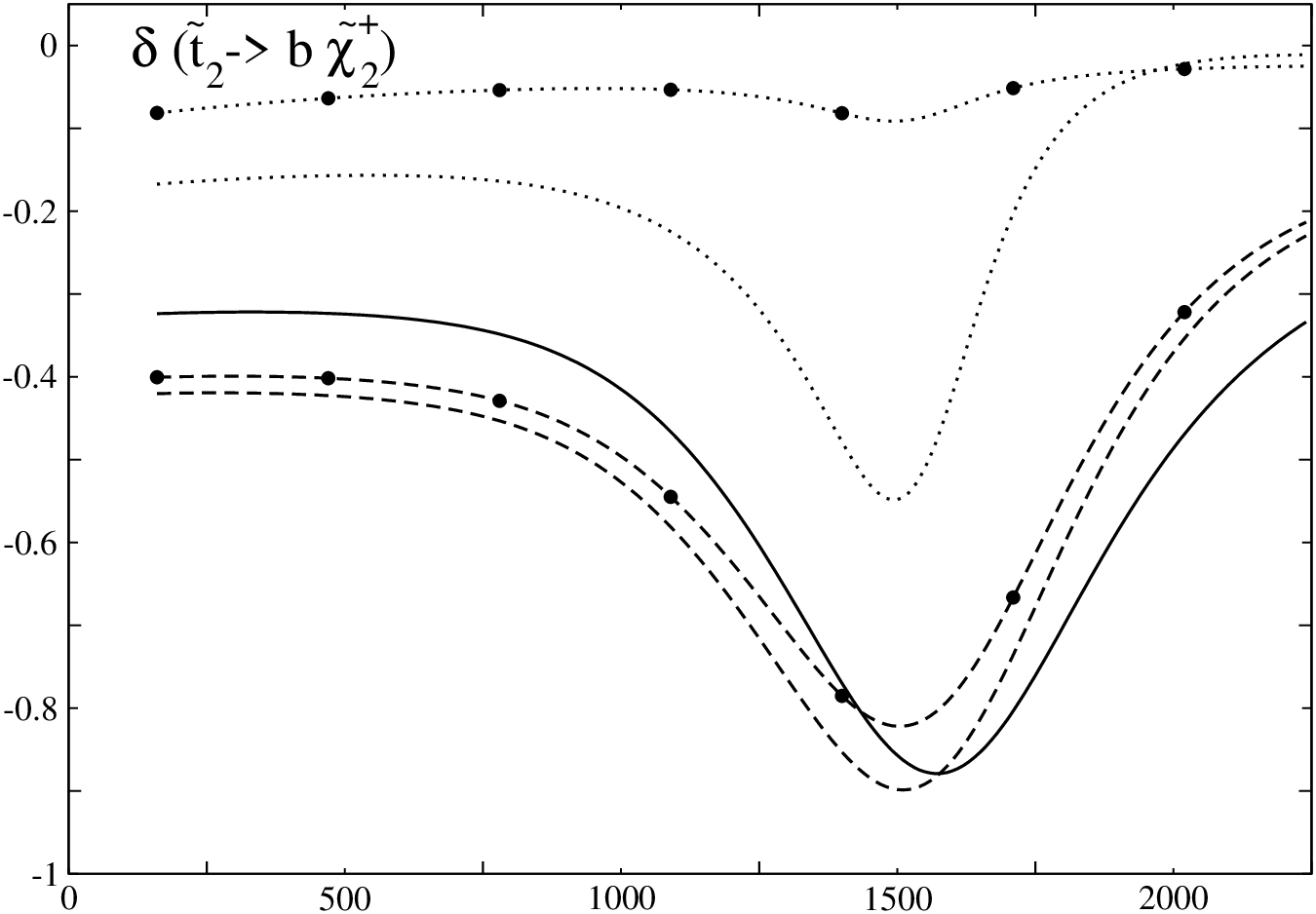}&
\includegraphics*[width=12.5cm]{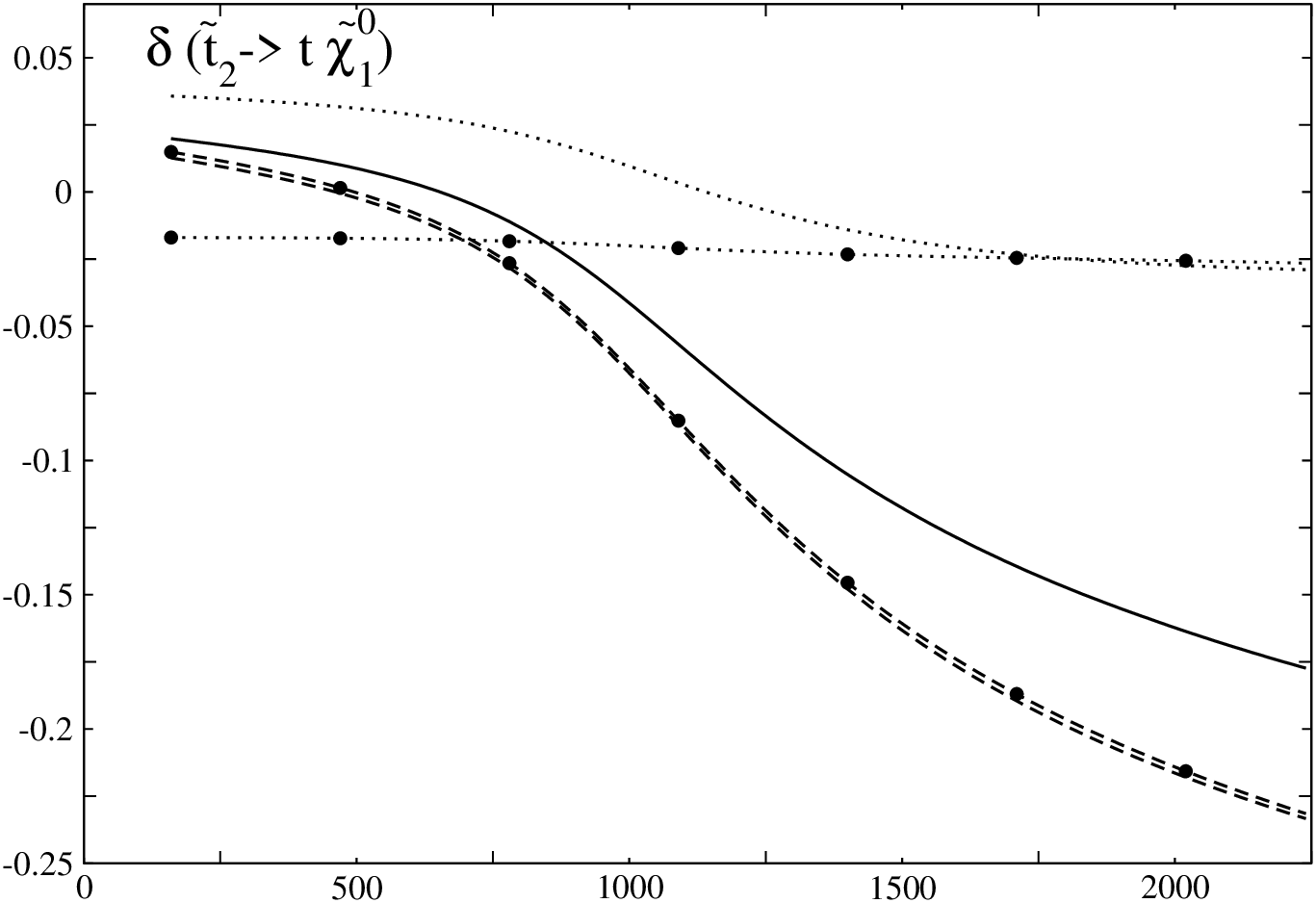}&
\includegraphics*[width=12.5cm]{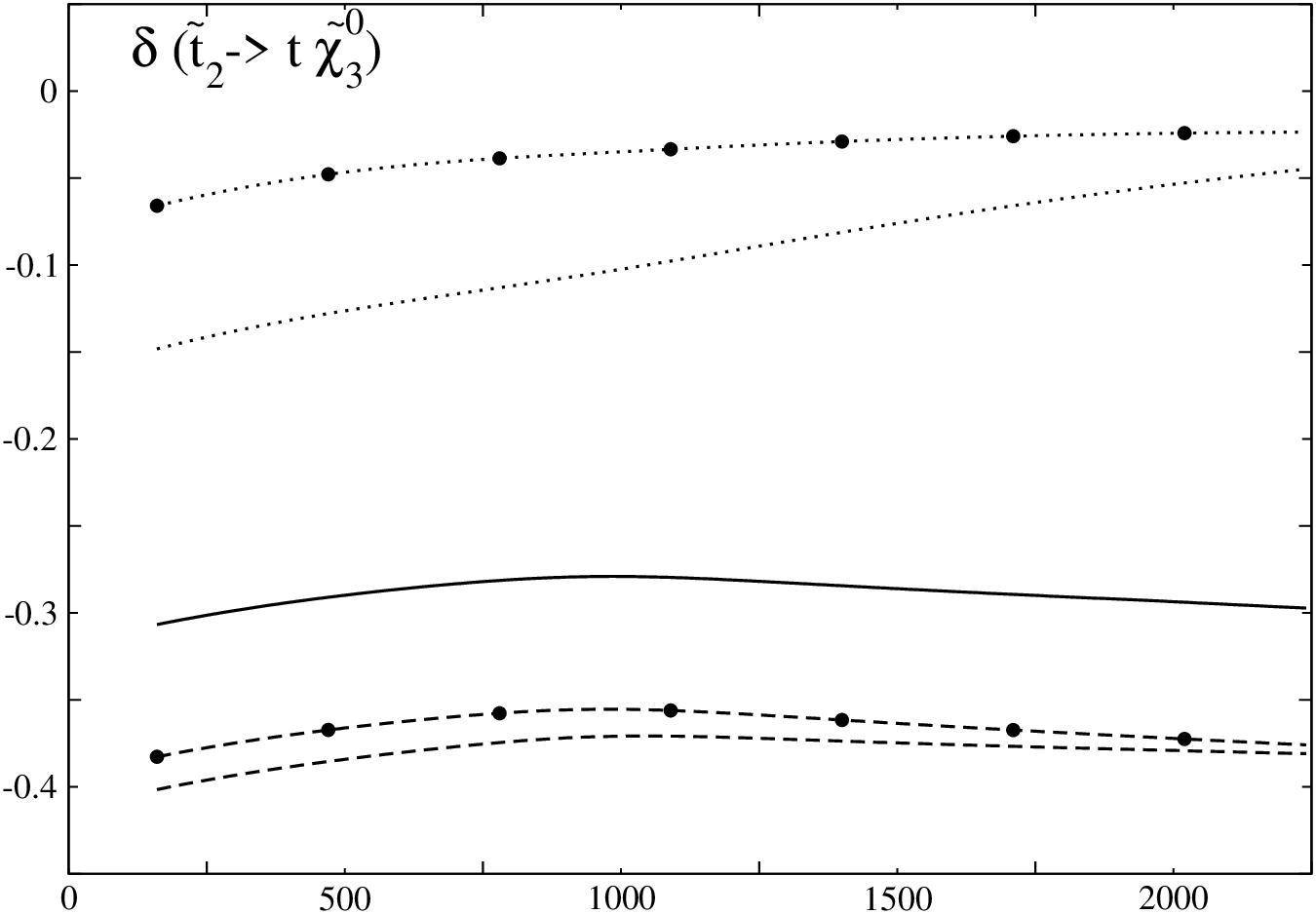}\\
\multicolumn{4}{c}{\huge $\MsfL\ ({\rm GeV})$}
\end{tabular}}}
\caption{Relative corrections of the
  squark decays into charginos and neutralinos, as a function of $\MsfL$. 
  We show the results  
  including only the effective Yukawa couplings,
  eq.~(\ref{eq:hbhbteff1}), and results including the
  $\log\mg/Q$ terms, eq.~(\ref{eq:logterms}) (lines with circles).
  The input parameters are given in eq.~(\ref{eq:SUSYparams}).}
\label{fig:msq}
\end{figure}

Finally Fig.~\ref{fig:msq} shows the evolution of 
the corrections to the partial decay
widths as a function of the $SU(2)_L$ squark mass
parameter $\MsfL$. The abrupt change which is seen at the
middle of the plots corresponds with the situation in which 
$\MsfL\simeq \MsfR$, and the physical states suffer
an abrupt change between left and right chirality. This explains the
difference in value and behaviour of the 
corrections in the regions of $\MsfL$ below and above that
point. 
This is also the situation, shown in Fig.~\ref{fig:stopdeccormsq1}, where
the partial decay widths can become zero, and the one-loop corrections
can become non-perturbative. In all situations the description including
the log-terms provides a better description of the radiative corrections
with a $\delta^{rem.}$~(\ref{eq:defremainder}) much smaller than with
the effective Yukawa couplings~(\ref{eq:hbhbteff1}) alone, and a much
softer variation, meaning that the description of
eq.~(\ref{eq:logterms}) is accurate for all values of the squark
mass. Let us remember, that by changing the SUSY parameter
$\MsfL$, the physical squark masses also change, and that 
since the renormalization scale is taken to be the decaying squark mass,
the log-terms of eq.~(\ref{eq:logterms}) effectively run also with
$\MsfL$ as $~\sim \log\mg/\MsfL$ (for
mostly-left-handed squarks) -- a contribution that can not be described
with the effective Yukawa couplings~(\ref{eq:hbhbteff1}). 

\section{Summary and conclusions}
\label{sec:conclusions}
We have proposed and analyzed an effective description of squark
interactions with charginos and neutralinos in the MSSM. We have applied
it to the partial decay widths of squarks into charginos and neutralinos.
We have compared it with the full one-loop corrections, and have
proposed a way to combine the effective description (which includes
higher order terms) with the complete one-loop description (which
includes all kinetic and mass-effects factors), providing an
\textit{improved} computation, $\Gamma^{imp.}$ (\ref{improved}). The
difference between 
the effective description and the \textit{improved} computation is 
encoded in the \textit{remainder} contribution, $\delta^{rem.}$
eq.~(\ref{eq:defremainder}), which gives a measure of the precision of
the effective description. 

The effective description includes the effective Yukawa
couplings~(\ref{eq:hbhbteff1}), which take into account the resummation
effects~\cite{Coarasa:1996qa,Carena:1999py,Guasch:2001wv,Guasch:2003cv}.
Note that the computation of the threshold
corrections~(\ref{eq:deltamq}) includes only the ($\tb$, $\cot\beta$)
proportional terms, since the terms that would be proportional to the
trilinear couplings ($A_b$, $A_t$) are actually
subleading~\cite{Guasch:2003cv}. This description produces large
\textit{remainder} $\delta^{Yuk.-rem.}$ corrections, and does not reproduce
the behaviour of the one-loop corrections as a function of several
parameters -- notably it is missing a $\log\mg$ term. Therefore, we
conclude that it does not reproduce
satisfactorily the one-loop corrections, and it is not a good
approximation.

We have computed the missing $\log\mg$ terms with the help of the
renormalization group -- eq.~(\ref{eq:logterms}) --, and found agreement
between the renormalization 
group analysis and the large mass expansion of the one-loop result
(\ref{eq:logterm1loop}). After including also the $\log\mg$-terms of
eq.~(\ref{eq:logterms}), the effective description produces a reasonable
approximation to the radiative-corrected partial decay widths of squarks
into charginos and neutralinos, as shown by a small absolute value of the
\textit{remainder} contributions $\delta^{rem.}$, and by a nearly-flat
behaviour of the corrections as a function of different parameters.

The origin of the logarithmic terms can be explained in different
(complementary) ways, depending on the kind of approximation that we
take. First of all, from the fundamental point of view, they are
\textit{non-decoupling} terms that appear due to the 
supersymmetry-breaking. Since we are testing SUSY relations (equality of
the gauge/Yukawa couplings to the gaugino/higgsino couplings), and SUSY
is broken (by the term $\mg$ among others), we have to find some effect
that tells us about the breaking of SUSY at that scale -- e.g. a
$\log\mg$-term. Second, from a fixed-order one-loop description point of
view, the SUSY relation appears because the UV-divergences of the loops
containing gluinos cancel with the UV-divergences of the loops
containing gluons, and the log-terms that accompany those loops combine
between them -- producing a $\log\mg$-term. And third, from a
renormalization group (and effective theory) point of view, the
gaugino/higgsino couplings run different than the gauge/Yukawa
couplings in the region where SUSY is broken, that is, for scales below
$\mg$. The running of the gaugino/higgsino couplings from the scale
$\mg$ to the chosen renormalization scale produces
$\log\mg/Q$-terms. Since the Yukawa couplings already include some QCD
running, whereas the EW-gauge couplings do not, the relation between the
gauge/gaugino and Yukawa/higgsino couplings has some
differences~(\ref{eq:logterms}). 

The presence of the non-decoupling
  $\log\mg$-terms implies a deviation of the equality between the
  higgsino/gaugino and Higgs/gauge couplings predicted by exact
  SUSY. This deviation is important, and has to be
  taken into account in the experimental measurement of SUSY
  relations. At the same time, it gives us access
  to information about heavy particles that can not be directly produced
  at the LHC/ILC. For these reasons it is important
  to include these effects in the computation for the predictions of
  squark observables at the LHC and the ILC.
The effective description of squark/chargino/neutra\-lino couplings given
by eqs.~(\ref{eq:hbhbteff1}), (\ref{eq:deltamq}), (\ref{eq:logterms}),
is simple to write, and to introduce in computer codes, it costs little
computational power, and provides a reasonable description for squark
decays into charginos and neutralinos, so it can be used in monte-carlo
generators and other computer programs that provide predictions for the
LHC and the ILC to improve their accuracy at a minimum cost. 

\section*{Acknowledgements}
J.G. and R.S.F. have been supported in part by MEC and FEDER under project 
FPA2007-66665C02-02, J.G. also by DURSI Generalitat de Catalunya under project
2005SGR00564; 
S.P. by a \textit{Ram{\'o}n y Cajal} contract from MEC (Spain) (PDRYC-2006-000930) and partially
by CICYT (FPA2006-2315) and DGIID-DGA (2008-E24/2);
R.S.F. also by a MEC FPI grant (BES-2005-8861).
This work has also been supported by the Spanish Consolider-Ingenio 2010
Programme CPAN (CSD2007-00042).
J.G. wishes to thank the hospitality of the Universidad de Zaragoza,
and S.P. wishes to thank the hospitality
of the Universitat de Barcelona, where part of this work was done.

\providecommand{\href}[2]{#2}

\end{document}